\documentclass{article}
\usepackage[a4paper,left=2cm,top=1.5cm,bottom=1.5cm,right=2cm]{geometry}
\usepackage[utf8]{inputenc}

\usepackage{color}
\usepackage{xcolor}
\usepackage{hyperref}
\definecolor{linkcolour}{rgb}{0,0.2,0.6}
\hypersetup{colorlinks,breaklinks,urlcolor=linkcolour, linkcolor=linkcolour, citecolor=blue}
\usepackage[onehalfspacing]{setspace}
\usepackage{graphicx}
\usepackage{caption}
\usepackage{subcaption}
\usepackage{multirow}
\usepackage{setspace} 
\usepackage{placeins}

\usepackage{authblk}

\makeatletter         
\renewcommand\maketitle{
{
\begin{center}
{\huge \setstretch{1.2}\@title \par}
\vspace{1cm}
{\large \textbf{The ECFA Early-Career Researcher (ECR) Panel}}
\\[1cm]
{\large \@date}
\\[1cm]
\begin{minipage}{0.82\textwidth}
\normalsize The European Committee for Future Accelerators (ECFA) Early-Career Researchers (ECR) Panel was invited by the ECFA Detector R\&D Roadmap conveners to collect feedback from the European ECR community.
A working group within the ECFA ECR panel held a Townhall Meeting to get first input, and then designed and broadly circulated a detailed survey to gather feedback from the larger ECR community.
A total of 473 responses to this survey were received, providing a useful overview of the experiences of ECRs in instrumentation training and related topics.
This report summarises the feedback received, and is intended to serve as an input to the ECFA Detector R\&D Roadmap process.
\end{minipage}
\end{center}
\vspace{1.5cm}
\begin{flushleft}
{The ECFA Early-Career Researcher (ECR) Panel: \href{mailto:ecfa-ecr-organisers@cern.ch}{ecfa-ecr-organisers@cern.ch}\\[0.5cm]\small  \@author}
\end{flushleft}}}
\makeatother

\RequirePackage[style=numeric,backend=biber,sorting=none]{biblatex}
\newcommand\Figure[1]{Figure~\ref{fig:#1}}
\newcommand\Table[1]{Table~\ref{tab:#1}}

\begin{document}

\title{Results of the 2021 ECFA Early-Career Researcher Survey on Training in Instrumentation}

\date{\today}


\author[1]{Anamika~Aggarwal}
\author[2]{Chiara~Amendola}
\author[3]{Liliana~Apolinario}
\author[*,4]{Jan-Hendrik~Arling}
\author[5]{Adi~Ashkenazi}
\author[6]{Kamil~Augsten}
\author[7]{Julien~Baglio}
\author[8]{Evelin~Bakos}
\author[5]{Liron~Barak}
\author[3]{Diogo~Bastos}
\author[9]{Bugra~Bilin}
\author[10]{Silvia~Biondi}
\author[11]{Neven~Blaskovic~Kraljevic}
\author[7]{Lydia~Brenner}
\author[12]{Francesco~Brizioli}
\author[13]{Antoine~Camper}
\author[14]{Alessandra~Camplani}
\author[15]{Xabier~Cid~Vidal}
\author[16]{Hüseyin~Dag}
\author[17]{Flavia~de~Almeida~Dias}
\author[18]{Eleonora~Diociaiuti}
\author[19]{Lennart~van~Doremalen}
\author[*,20]{Katherine~Dunne}
\author[21]{Filip~Erhardt}
\author[22]{Pedro~Fernández~Manteca}
\author[23]{Andrei~Alexandru~Geanta}
\author[23]{Stefan~Alexandru~Ghinescu}
\author[7]{Loukas~Gouskos}
\author[24]{Andrej~Herzan}
\author[25]{Viktoria~Hinger}
\author[26]{Bojan~Hiti}
\author[*,27]{Armin~Ilg}
\author[28]{Gianluca~Inguglia}
\author[*,29]{Adrián~Irles}
\author[4]{Hendrik~Jansen}
\author[30]{Kateřina~Jarkovská}
\author[31]{Lucia~Keszeghova}
\author[32]{Henning~Kirschenmann}
\author[33]{Sotiroulla~Konstantinou}
\author[34]{Magdalena~Kuich}
\author[35]{Neelam~Kumari}
\author[6]{Katarína~Křížková~Gajdošová}
\author[36]{Aleksandra~Lelek}
\author[37]{Jeanette~Lorenz}
\author[3]{Ana~Luisa~Carvalho}
\author[38]{Jakub~Malczewski}
\author[18]{Giada~Mancini}
\author[37]{Alexander~Mann}
\author[39]{Laura~Martikainen}
\author[40]{Émilie~Maurice}
\author[41]{Seán~Mee}
\author[*,42]{Predrag~Milenovic}
\author[43]{Vukasin~Milosevic}
\author[14]{Zuzana~Moravcova}
\author[44]{Laura~Moreno~Valero}
\author[45]{Louis~Moureaux}
\author[46]{Heikki~Mäntysaari}
\author[47]{Nikiforos~Nikiforou}
\author[4]{Younes~Otarid}
\author[7]{Alex~Pearce}
\author[7]{Michael~Pitt}
\author[23]{Vlad-Mihai~Placinta}
\author[48]{Giulia~Ripellino}
\author[49]{Bryn~Roberts}
\author[50]{Luka~Šantelj}
\author[*,51]{Steven~Schramm}
\author[*,52]{Mariana~Shopova}
\author[53]{Kirill~Skovpen}
\author[50]{Aleks~Smolkovič}
\author[54]{Gamze~Sokmen}
\author[55]{Paweł~Sznajder}
\author[56]{Abigail~Victoria~Waldron}
\author[*,57]{Sarah~Williams}
\author[58]{Valentina~Zaccolo}
\author[59]{Manuel~Zeyen}

\affil[*]{ECFA ECR Detector R\&D Working Group Member\texorpdfstring{\protect\\\,}{}}
\affil[1]{Institute for Mathematics, Astrophysics and Particle Physics, Radboud University/Nikhef, Nijmegen; Netherlands}
\affil[2]{IRFU, CEA, Université Paris-Saclay, Gif-sur-Yvette; France}
\affil[3]{Laboratório de Instrumentação e F\'isica Experimental de Part\'iculas - LIP, Lisboa; Portugal}
\affil[4]{Deutsches Elektronen-Synchrotron DESY, Hamburg; Germany}
\affil[5]{Raymond and Beverly Sackler School of Physics and Astronomy, Tel Aviv University, Tel Aviv; Israel}
\affil[6]{Faculty of Nuclear Sciences and Physical Engineering, Czech Technical University in Prague, Prague; Czech Republic}
\affil[7]{CERN, Geneva; Switzerland}
\affil[8]{Institute of Physics, University of Belgrade, Belgrade; Serbia}
\affil[9]{FNRS Scientific Collaborator, Université Libre de Bruxelles, Bruxelles; Belgium}
\affil[10]{Universita’ di Bologna, Dipartimento di Fisica and INFN Sezione di Bologna; Italy}
\affil[11]{European Spallation Source, Lund; Sweden}
\affil[12]{University of Perugia and INFN Sezione di Perugia; Italy}
\affil[13]{Department of Physics, University of Oslo, Oslo; Norway}
\affil[14]{Niels Bohr Institute, University of Copenhagen, Copenhagen; Denmark}
\affil[15]{Instituto Galego de F\'isica de Altas Enerx\'ias (IGFAE), Universidade de Santiago de Compostela, Santiago de Compostela; Spain}
\affil[16]{Department of Physics, Bursa Technical University, Bursa; Turkey}
\affil[17]{Nikhef National Institute for Subatomic Physics and University of Amsterdam, Amsterdam; Netherlands}
\affil[18]{INFN e Laboratori Nazionali di Frascati, Frascati; Italy}
\affil[19]{Institute for Gravitational and Subatomic Physics (GRASP), Utrecht University/Nikhef, Utrecht; Netherlands}
\affil[20]{Department of Physics, Stockholm University; Sweden}
\affil[21]{Physics department, Faculty of science, University of Zagreb, Zagreb; Croatia}
\affil[22]{Instituto de F\'isica de Cantabria (IFCA), CSIC-Universidad de Cantabria, Santander; Spain}
\affil[23]{Horia Hulubei National Institute of Physics and Nuclear Engineering, Bucharest-Magurele; Romania}
\affil[24]{Slovak Academy of Sciences, Bratislava; Slovakia}
\affil[25]{Paul Scherrer Institut, Villigen; Switzerland}
\affil[26]{Department of Experimental Particle Physics, Jožef Stefan Institute and Department of Physics, University of Ljubljana, Ljubljana; Slovenia}
\affil[27]{Albert Einstein Center for Fundamental Physics and Laboratory for High Energy Physics, University of Bern, Bern; Switzerland}
\affil[28]{Institute of High Energy Physics, Austrian Academy of Sciences, Vienna; Austria}
\affil[29]{IFIC, Universitat de Val\`encia and CSIC; Spain}
\affil[30]{Faculty of Mathematics and Physics, Charles University, Prague; Czech Republic}
\affil[31]{Faculty of Mathematics, Physics and Informatics, Comenius University, Bratislava; Slovakia}
\affil[32]{Department of Physics, University of Helsinki, Helsinki; Finland}
\affil[33]{University of Cyprus, Nicosia; Cyprus}
\affil[34]{University of Warsaw, Warsaw; Poland}
\affil[35]{CPPM, Aix-Marseille Université, CNRS/IN2P3, Marseille; France}
\affil[36]{Universiteit Antwerpen, Antwerpen; Belgium}
\affil[37]{Fakultät für Physik, Ludwig-Maximilians-Universität München, München; Germany}
\affil[38]{Henryk Niewodniczanski Institute of Nuclear Physics Polish Academy of Sciences, Kraków; Poland}
\affil[39]{Helsinki Institute of Physics, Helsinki; Finland}
\affil[40]{Laboratoire Leprince-Ringuet, CNRS/IN2P3, Ecole Polytechnique, Institut Polytechnique de Paris, Palaiseau; France}
\affil[41]{Institute for Physics, University of Graz, Graz; Austria}
\affil[42]{University of Belgrade: Faculty of Physics and VINCA Institute of Nuclear Sciences, Belgrade; Serbia}
\affil[43]{Institute of High Energy Physics, Beijing; China}
\affil[44]{Institut für Theoretische Physik, Westfälische Wilhelms-Universität Münster, Münster; Germany}
\affil[45]{FRIA Grantee and Université Libre de Bruxelles, Bruxelles; Belgium}
\affil[46]{Department of Physics, University of Jyväskylä and Helsinki Institute of Physics, University of Helsinki; Finland}
\affil[47]{Department of Physics, University of Texas at Austin, Austin TX; United States of America}
\affil[48]{Department of Physics, Royal Institute of Technology, Stockholm; Sweden}
\affil[49]{Department of Physics, University of Warwick, Coventry; United Kingdom}
\affil[50]{Jožef Stefan Institute, University of Ljubljana, Ljubljana; Slovenia}
\affil[51]{Département de Physique Nucléaire et Corpusculaire, Université de Genève, Genève; Switzerland}
\affil[52]{Institute for Nuclear Research and Nuclear Energy, Bulgarian Academy of Sciences, Sofia; Bulgaria}
\affil[53]{Ghent University, Ghent; Belgium}
\affil[54]{Middle East Technical University, Physics Department, Ankara; Turkey}
\affil[55]{National Centre for Nuclear Research (NCBJ), Warsaw; Poland}
\affil[56]{Blackett Laboratory, Imperial College London, London; United Kingdom}
\affil[57]{Cavendish Laboratory, University of Cambridge, Cambridge; United Kingdom}
\affil[58]{Dipartimento di Fisica dell’Università and Sezione INFN, Trieste; Italy}
\affil[59]{Institute for Particle Physics and Astrophysics, ETH Zürich, Zürich; Switzerland}

\maketitle

\clearpage
\section{Introduction}

The European Committee for Future Accelerators (ECFA) Early-Career Researchers (ECR) Panel was formed in late 2020, in order to represent European ECR members in ECFA-related discussions~\cite{ECFAECRPanel}.  In this context, ECR refers to a diverse range of individuals, from students to non-tenured academics, and with backgrounds ranging from engineering to physics.  As of writing, there are 75 ECFA ECR Panel members, representing 29 ECFA member countries and large European accelerator facilities.

In Spring 2021, the ECFA ECR Panel was invited to provide input to the ECFA Detector R\&D Roadmap process, and particularly the ECFA R\&D Training Symposium that took place on April 30, 2021~\cite{TrainingSymposium}.  In preparation for this event, the ECFA ECR Panel formed a dedicated working group aimed at Detector R\&D related matters, which proceeded to organize the response.  An initial Townhall Meeting was held on April 7, 2021 with the intent of gathering feedback on ECR priorities for discussion~\cite{Townhall}, after which a survey was designed and distributed among the ECR community, with the goal of collecting their opinion about training and related issues in instrumentation.

The preliminary results of this survey were presented at the aforementioned symposium~\cite{SlidesAtSymposium}.  This report provides a more extensive analysis of the survey replies, and is intended to provide input to the ECFA Detector R\&D Roadmap process.  The report is structured as follows: the survey strategy and demographics are described first, followed by dedicated discussions on ECR general training experiences, recognition, networking, engineer perspectives, and open-form suggestions that were received.  A summary is provided, followed by a pair of appendices containing the summaries of the responses to individual questions and raw cross-analysis results.

\section{Design and distribution of the survey}

The survey was designed by the members of ECFA ECR Detector R\&D working group, taking into account the issues raised in the Townhall Meeting. It was circulated among email lists representing groups such as the LHC experiments, national ECRs, and detector R\&D collaboration lists. A total of 473 answers were recorded; the raw results of the survey are provided in Appendix A. 80\% of the participants were employed or undergoing education in Europe, 13\% in North America, 6\% from Asia, and 1\% were from South America.

\begin{figure}[h!]
    \centering
    \includegraphics[width=0.75\textwidth]{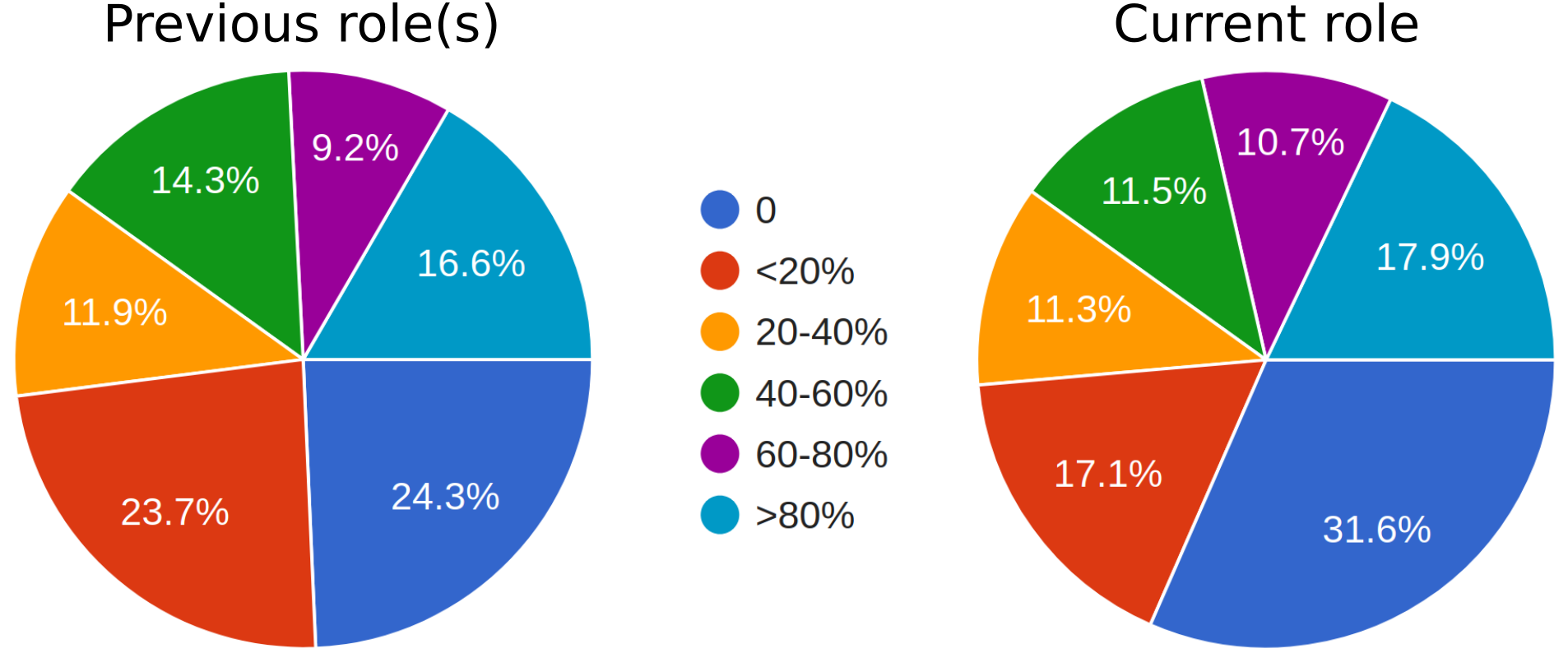}
    \caption{Pie charts showing the fraction of time spent working on instrumentation in participants' previous role(s) (left) and current role (right). There were 469 responses to both questions.}
    \label{fig:timeInInst}
\end{figure}

The advertisement mail mentioned that the survey was targeted at all ECRs, regardless of whether they were involved in instrumentation work or not, to avoid ‘selection bias’ and potentially identify barriers that were preventing ECR involvement in instrumentation work. The resulting fraction of time the participants invested in instrumentation work both in the past and at present features around 10\% or more of the participants in each category, as can be seen in \Figure{timeInInst}. The survey was therefore successful in reaching ECRs with a range of different commitment-levels to instrumentation work. Engineers were also explicitly encouraged to participate.

Most of the survey questions could be answered with \textit{yes} / \textit{no} / (\textit{not sure}), \textit{strongly agree} / \textit{agree} / \textit{disagree} / \textit{strongly disagree}, or on a scale from 1 (strongly disagree) to 5 (strongly agree). Where appropriate, an \textit{other} answer was also possible, where the respondents could enter their own custom response.

Questions related to diversity and inclusion were designed not to focus exclusively on specific categories such as gender, ethnicity, etc., but rather asked the participants whether they saw their field as diverse and if they identified as an underrepresented minority in their field, by their own definitions of minority. This was intentionally done to account for any possible kind of discrimination/under-representation in an inclusive manner. The questions about the impact of identifying as part of an underrepresented minority however allowed them to write an answer explaining their individual situation, if they felt comfortable doing so.

\section{ECR experiences of instrumentation training}

The survey asked participants about their experiences with instrumentation work and training, as well as any experiences they have had delivering training. The survey questions were purposefully open-ended regarding the definition of ‘training,’ to reflect how training is enacted in real-life; for example, some commented they received no ‘formal training’ and instead learned on the job. Detailed breakdowns of the results are provided in the appendices. The following are a few key points to highlight:

\begin{itemize}
    \item The potential of peer-to-peer training: of 172 respondents, 79\% either agreed or strongly agreed with the statement that they found the peer-to-peer training they had participated in useful, and 83\% indicated agreement that they would participate in peer-to-peer training if it were available.

    \item Transferable skills: when asked “On a range from 1-5, how applicable do you think the skills/experience you have gained are to the broader physics community”, 62\% responded either 4 or 5.

    \item Support for Open Source Software (OSS) tools: 71\% of 334 respondents indicated that their instrumentation work involved using open source software tools, whilst 70\% of 330 respondents said they had not received training for such tools. The use of OSS in PCB design, FPGA development, and ASIC design was highlighted by one respondent as a way to increase participation of marginalised groups who are members of institutions without the funding necessary to acquire expensive licenses.
    
    \item Possibilities for remote training/work in instrumentation: 39\% of 331 respondents had experience with remote training or work in instrumentation outside of their institute, and 61\% out of 176 responded ‘agree’ or ‘strongly agree’ to the statement that their experience of remote training/work in instrumentation had been positive. However it should be acknowledged that for some areas in-person training is more efficient and can ease networking opportunities, the importance of which is discussed later in this report.
\end{itemize}

\begin{figure}[h!]
\centering
\begin{subfigure}[b]{.45\linewidth}
\includegraphics[width=\linewidth]{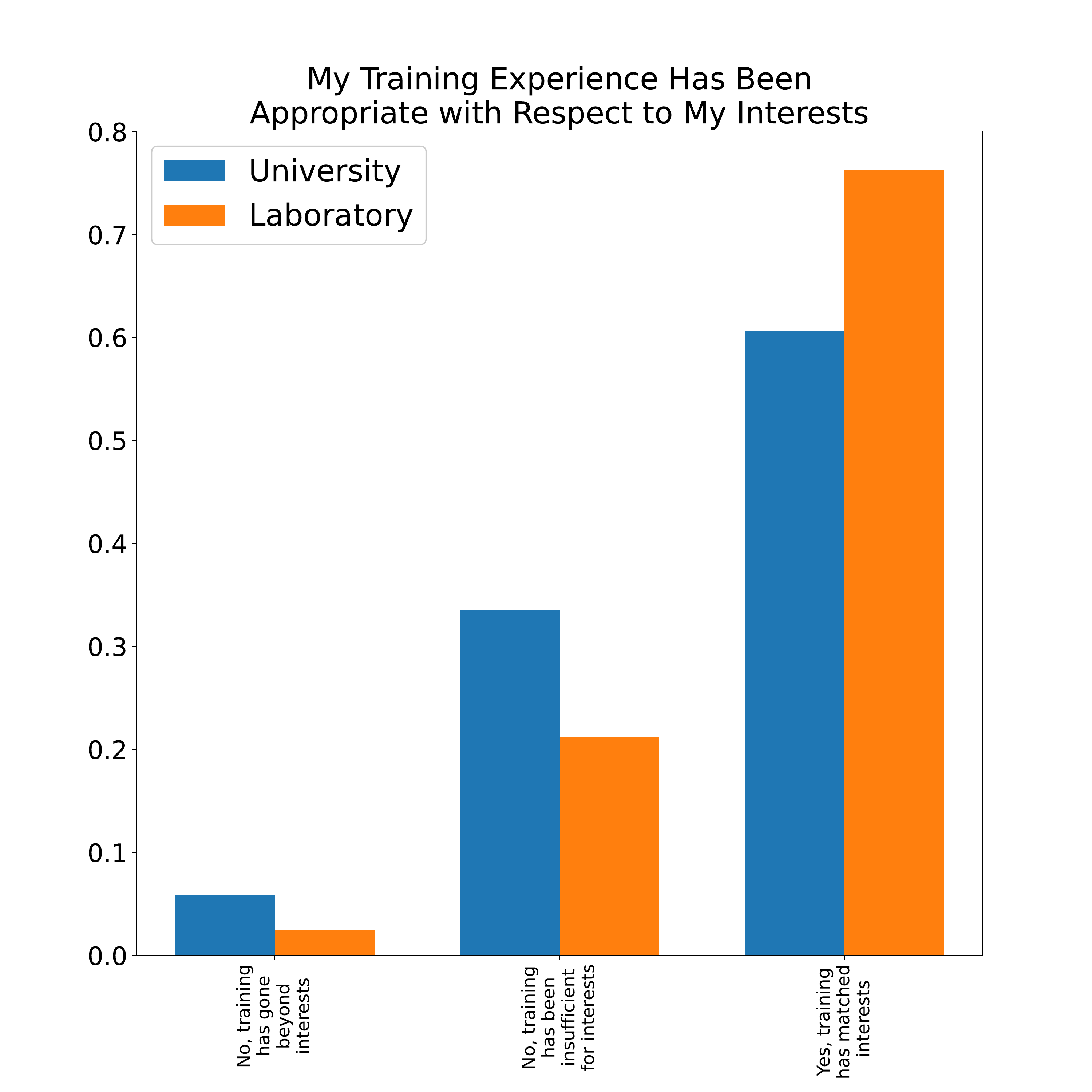}
\caption{}
\end{subfigure}
\begin{subfigure}[b]{.45\linewidth}
\includegraphics[width=\linewidth]{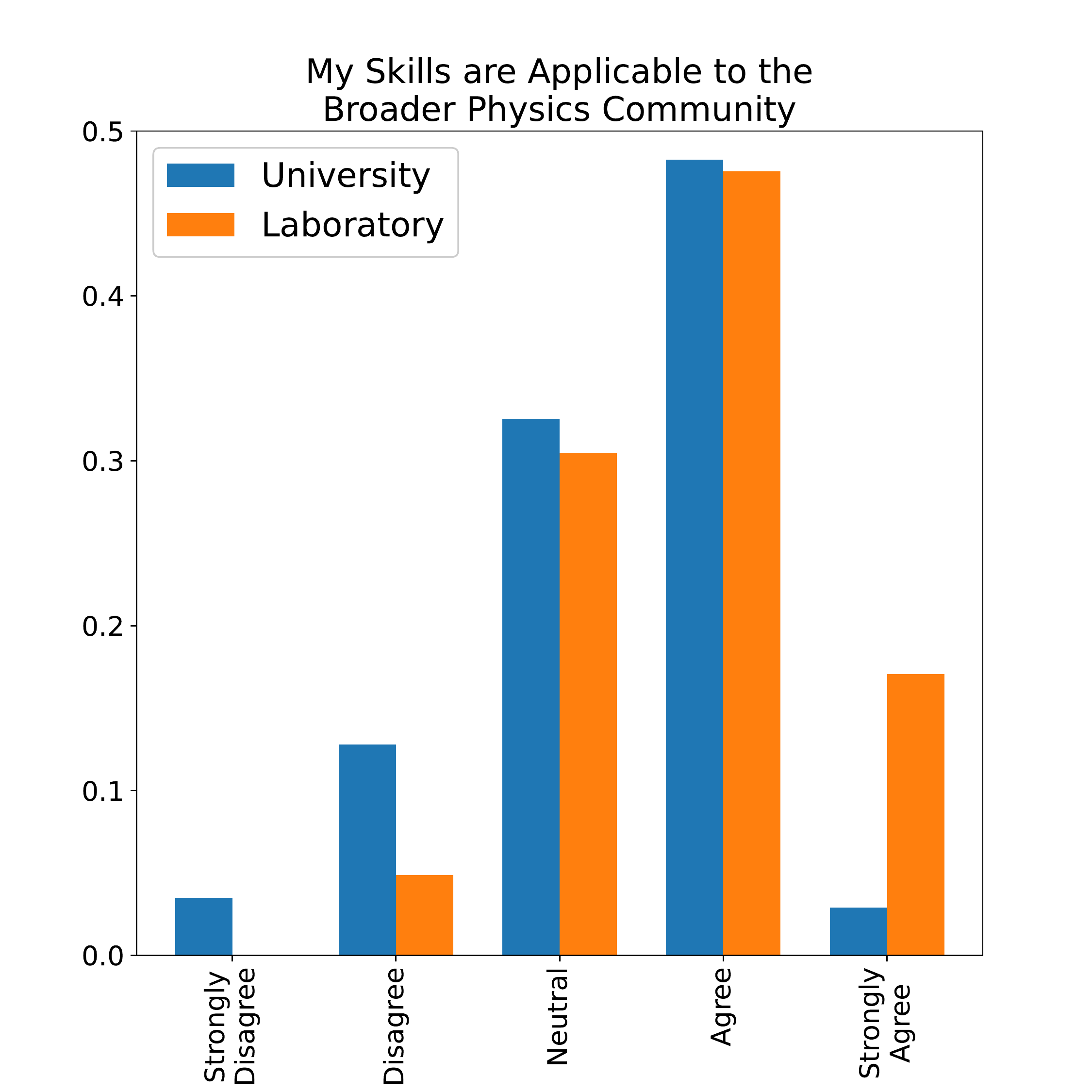}
\caption{}
\end{subfigure}
\caption{Comparison of the (normalised) responses to two questions on instrumentation training between participants employed or studying at a university compared to those at a laboratory. The number of respondents to each question, and the overall summary of the results, can be found in Appendix A on pages~\pageref{fig:training_experience_appropriate_to_interests}(a) and \pageref{fig:application_to_broader_community}(b). Figures with the raw number of responses for certain populations are found in Appendix B.} 
\label{fig:UniversityVsLab}
\end{figure}

Further analysis of the responses to some of the questions for particular groups of participants will now be provided. Firstly, in the ECFA ‘Training in Instrumentation’ symposium, one question raised was if there were significant differences in participants’ experiences depending on whether they are employed/studying at a university, or instead at a national/research laboratory. Of the survey respondents, 68\% were based at a university. Of the 32\% in ‘other’ institutions, 72\% were based at a research laboratory.  \Figure{UniversityVsLab} compares the responses of those at universities, to those at laboratories, for two questions about their training experiences. Both are normalised to the number of respondents in each category for that question. 

Another interesting question to consider is one of the open-form questions where participants were asked how they view instrumentation work. They could select any number of the following statements, as well as providing their own personal responses. \Table{UniversityVsLab} shows, out of 470 respondents, how many selected each of the five selected categories. The distributions of these responses between those at a university and all other institutions are provided, where it is interesting to look for cases where specific replies do not agree with the overall fraction of replies from universities (68\%) versus other institutions (32\%).  Deviations from a 68\%:32\% reply balance indicate differences in feelings related to instrumentation work for different employment types, although it is important to note that some of the categories suffer from low statistics.
\begin{itemize}
    \item Of those that think negatively about instrumentation work (i.e.~that they avoid it or its a liability) proportionately more are at a university.
    
    \item Of those who think positively (i.e.~that they enjoy/benefit from it), proportionately more are at other institutes ($\sim$70\% of which are national/research laboratories).
    
    \item Those who had instrumentation training in their MSc.~or PhD. (262 responses) were more likely to view work in instrumentation as a benefit to their careers and something they enjoy in comparison to those who did not (206 responses). 
\end{itemize}

\begin{table}[h!]
    \centering
     \setstretch{1.25}
     \setlength{\tabcolsep}{4.5pt}
     \begin{tabular}{c|rc|rc|rc|rc|rc}
        I view work in & \multicolumn{2}{|c|}{...something I} & \multicolumn{2}{|c|}{...a liability} & \multicolumn{2}{|c|}{...something I} & \multicolumn{2}{|c|}{...something} & \multicolumn{2}{|c}{...a benefit}
        \\
        instrumentation as... & \multicolumn{2}{|c|}{don't think about} & \multicolumn{2}{|c|}{to my career} & \multicolumn{2}{|c|}{would rather avoid} & \multicolumn{2}{|c|}{I enjoy} & \multicolumn{2}{|c}{to my career}
         \\[0.1cm]\hline
         Total responses & \hspace{0.3cm}65 & & \hspace{0.2cm}67 & & \hspace{0.3cm}34 & & \hspace{0.2cm}242 & & \hspace{0.1cm}307 &
         \\[0.25cm]\hline
         University & 45 & (69\%) & 48 & (72\%) & 27 & (79\%) & 156 & (64\%) & 199 & (65\%)
         \\[0.25cm]
         Other institutions & 20 & (31\%) & 19 & (28\%) & 7 & (21\%) & 86 & (36\%) & 108 & (35\%)
         \\[0.25cm]\hline
         Had instrumentation & \multirow{2}{*}{21} & \multirow{2}{*}{(32\%)} & \multirow{2}{*}{42} &  \multirow{2}{*}{(63\%)} & \multirow{2}{*}{9} & \multirow{2}{*}{(26\%)} & \multirow{2}{*}{155} & \multirow{2}{*}{(64\%)} & \multirow{2}{*}{190} & \multirow{2}{*}{(62\%)}
         \\
         training in MSc./PhD. & & & & & & & & & & \\[0.25cm]
         No instrumentation & \multirow{2}{*}{44} & \multirow{2}{*}{(68\%)} & \multirow{2}{*}{25} &  \multirow{2}{*}{(37\%)} & \multirow{2}{*}{24} & \multirow{2}{*}{(70\%)} & \multirow{2}{*}{84} & \multirow{2}{*}{(35\%)} & \multirow{2}{*}{114} & \multirow{2}{*}{(37\%)}
         \\
         training in MSc./PhD. & & & & & & & & & &
    \end{tabular}
    \caption{Responses to the statement “I view instrumentation work as...” distributed between participants at a university compared to other institutes, where here ‘other’ refers to all other institutions, 72\% of which were national/research laboratories.}
    \label{tab:UniversityVsLab}
\end{table}

\begin{figure}[h!]
\centering
\begin{subfigure}[b]{.45\linewidth}
\includegraphics[width=\linewidth]{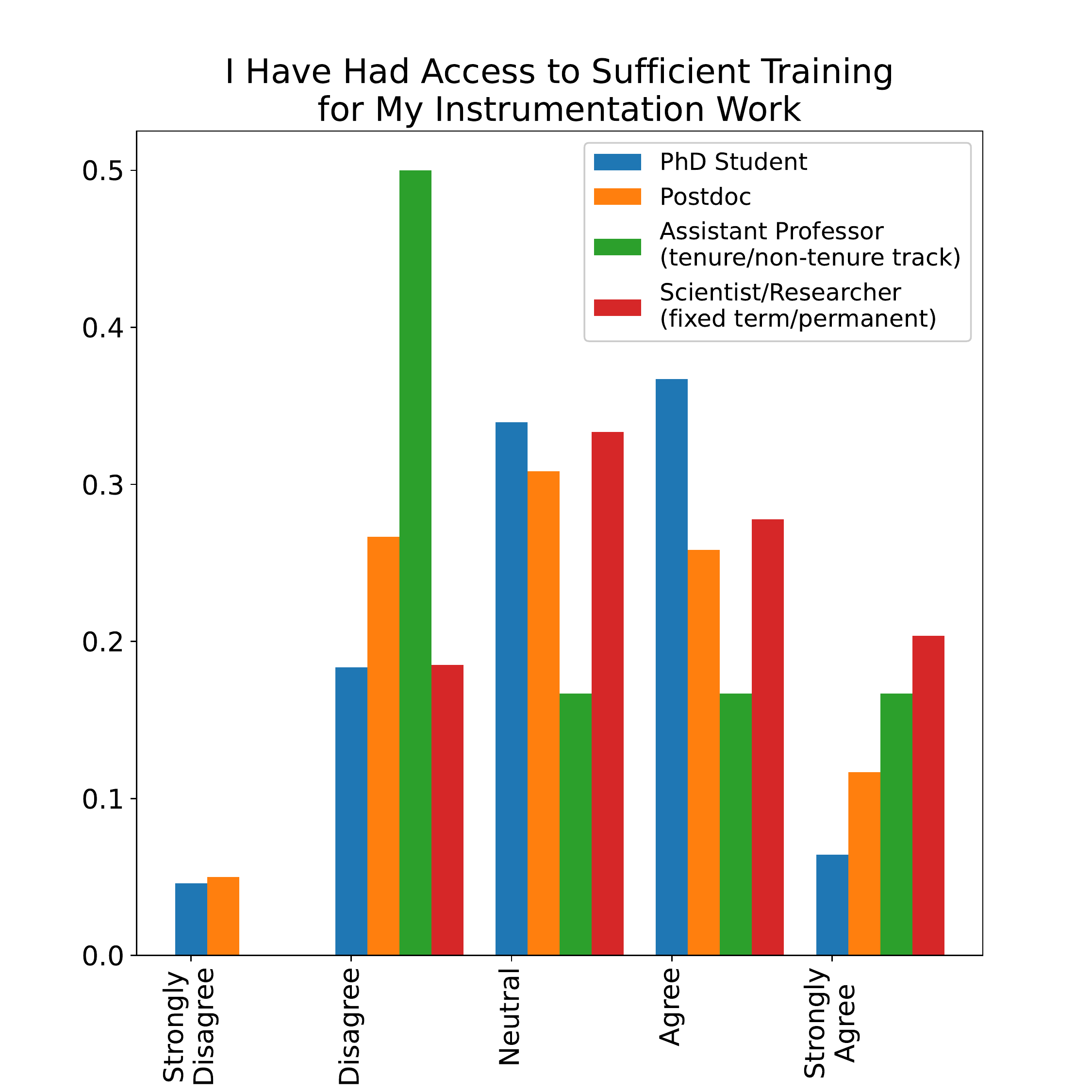}
\caption{}
\end{subfigure}
\begin{subfigure}[b]{.45\linewidth}
\includegraphics[width=\linewidth]{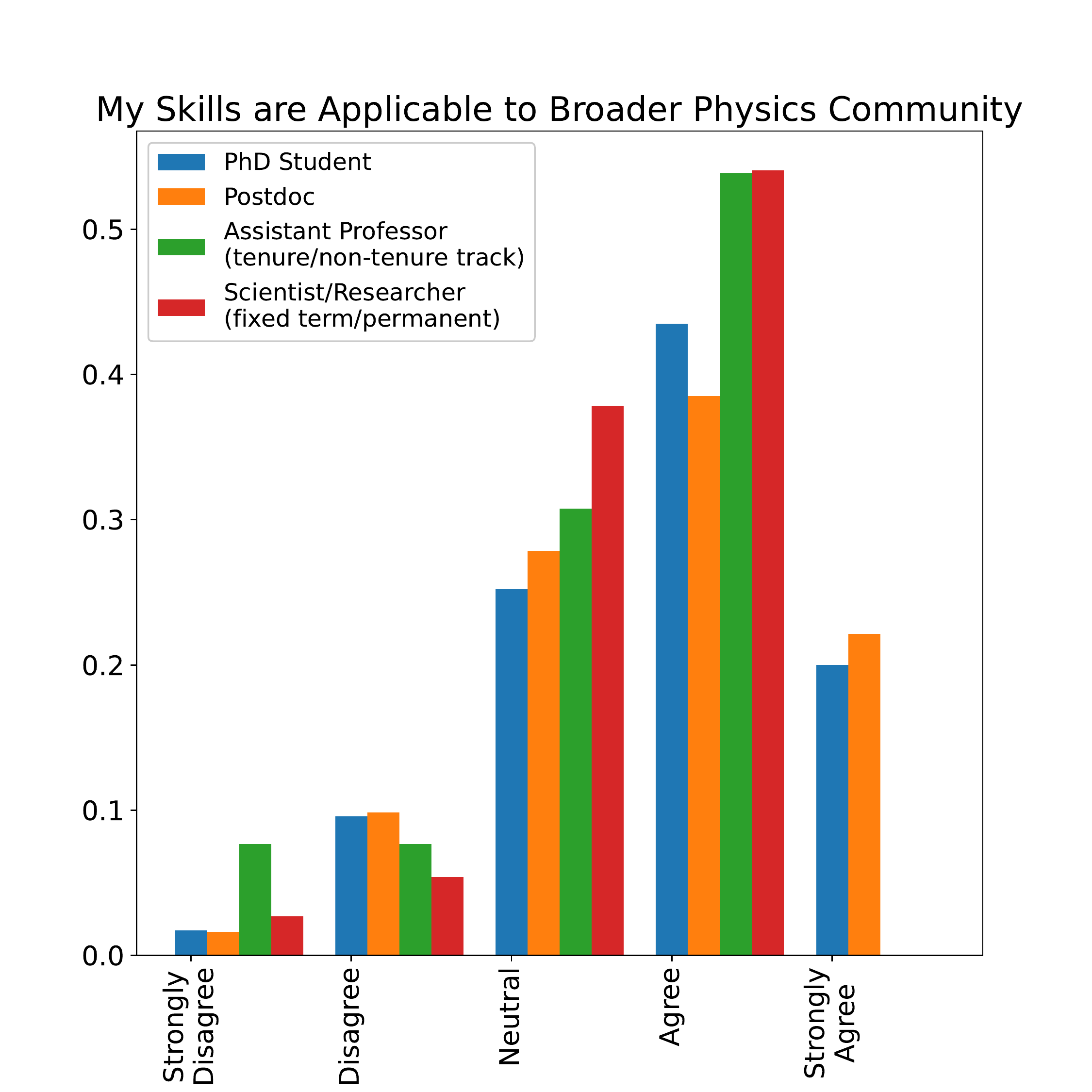}
\caption{}
\end{subfigure}
\begin{subfigure}[b]{.45\linewidth}
\includegraphics[width=\linewidth]{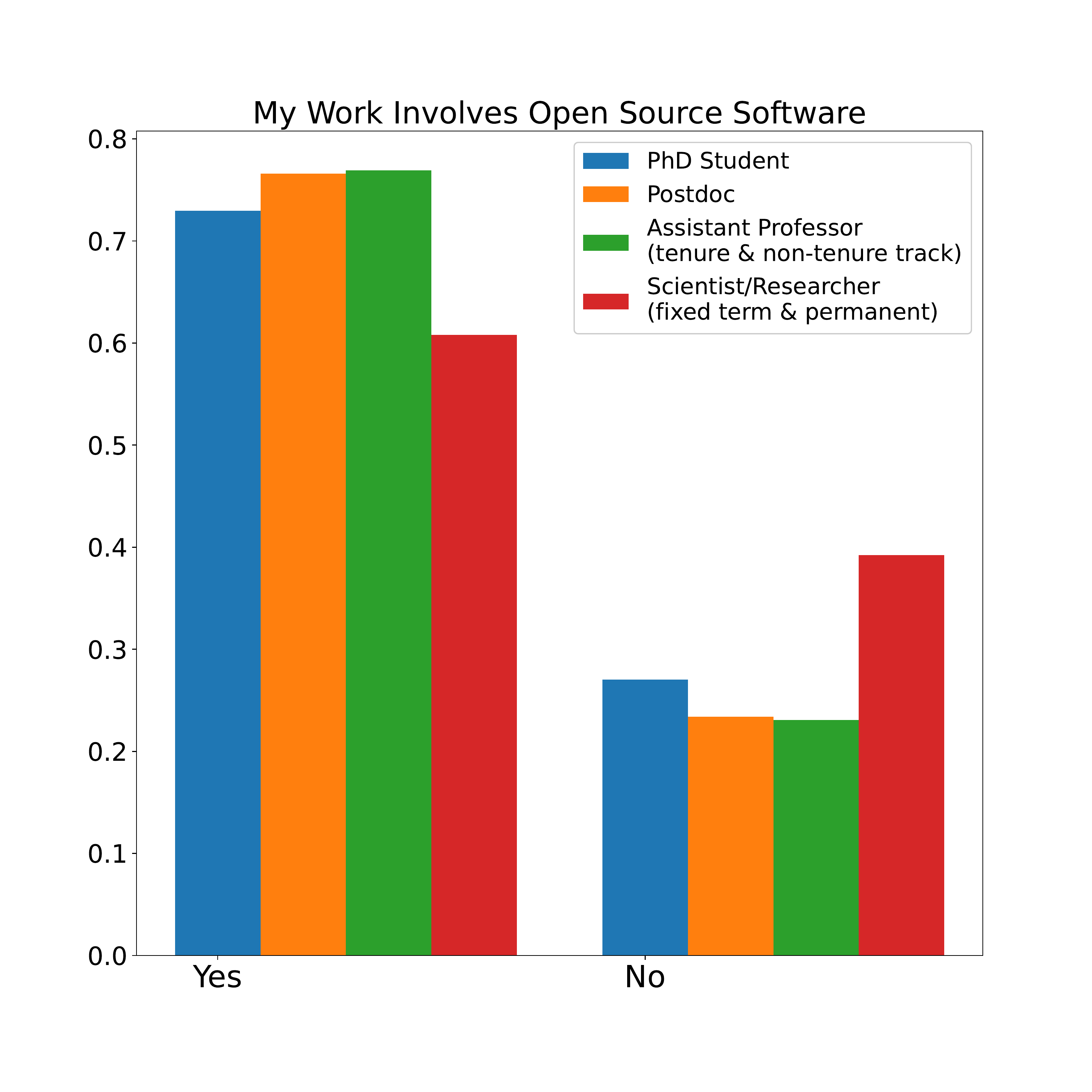}
\caption{}
\end{subfigure}
\begin{subfigure}[b]{.45\linewidth}
\includegraphics[width=\linewidth]{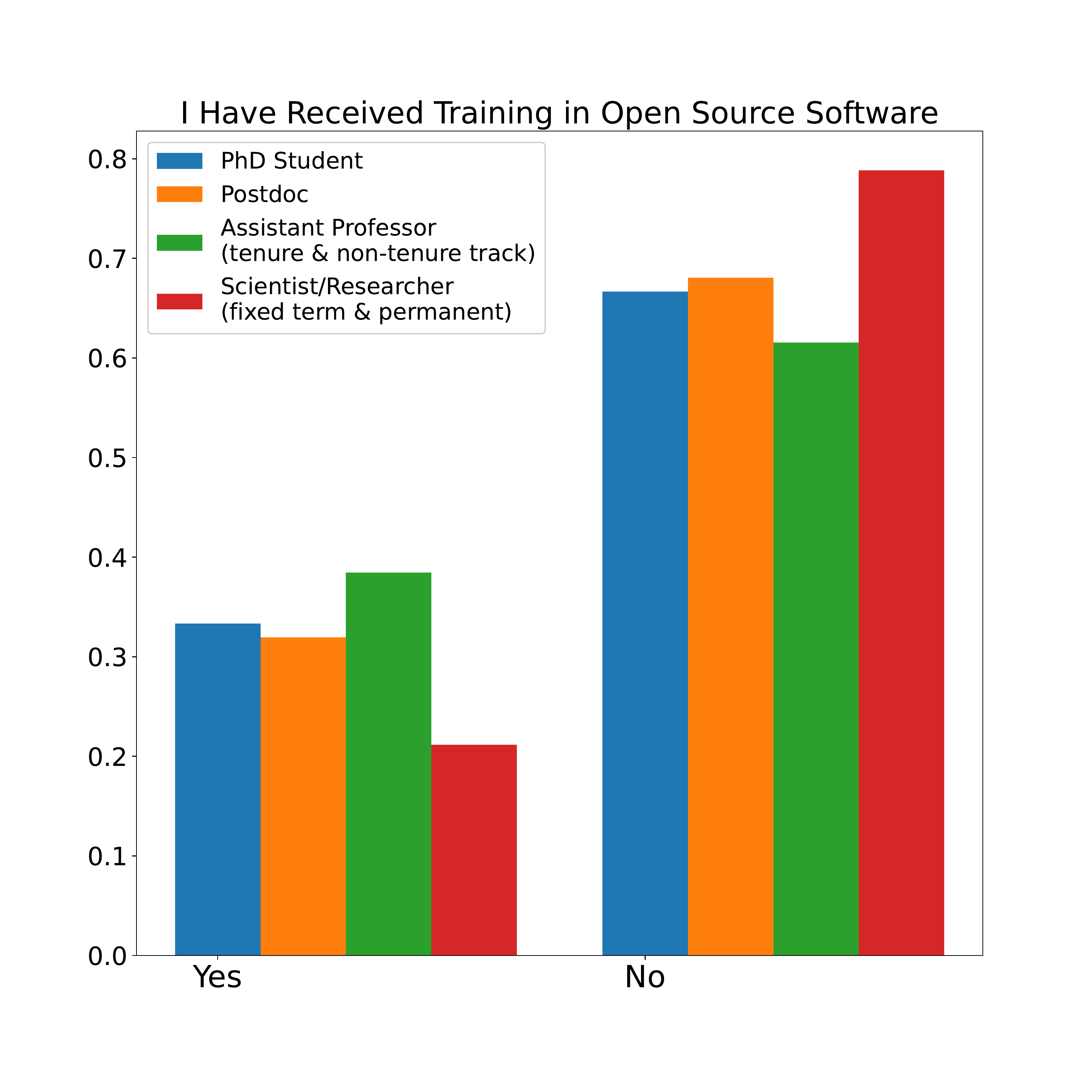}
\caption{}
\end{subfigure}
\caption{Comparisons of (normalised responses) for several questions related to instrumentation training between different career stages of the ECRs that responded to the survey. The number of respondents to each question, and the overall summary of the results, can be found in Appendix A on pages~\pageref{fig:access_to_training}(a), \pageref{fig:application_to_broader_community}(b), and \pageref{fig:open_source}(c,d). Figures with the raw number of responses for certain populations are found in Appendix B.} 
\label{fig:InstTraining}
\end{figure}

Further analysis has also been performed between different career stages of the ECRs that responded to the survey. The plots in \Figure{InstTraining} are comparisons of the answers from the four most represented career categories: PhD Students, Postdocs, Assistant Professors (tenure and non-tenure track) and Scientists/Researchers (fixed term and permanent). The answers are normalised for comparison as the number of responses is different for each population.  Additional analysis of these categories (both career stage and institution) will be provided in the subsequent sections of this report which will focus in turn on diversity and inclusion issues, recognition, and networking.

\section{Diversity and inclusion issues}

\begin{figure}
\centering
\begin{subfigure}[b]{.45\linewidth}
\includegraphics[width=\linewidth]{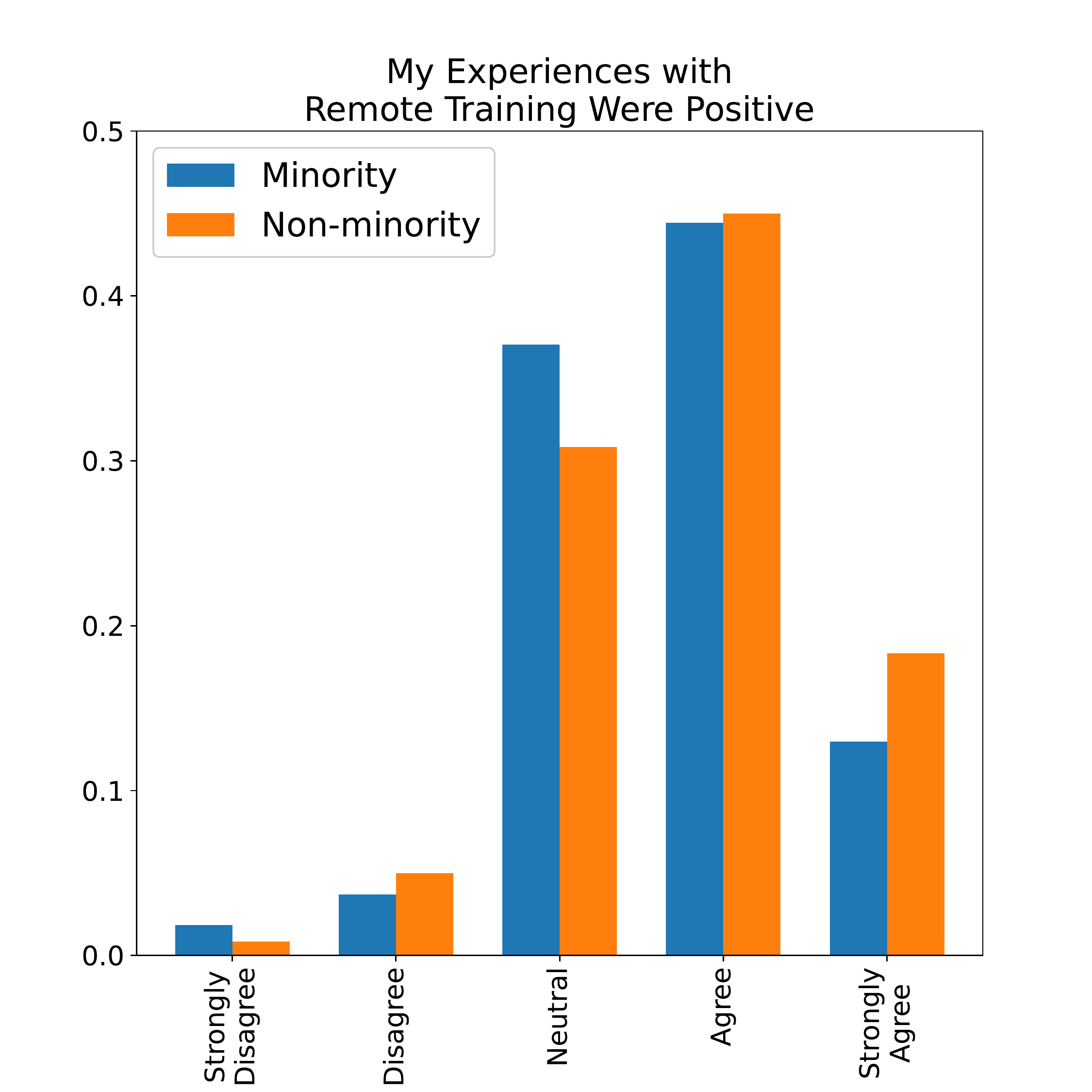}
\caption{}
\end{subfigure}
\begin{subfigure}[b]{.45\linewidth}
\includegraphics[width=\linewidth]{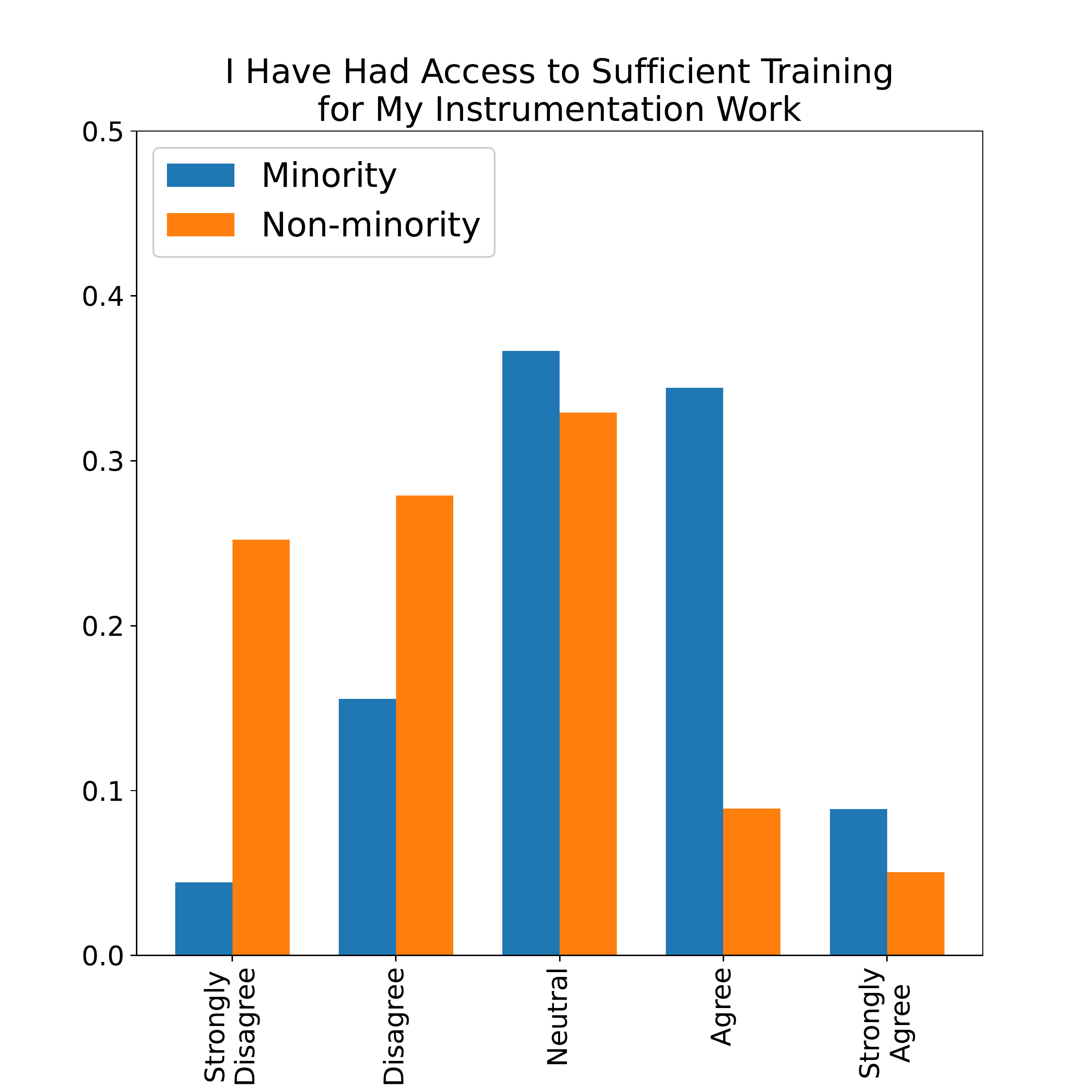}
\caption{}
\end{subfigure}
\begin{subfigure}[b]{.45\linewidth}
\includegraphics[width=\linewidth]{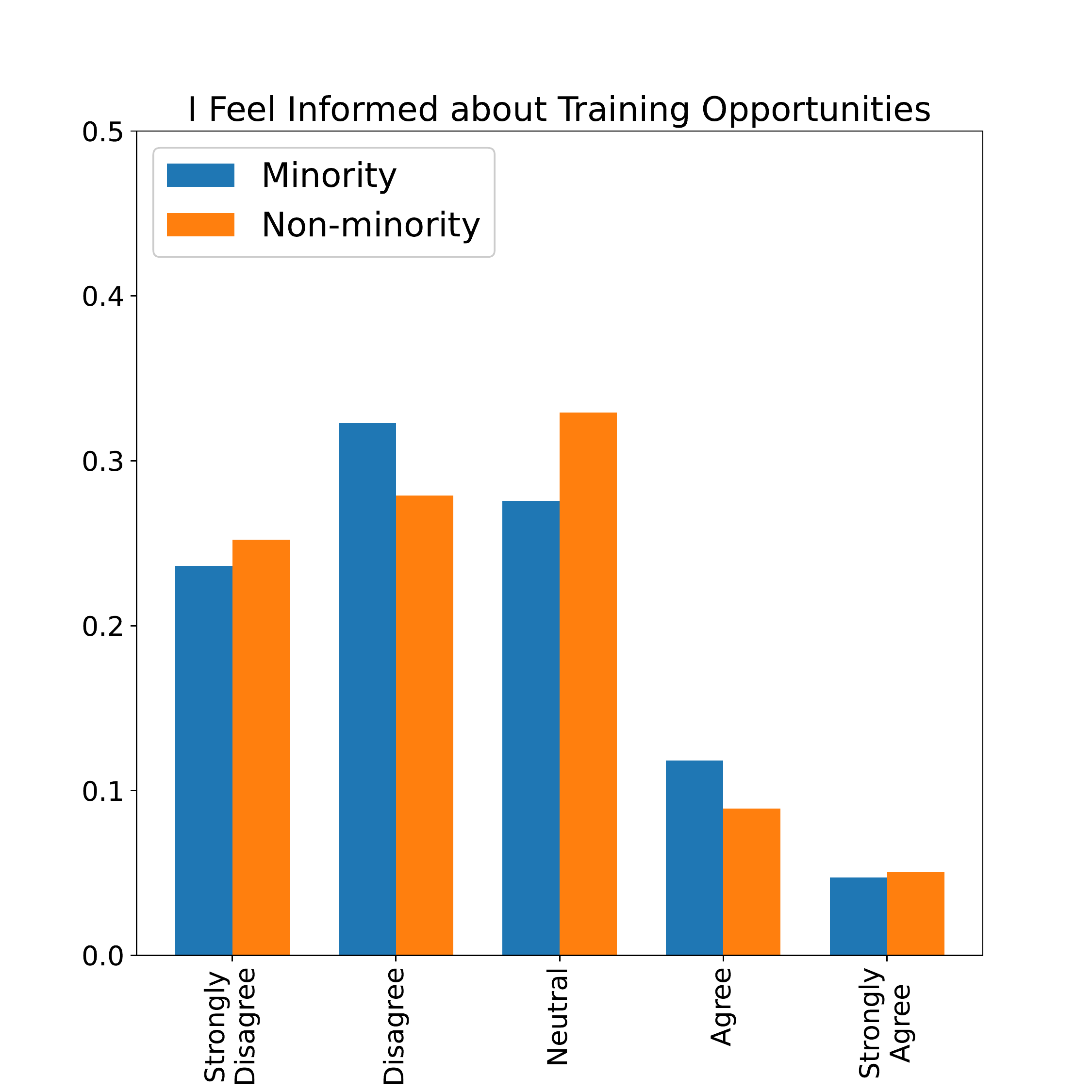}
\caption{}
\end{subfigure}
\begin{subfigure}[b]{.45\linewidth}
\includegraphics[width=\linewidth]{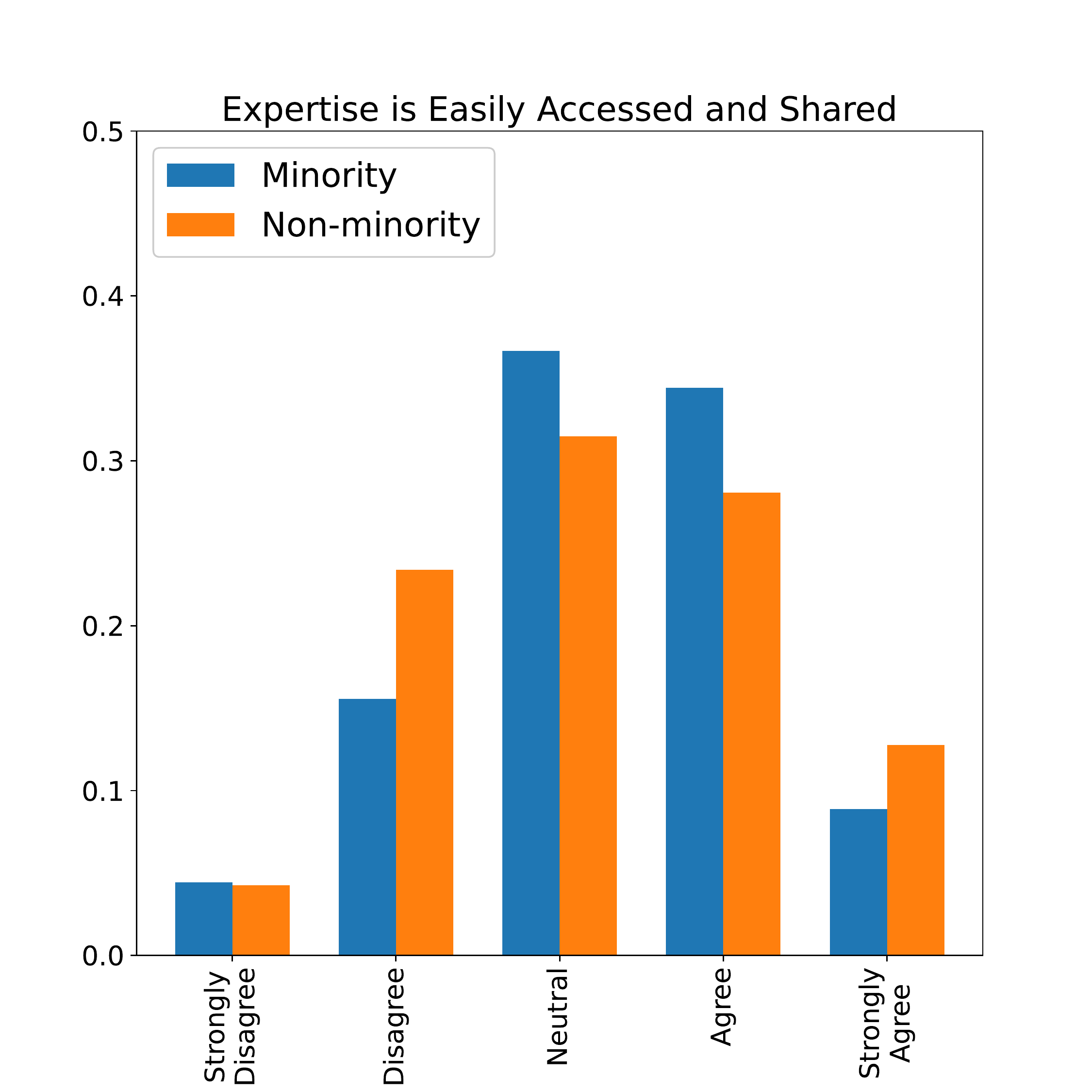}
\caption{}
\end{subfigure}
\begin{subfigure}[b]{.45\linewidth}
\includegraphics[width=\linewidth]{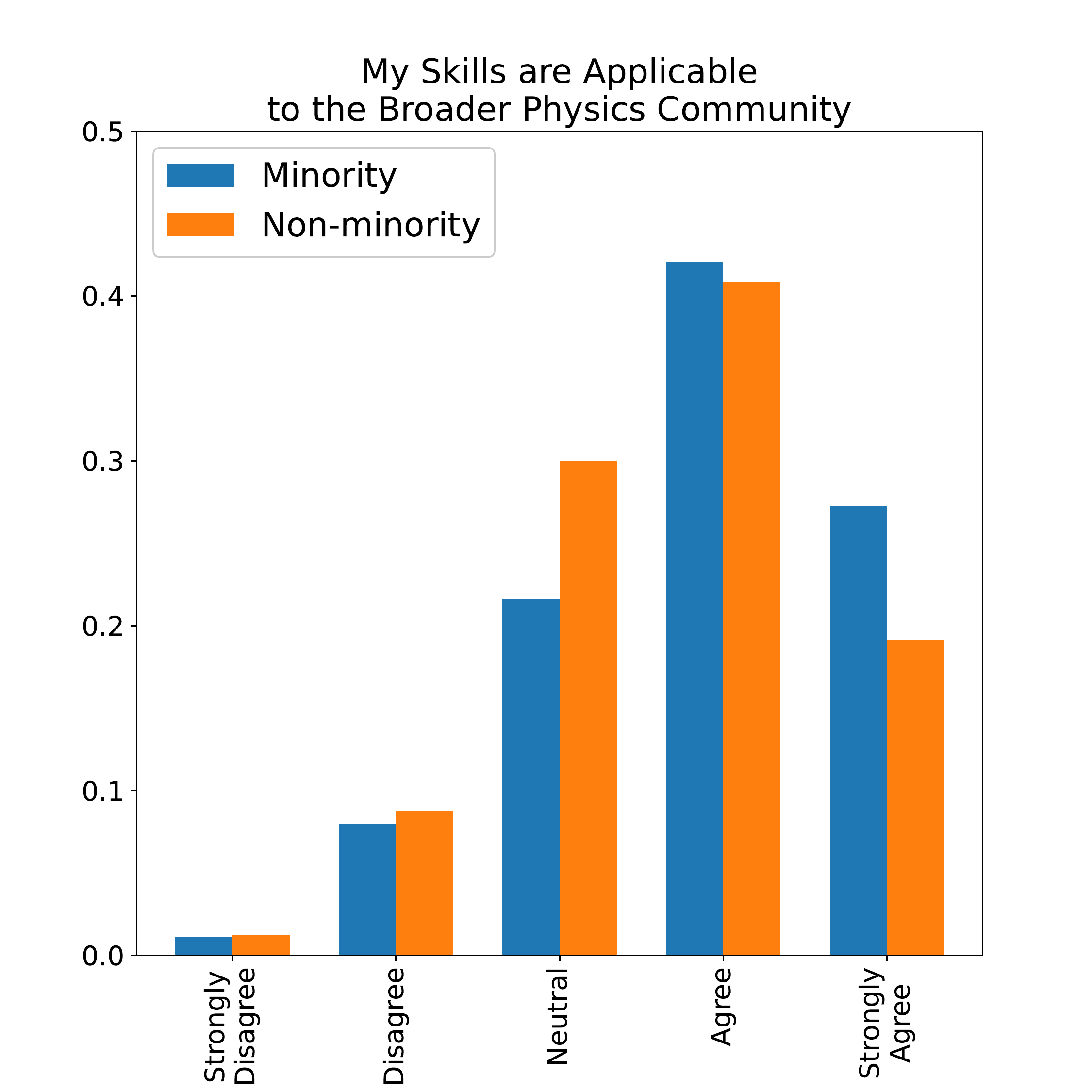}
\caption{}
\end{subfigure}
\begin{subfigure}[b]{.45\linewidth}
\includegraphics[width=\linewidth]{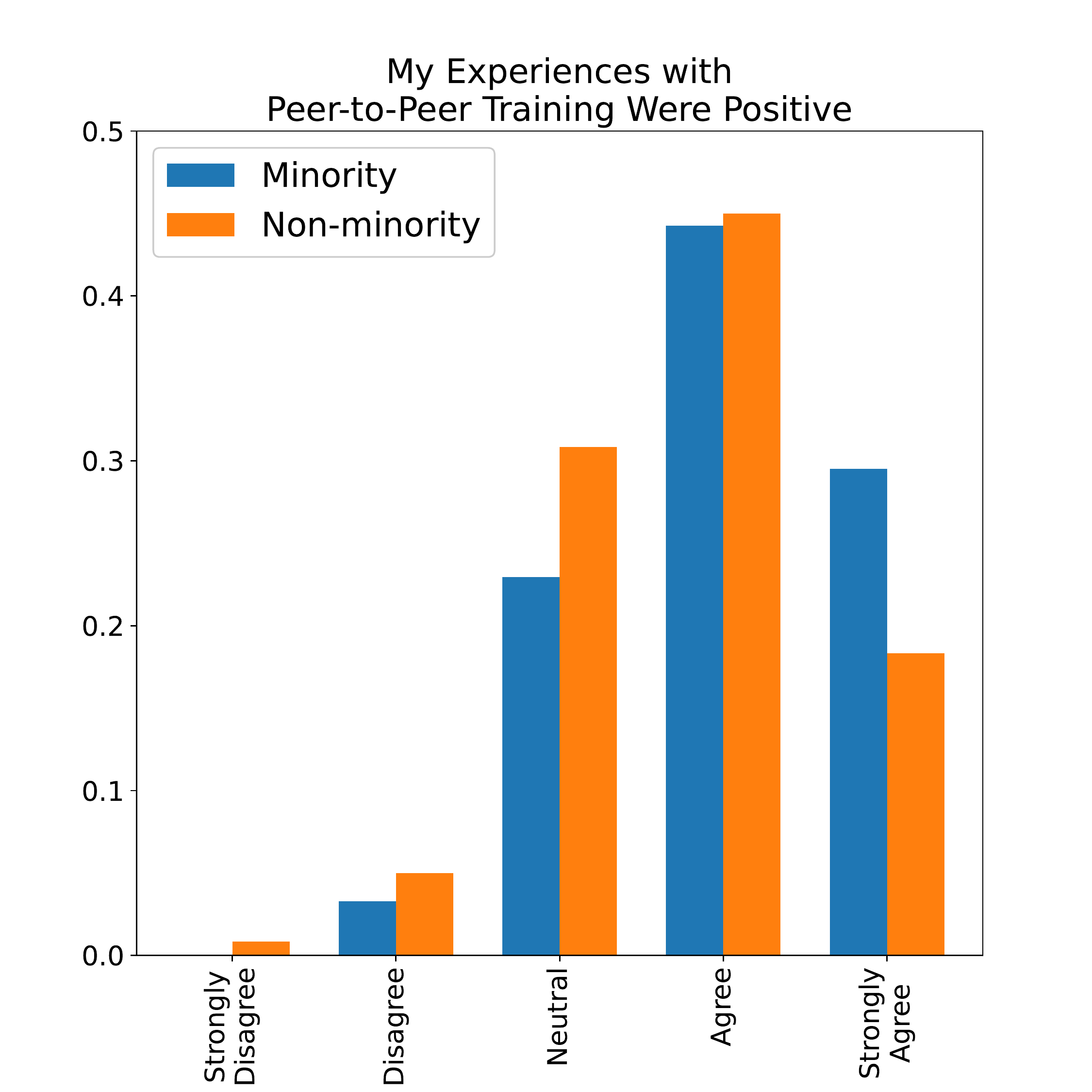}
\caption{}
\end{subfigure}
\caption{Comparisons of (normalised responses) for several questions related to instrumentation training for the respondents who identified as a minority in their field and those who didn’t. The number of respondents to each question, and the overall summary of the results, can be found in Appendix A on pages~\pageref{fig:remote}(a), \pageref{fig:sufficient}(b), \pageref{fig:informed}(c), \pageref{fig:expertise}(d), \pageref{fig:application_to_broader_community}(e), \pageref{fig:peer}(f). Figures with the raw number of responses for certain populations are found in Appendix B.} 
\label{fig:MinorityReplies}
\end{figure}

The importance of supporting diversity and inclusion within HEP was strongly highlighted in the recent European Strategy Update~\cite{ESU}, and several questions in the survey explored these topics further. 28\% of 465 respondents said that they identify as an underrepresented minority in their field. \Figure{MinorityReplies} shows a comparison of responses to experiences in instrumentation training between those who identified themselves as an underrepresented minority and those who did not. The results are normalised for comparison as the number of responses differed for each population and each question.

One question in the survey asked participants to provide free-form comments on how their identity had impacted their training experience. Of 29 responses to this question, key points raised include:

\begin{itemize}
    \item Gender identity: this was referred to specifically in ten responses, which raised issues including assumptions being made that women wouldn’t want to get involved in instrumentation due to the nature of the work, sexism in the workplace, women being treated differently in meetings to male colleagues, and the under-representation of women as speakers in training events.
    
    \item The lack of facilities in some geographical regions for instrumentation training (which then means researchers must travel abroad if remote training is not an option, and which is only possible if their institutes can support such travel).
    
    \item Challenges associated with ‘unconscious bias’.
    
    \item One respondent mentioned discrimination and harassment due to their LGBTQ+ identity.
\end{itemize}

Several questions on the survey also asked respondents whether they see both their immediate working group, and the broader experiment/collaboration in which they work, as diverse. Of 464 respondents, 24\% either disagreed or strongly disagreed with the statement that their immediate working group is diverse, compared to 17\% for their broader experiment/collaboration. In each case those answering in these categories were encouraged to share free-form comments on the matter. There were 62 responses for the question related to their immediate working group:

\begin{itemize}
    \item Many cited that diversity related to gender and/or ethnicity is a problem.
    
    \item Several mentioned concerns about ‘geographical’ diversity.
    
    \item Some comments on (lack of) diversity of training experiences, which can make it difficult to get exposure to new areas of work.
\end{itemize}

39 respondents provided comments on why they disagreed or strongly disagreed with the thought of their experiment/collaboration being regarded as diverse. Comments included:

\begin{itemize}    
    \item Most members of the experiment seem to share the same technical skills.
    
    \item Alternative career paths are not valued.
    
    \item Some mentioned that their broader collaboration is on the whole more diverse than their immediate working group.
    
    \item As in the previous questions, some cited lack of diversity in terms of gender, ethnicity, and other minority groups.
\end{itemize}

\section{Recognition}

\begin{figure}[h!]
\centering
\begin{subfigure}[b]{.45\linewidth}
\includegraphics[width=\linewidth]{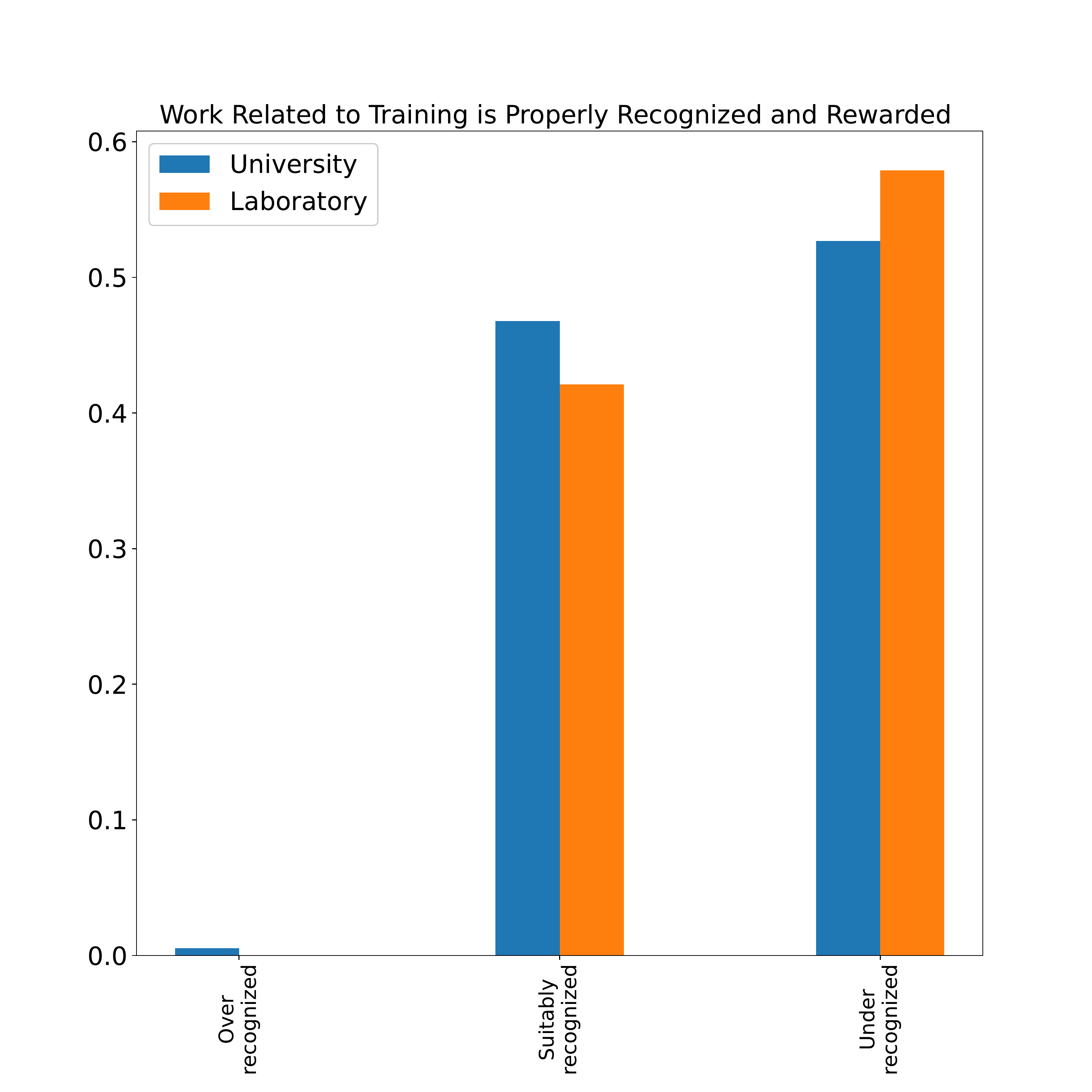}
\caption{}
\end{subfigure}
\begin{subfigure}[b]{.45\linewidth}
\includegraphics[width=\linewidth]{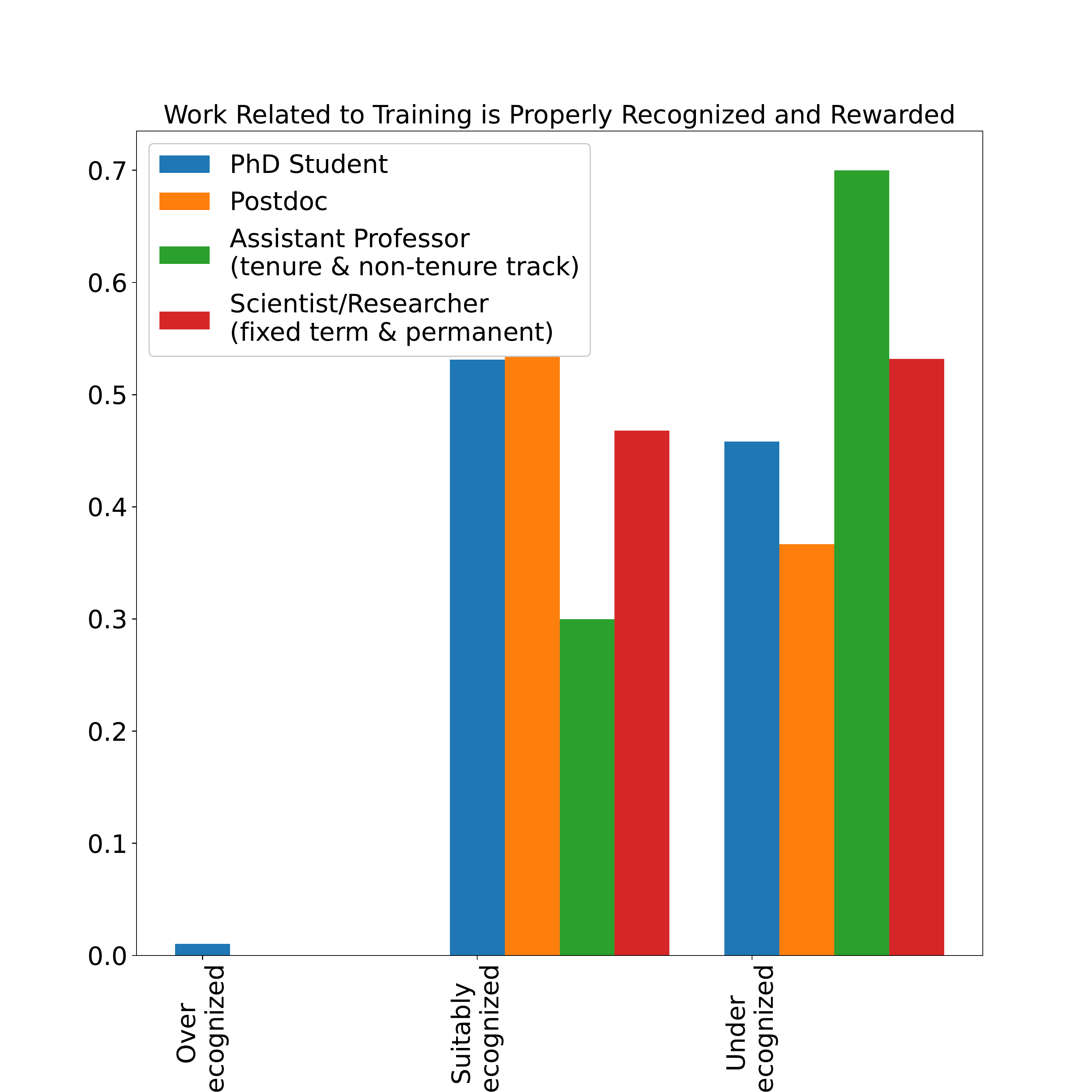}
\caption{}
\end{subfigure}
\begin{subfigure}[b]{.45\linewidth}
\includegraphics[width=\linewidth]{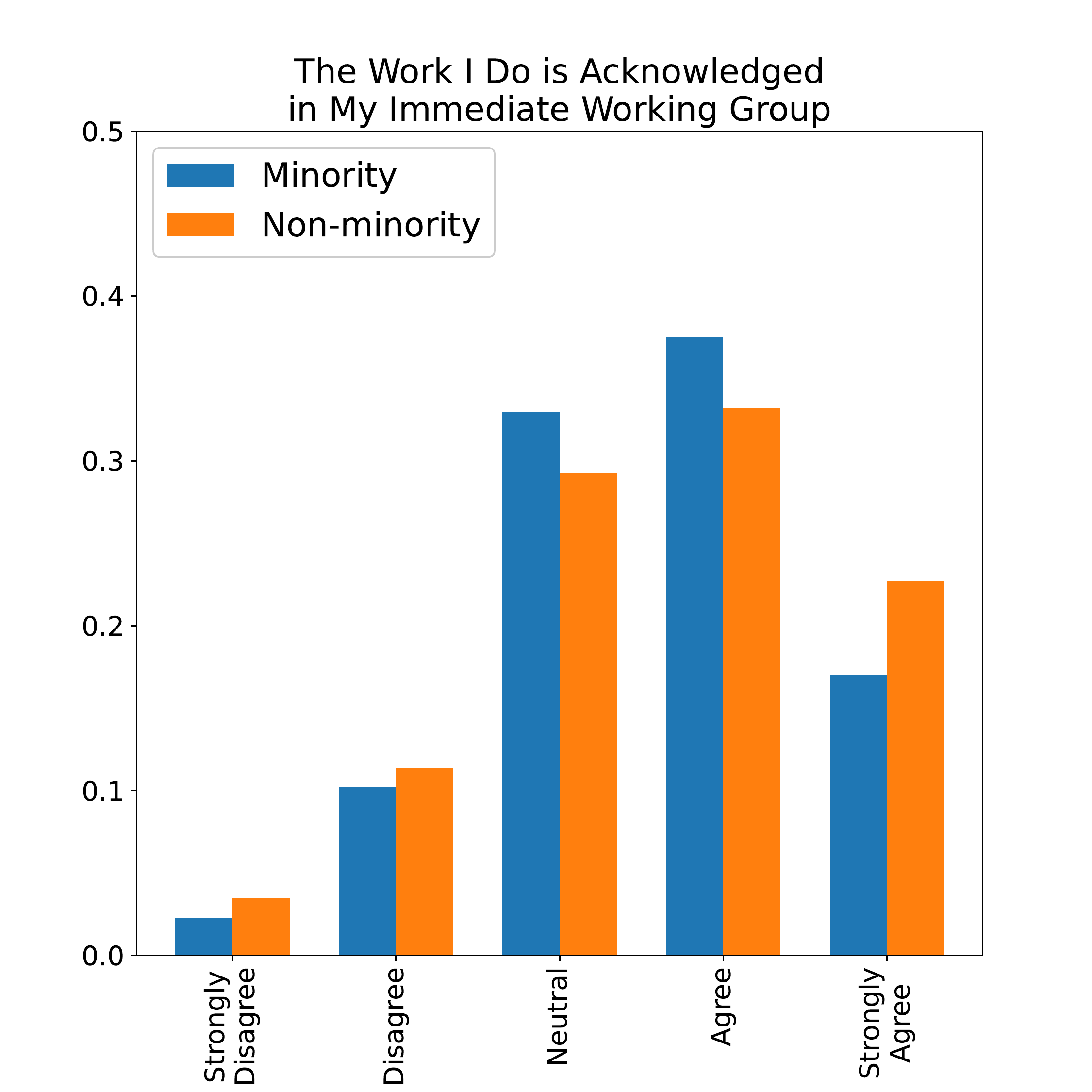}
\caption{}
\end{subfigure}
\begin{subfigure}[b]{.45\linewidth}
\includegraphics[width=\linewidth]{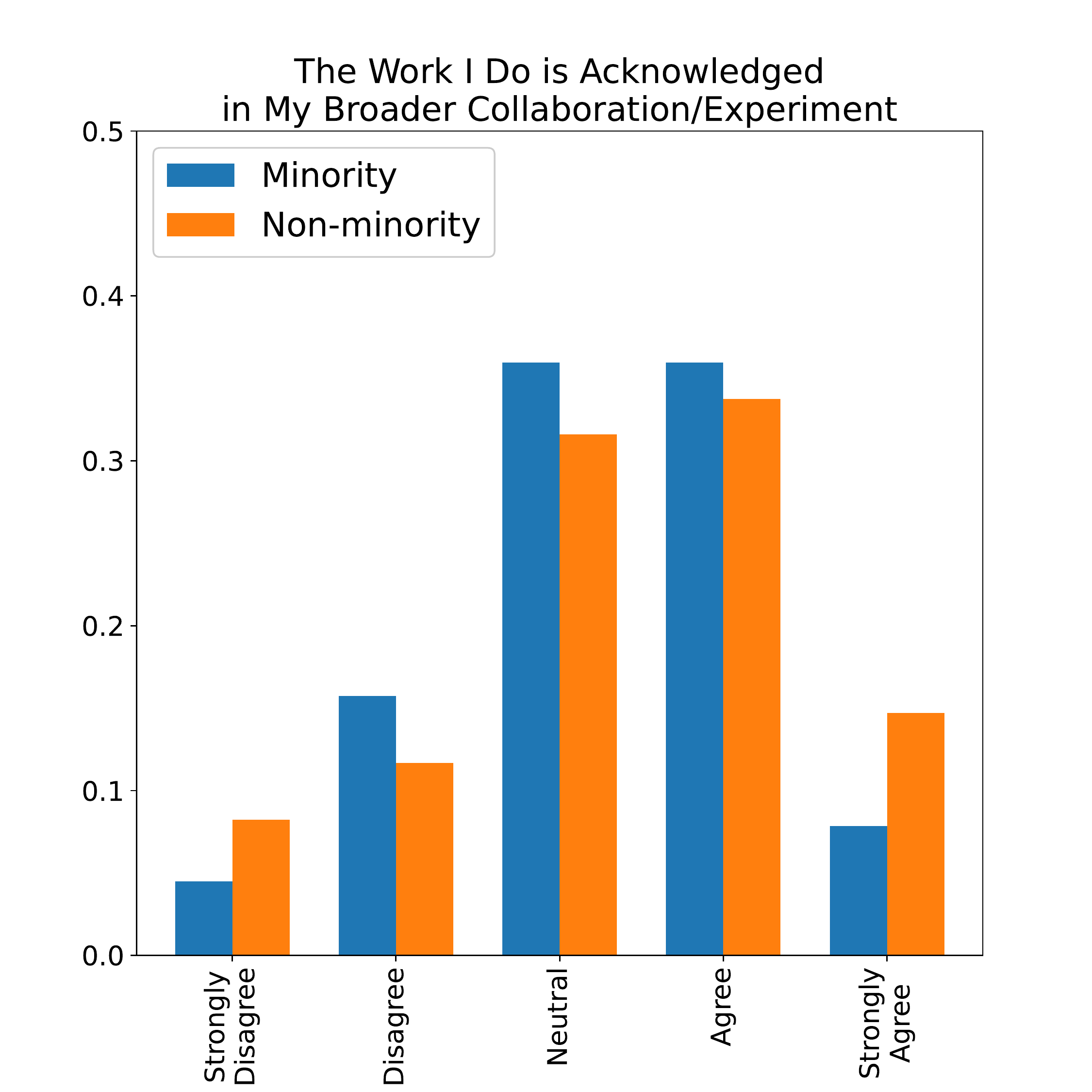}
\caption{}
\end{subfigure}
\caption{Comparisons of (normalised responses) for several questions related to recognition for instrumentation work and training. The top plots refer to the question “Do you feel that work relating to training is properly recognised and rewarded in your instrumentation group?” which had 295 responses in total. The results are compared for laboratory-based and university-based respondents on the left, and for different ECR categories on the right. The bottom plots compare the normalised responses to two questions for respondents that did and didn’t identify as minorities. The left plot corresponds to the level of agreement with the statement “I feel the work I do/did in instrumentation is acknowledged in my immediate working group” (322 responses) and the right to the statement “I feel the work I do/did in instrumentation is acknowledged in my collaboration/experiment/broader working group” (325 responses). The number of respondents to each question, and the overall summary of the results, can be found in Appendix A on pages~\pageref{fig:recognized}(a, b) and \pageref{fig:app-ack-glob}(c, d). Figures with the raw number of responses for certain populations are found in Appendix B.} 
\label{fig:Recognition}
\end{figure}

The survey asked questions about both recognition for work in instrumentation, and recognition for the work involved in designing and delivering training programmes. The responses differentiated by career category and also by institution type are shown in \Figure{Recognition} for several questions including whether work relating to training is properly recognised and rewarded in their instrumentation group, and whether their work in instrumentation is acknowledged in both their working group, and in the broader experiment/collaboration. The answers are normalised for comparison as the number of responses is different for each population and question. The following are some interesting points:

\begin{itemize}
    \item 53\% of 295 respondents answered that ‘Training is Suitably Recognised’ to the question ‘Do you feel that work relating to training is properly recognised and rewarded in your instrumentation group?’, while the remainder felt that ‘Training is Under-Recognised’.
    
    \item For those who indicated they worked at a Laboratory, 58\% responded that ‘Training is Under-Recognised’, while the remainder felt that ‘Training is Suitable Recognised’.
    
    \item 14\% of respondents indicated they either Disagreed or Strongly Disagreed that their work was acknowledged in their immediate working group, vs 20\% who Disagreed or Strongly Disagreed that their work was acknowledged in their broader experiment/collaboration.
\end{itemize}

The survey section on recognition allowed for open comments on recognition. A few highlights of the points raised are provided below:

\begin{itemize}
    \item In general, there is a tendency in the comments to mention the underestimation of the complexities of the instrumentation work by the non-experts of the group, and it is even in some cases belittled (sic) or taken for granted. One respondent indicated they are not seen as real physicists.
    
    \item Instrumentation has a smaller rate of publications than other areas (i.e.~theory or experiment data analysis). This can result in instrumentation having a smaller specific weight in the group, especially in terms of recognition.
    
    \item There is a lack of permanent positions for detector physicists. One respondent indicated that they were given feedback they had done too much instrumentation work and not enough analysis in a post-interview follow up. At the Training symposium, it was brought up that tracking data on the number of permanent positions that are allocated for work primarily or exclusively related to instrumentation would provide useful insight.
    
    \item There is a lack of opportunities for leadership experience for junior members. Comments in this regard mentioned that presentations are often given on behalf of a working group, so individual contributions are not visible. Junior members are not usually invited to organise workshops, with these tasks being relegated to senior members.
\end{itemize}

\section{Networks for Instrumentation}

When asked whether they were satisfied with the networking opportunities available to ECRs in instrumentation, only 22\% agreed or strongly agreed, meaning that 78\% do not feel positively about their networking opportunities. 26 respondents provided comments on why they disagreed or strongly disagreed with the statement:

\begin{itemize}
    \item Many comments highlighted the difficulty accessing networks in instrumentation relative to other areas.
    
    \item One comment suggested providing more training on ‘how to network’.
    
    \item Several noted additional challenges due to the COVID-19 pandemic.
\end{itemize}

Participants were also asked for comments on ideas for improving the networking opportunities for ECRs; replies included:

\begin{itemize}
    \item A suggestion to make instrumentation work less ‘top-down’, so ECRs get more exposure to senior collaborators.
    
    \item More workshops/summer schools devoted to instrumentation to connect those working in different experiments/user facilities.
    
    \item Increase links between instrumentation projects and those providing the training, so training is more applicable to the ongoing work.
    
    \item Dedicated discussion forums for instrumentation.
\end{itemize}

\begin{figure}[h!]
\centering
\begin{subfigure}[b]{.45\linewidth}
\includegraphics[width=\linewidth]{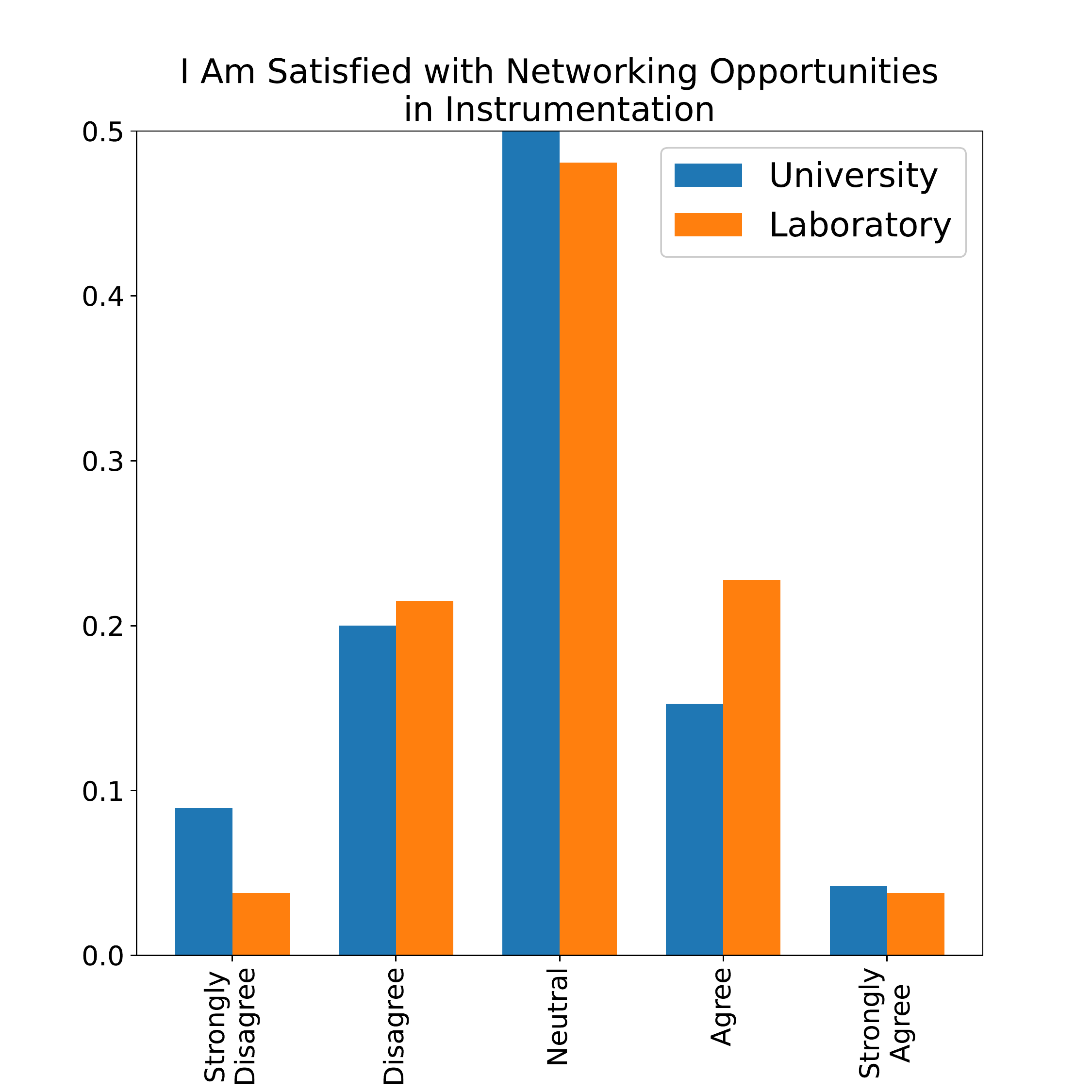}
\caption{}
\end{subfigure}
\begin{subfigure}[b]{.45\linewidth}
\includegraphics[width=\linewidth]{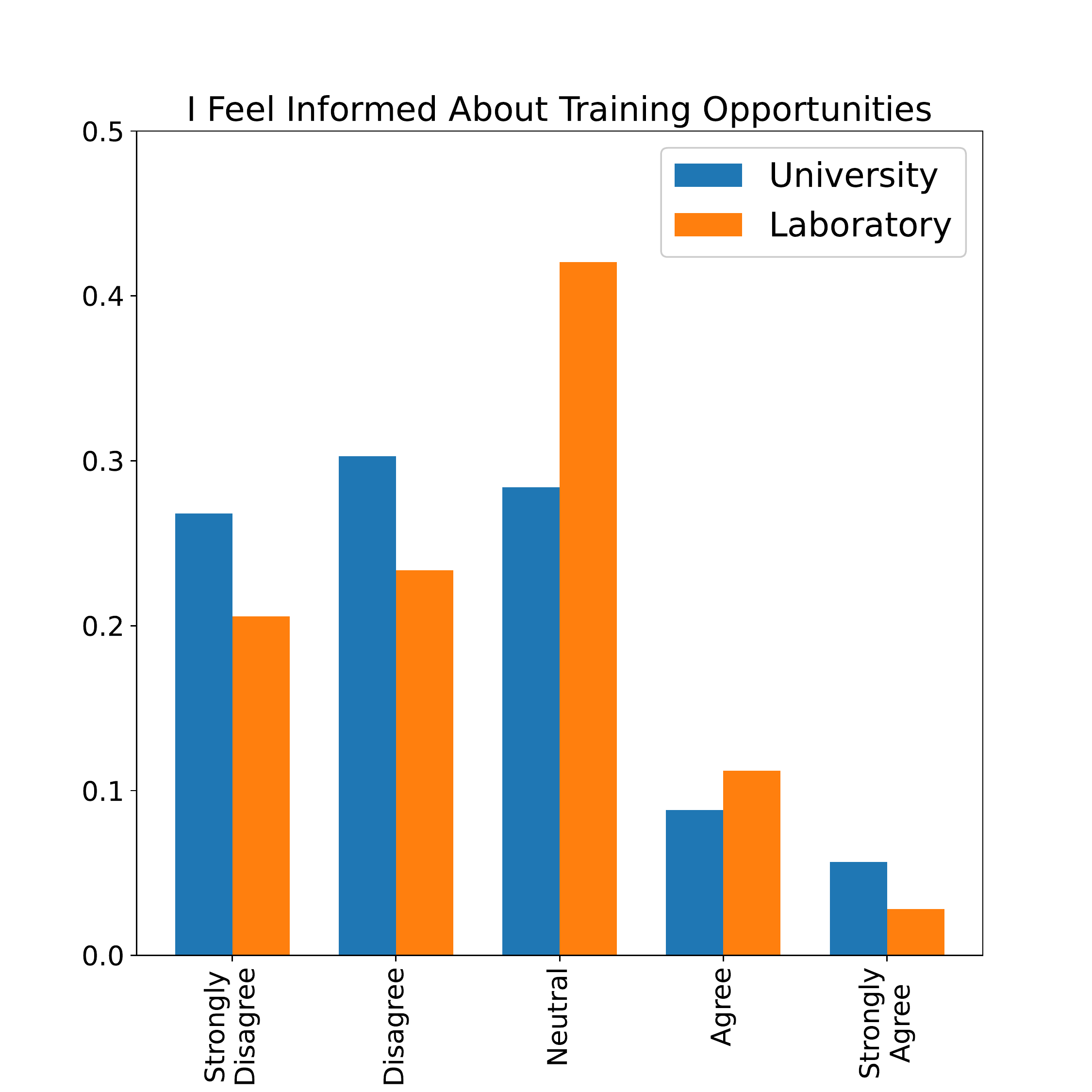}
\caption{}
\end{subfigure}
\\[-0.5cm]
\begin{subfigure}[b]{.45\linewidth}
\includegraphics[width=\linewidth]{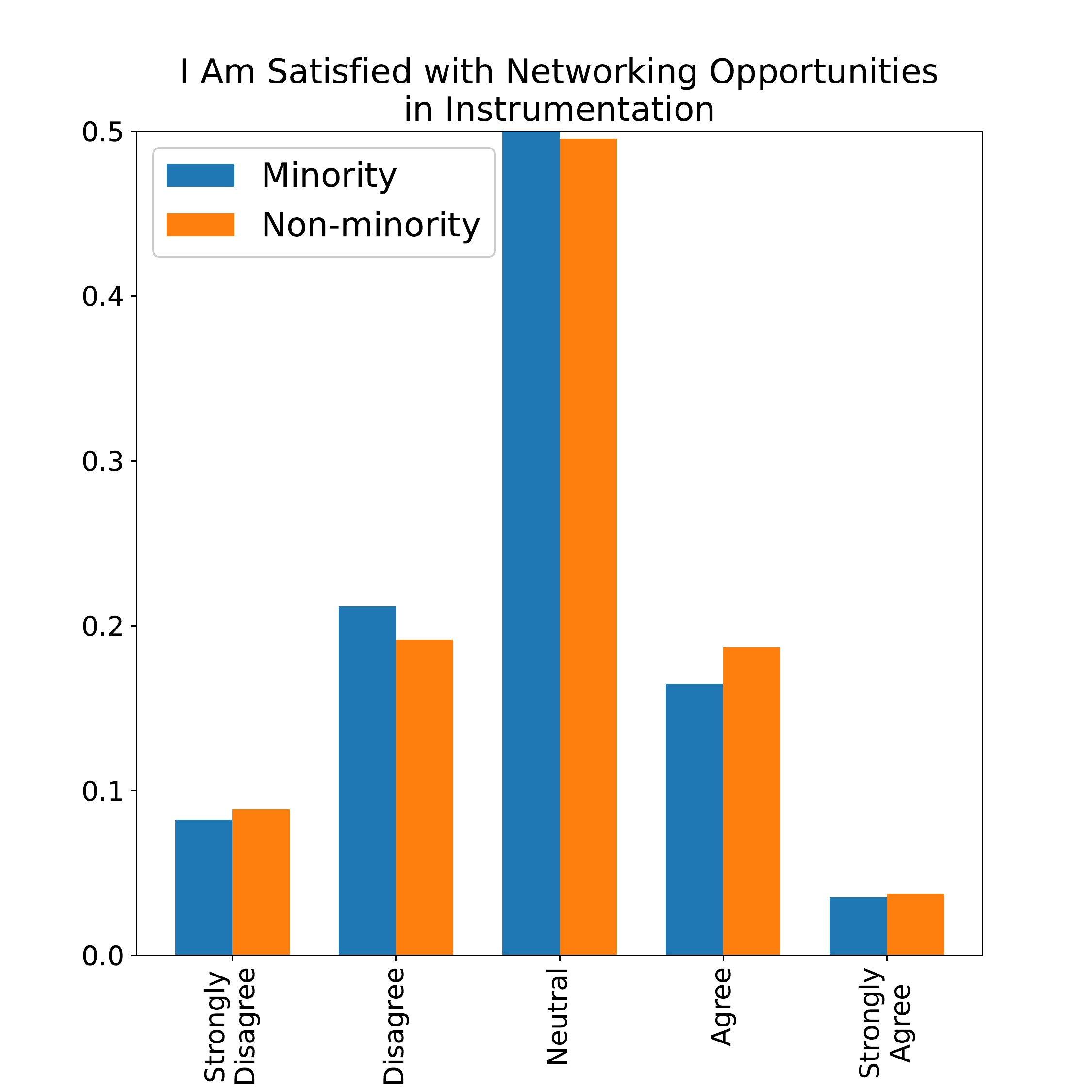}
\caption{}
\end{subfigure}
\begin{subfigure}[b]{.45\linewidth}
\includegraphics[width=\linewidth]{BodyFigures/informed-min.pdf}
\caption{}
\end{subfigure}
\\[-0.5cm]
\begin{subfigure}[b]{.45\linewidth}
\includegraphics[width=\linewidth]{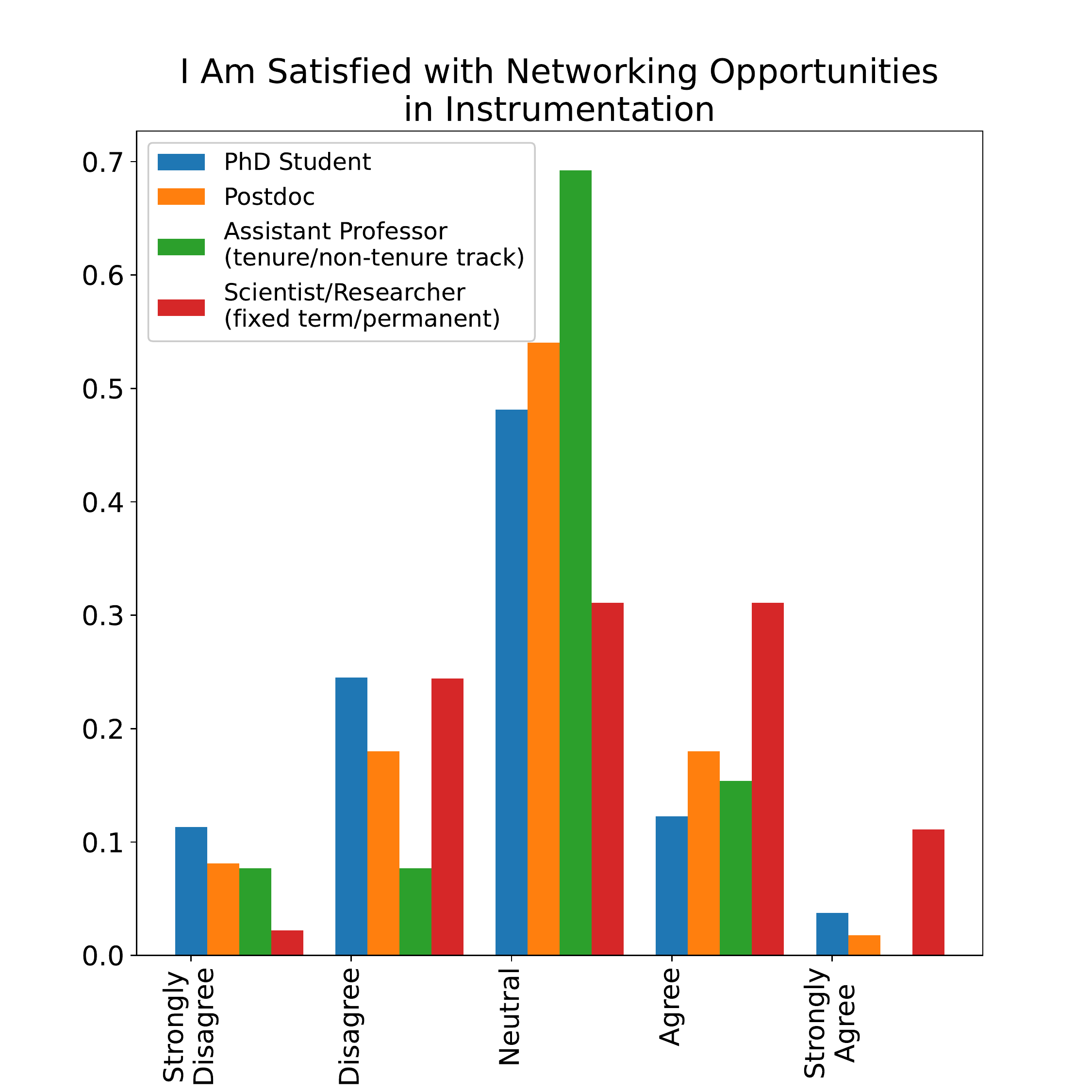}
\caption{}
\end{subfigure}
\begin{subfigure}[b]{.45\linewidth}
\includegraphics[width=\linewidth]{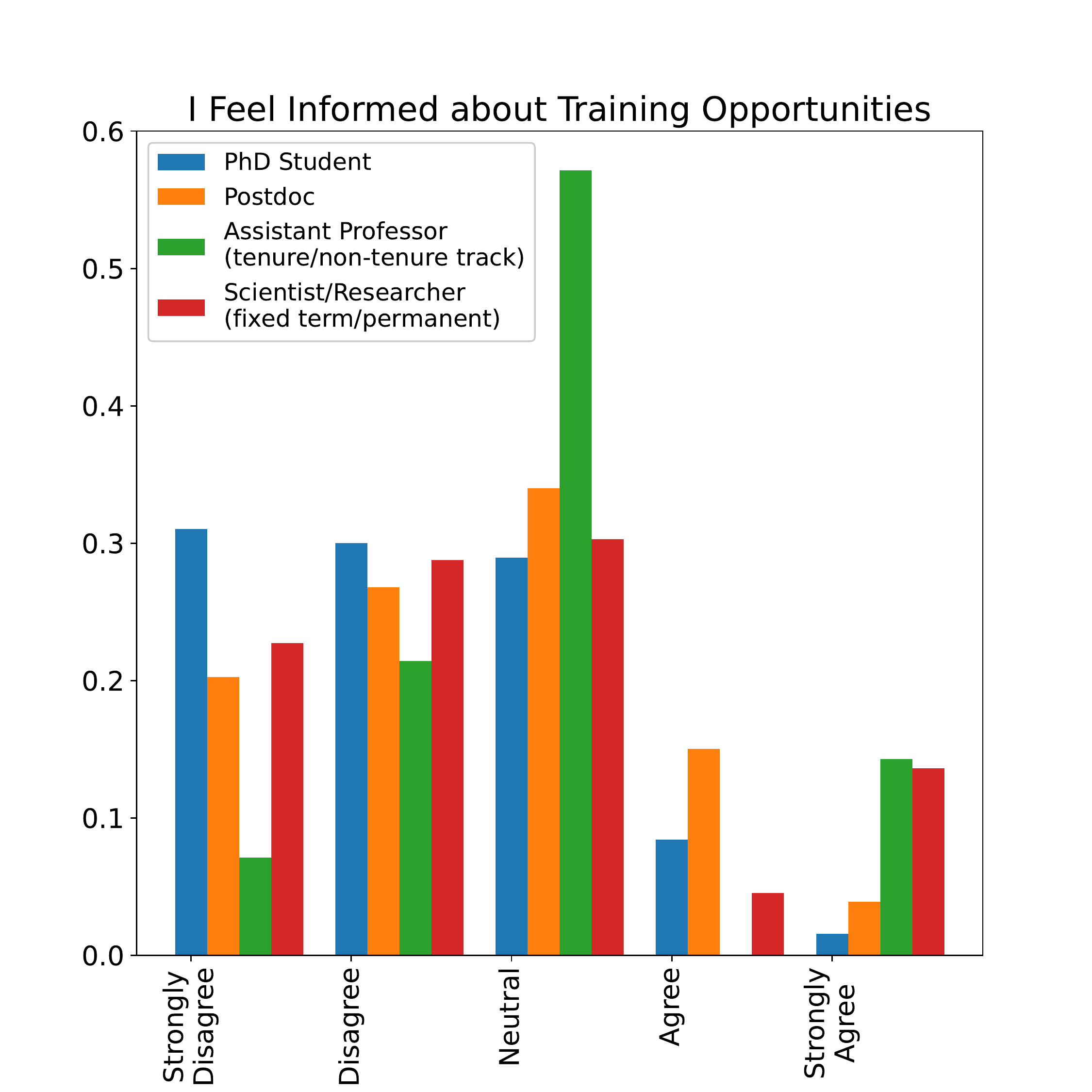}
\caption{}
\end{subfigure}
\caption{Comparison of normalised responses to the networking questions. The plots on the left refer to the statement “I am satisfied with networking opportunities for early career researchers in instrumentation that are available to me” (304 responses), and the right hand plots show how well-informed respondents feel about training opportunities (471 responses). The top row compares university-based and laboratory based respondents, the middle row is those that did and didn’t identify as a minority, and the bottom row differentiates between ECR career stages. Figures with the raw number of responses for certain populations are found in Appendix B.} 
\label{fig:Networking}
\end{figure}

\Figure{Networking} shows comparisons between the responses to questions related to networking between university- and non-university based respondents, those that did and didn’t identify as a minority group, and for the different ECR career stages.

\section{Engineers working in instrumentation}

During the Townhall, it was noted that no engineers were present in the discussion. There was a discussion as to whether this was because of a lack of convening power from the ECR-ECFA panel, or if it is a more structural issue. To try to mitigate this issue, the survey advertisement directly requested input from engineers, and there were questions explicitly defined for only engineers. Nonetheless, among 473 survey participants, only 25 identified themselves as working in engineering. In addition to these 25, 12 more participants identified themselves as working in computer engineering (software/firmware development) related tasks.
The replies of engineers to the engineer-specific questions were very interesting, as discussed in the talk and provided in Appendix A. However, we have not further cross-analysed the engineering replies given the low number of responses we received.  We would encourage the community to make an effort to further include ECR engineer input in the discussion, and hope that the people concerned will have provided feedback through other channels.

\FloatBarrier

\section{Open Suggestions from Survey Respondents}

The final section of the survey gave respondents the opportunity to provide open suggestions on how to improve ECR experiences within instrumentation training. Suggestions included:

\begin{itemize}
    \item A respondent brought up an issue that was iterated first at the Townhall discussion and reiterated at the Training Symposium, which is that experiments with shorter timelines give trainees the opportunity for continuity between R\&D and detector/experiment commissioning and operation. 
    
    \item A respondent pointed out in particular that neutrino experiments and smaller experiments are unique from larger collaborations in that researchers are not divided into analysis and instrumentation tracks, instead these tasks are more integrated in the work of any individual researcher.
    
    \item Young researchers would benefit from open laboratories at large facilities so they could take training in instrumentation.
    
    \item The lack of accessibility to training and facilities for under-resourced regions was highlighted.
    
    \item A common theme is a desire to shift norms around hiring practices, with a general belief that instrumentation work is undervalued.
\end{itemize}

\section{Summary}

This report summarises the survey circulated to Early-Career Researchers on their experiences and views in training in instrumentation, as well as a Townhall discussion amongst early career members on April 7, 2021. These efforts were instigated as a response to an invitation by the ECFA Detector R\&D Roadmap Training Task Force coordinators for junior researchers to provide input to the roadmap. This report is supplemental to the presentation given at the Roadmap Symposium on Training, on April 30, 2021. The authors have refrained from making recommendations at this stage and reported only highlights of the results, while the full set of summary plots are provided in Appendix A and raw cross-analysis results can be found in Appendix B. We hope that this report will provide fruitful input to the ongoing ECFA Detector R\&D Roadmap, and are happy to engage in further discussions within the community to improve training provisions for instrumentation work.

\addcontentsline{toc}{section}{Bibliography}
\printbibliography[title=References]

\clearpage
\addcontentsline{toc}{section}{Appendix A}
\section*{Appendix A: Raw survey summary results}

The following plots are raw summaries of replies to each survey question, other than free-form questions.  In some cases, people were allowed to provide customised replies on top of selecting existing reply categories.  In such cases, replies have been grouped into either existing or new categories based on commonality of content.  This was done both to condense the number of reply categories into a reasonable multiplicity to be shown on the plot, and to avoid listing information that could potentially lead to identifying specific respondents.

\vspace{0.5cm}
\begin{figure}[h!]
    \centering
    \includegraphics[height=0.25\textheight,width=0.9\textwidth,keepaspectratio]{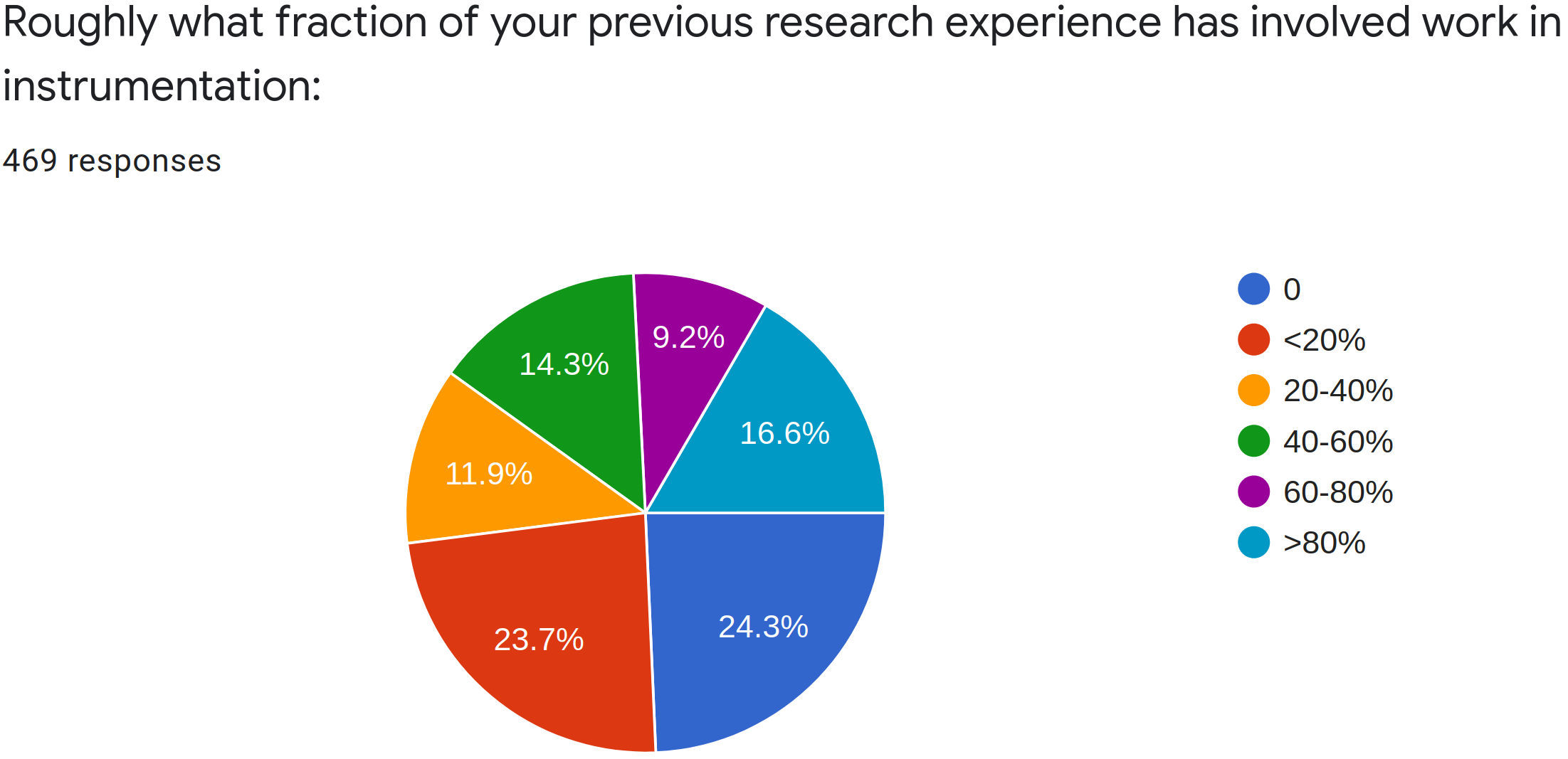}
    \caption{}
\end{figure}
\begin{figure}[h!]
    \centering
    \includegraphics[height=0.25\textheight,width=0.9\textwidth,keepaspectratio]{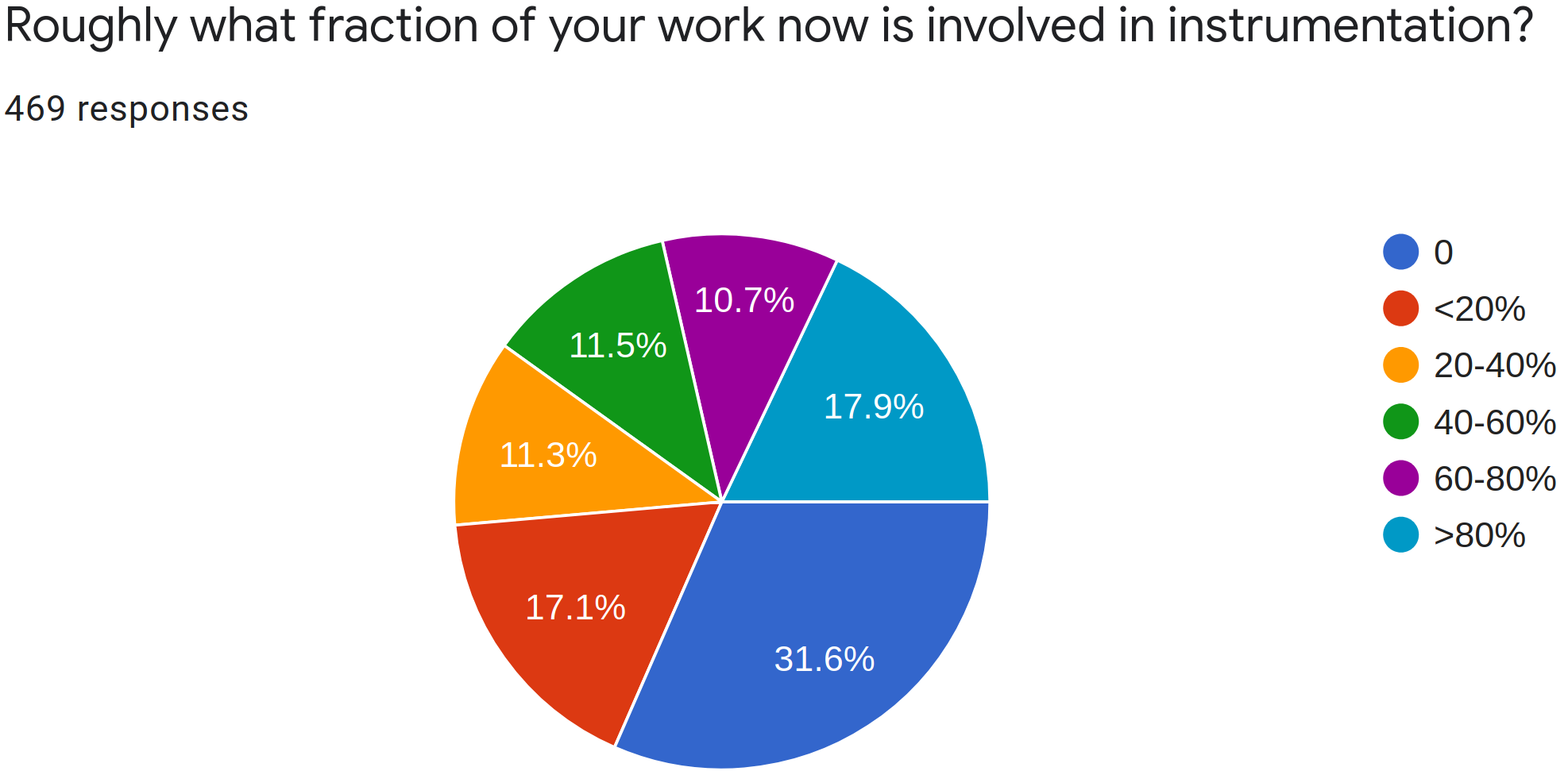}
    \caption{}
\end{figure}
\begin{figure}[h!]
    \centering
    \includegraphics[height=0.25\textheight,width=0.9\textwidth,keepaspectratio]{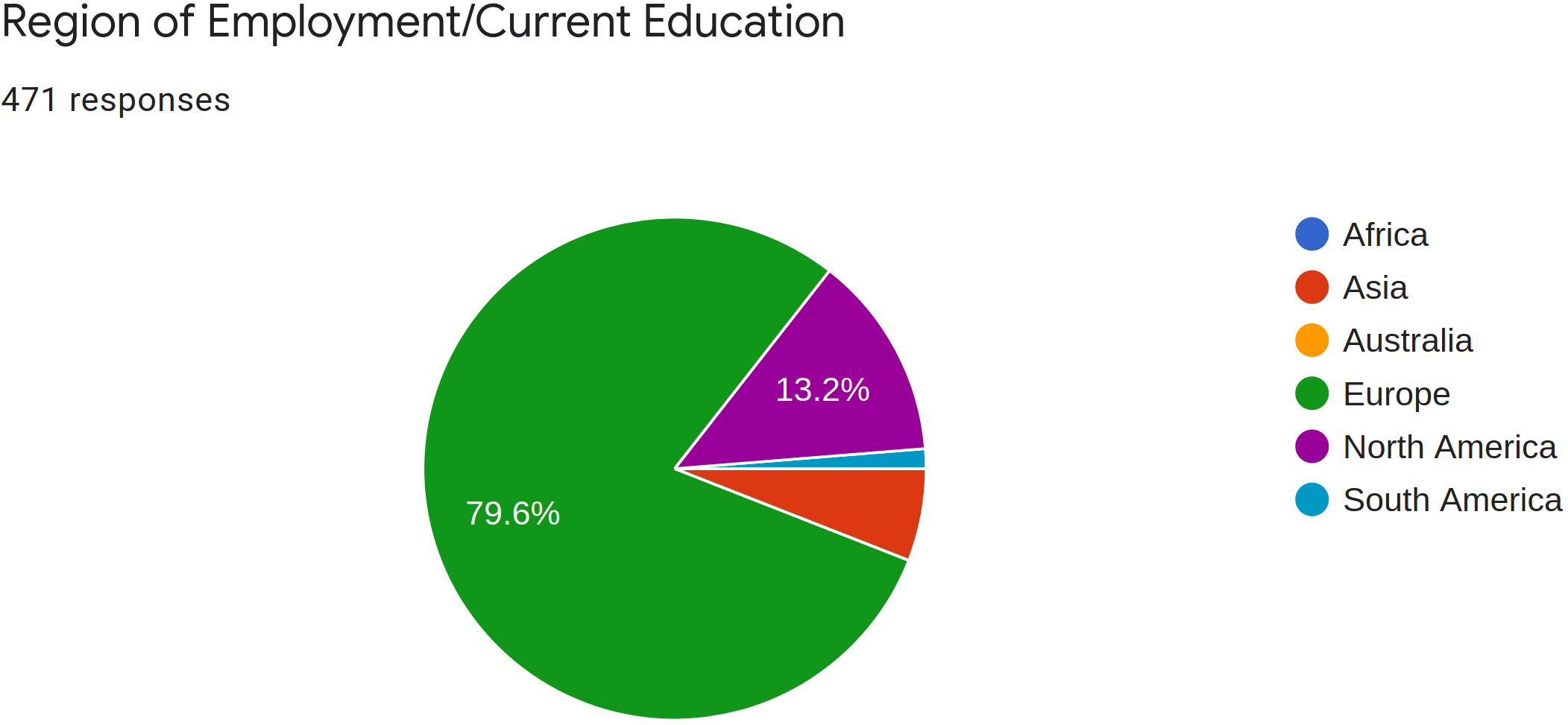}
    \caption{}
\end{figure}

\begin{figure}[h!]
    \centering
    \includegraphics[width=\textwidth]{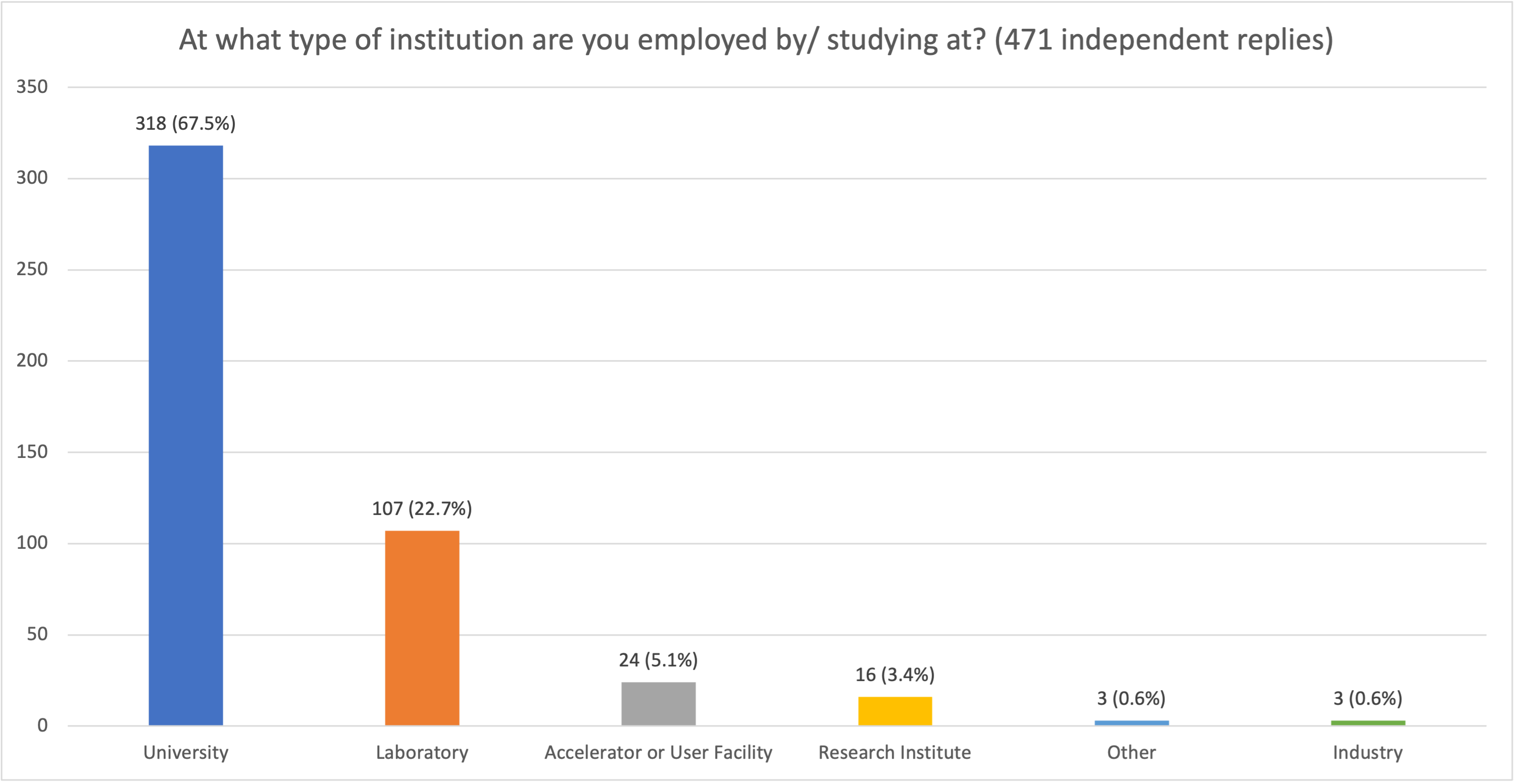}
    \caption{}
\end{figure}
\begin{figure}[h!]
    \centering
    \includegraphics[width=\textwidth]{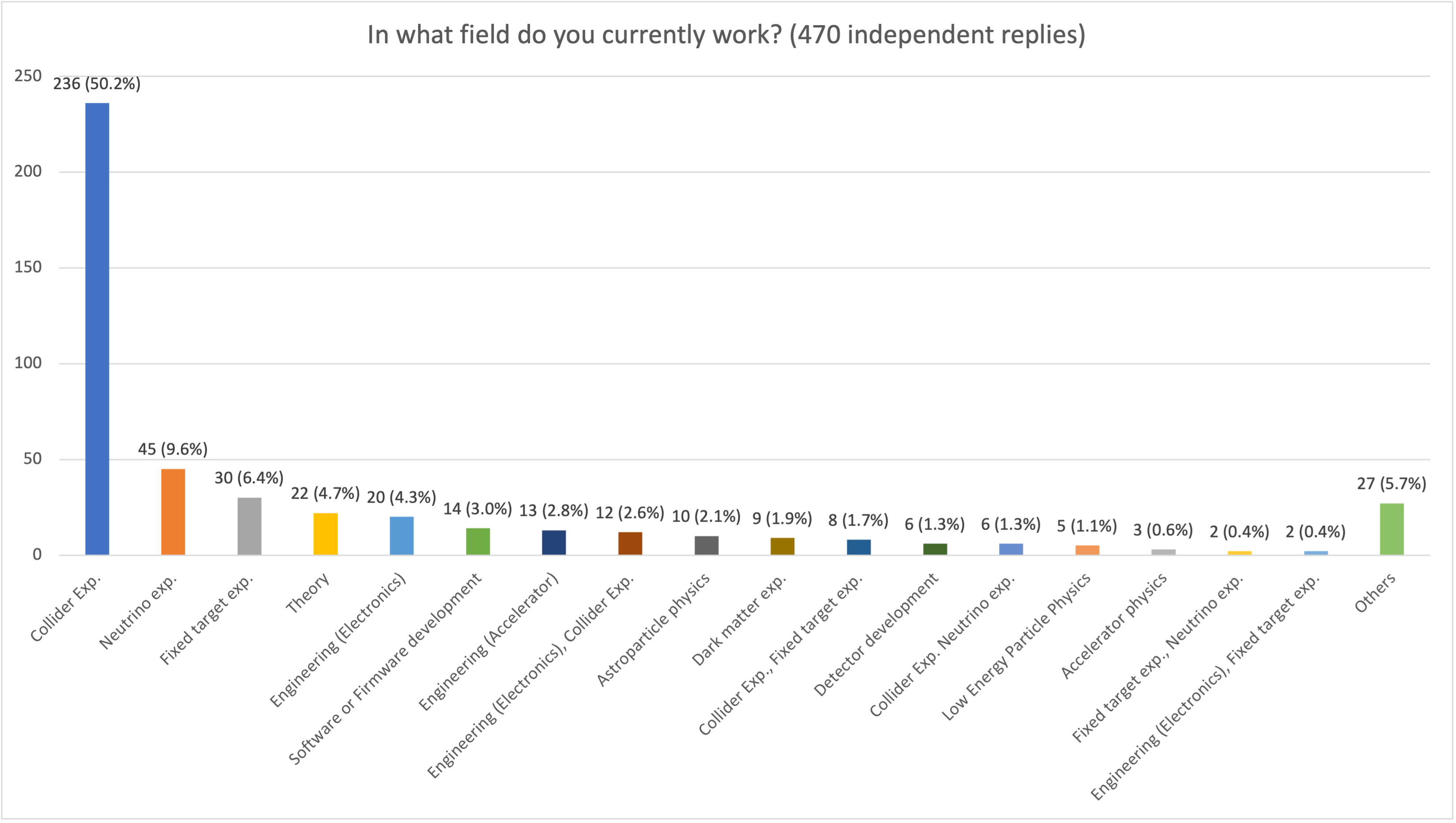}
    \caption{}
\end{figure}

\begin{figure}[h!]
    \centering
    \includegraphics[width=\textwidth]{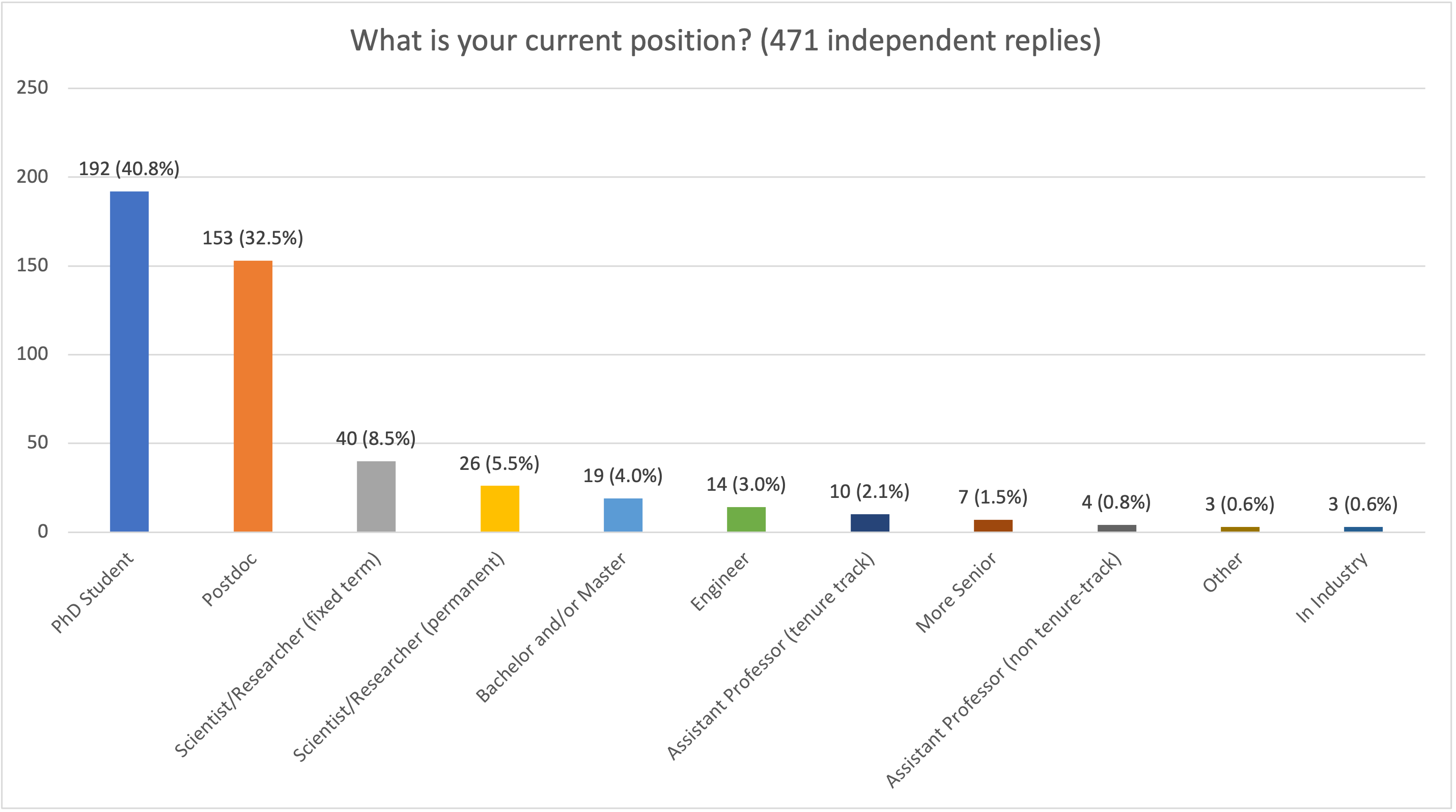}
    \caption{}
\end{figure}
\begin{figure}[h!]
    \centering
    \includegraphics[height=0.25\textheight,width=0.9\textwidth,keepaspectratio]{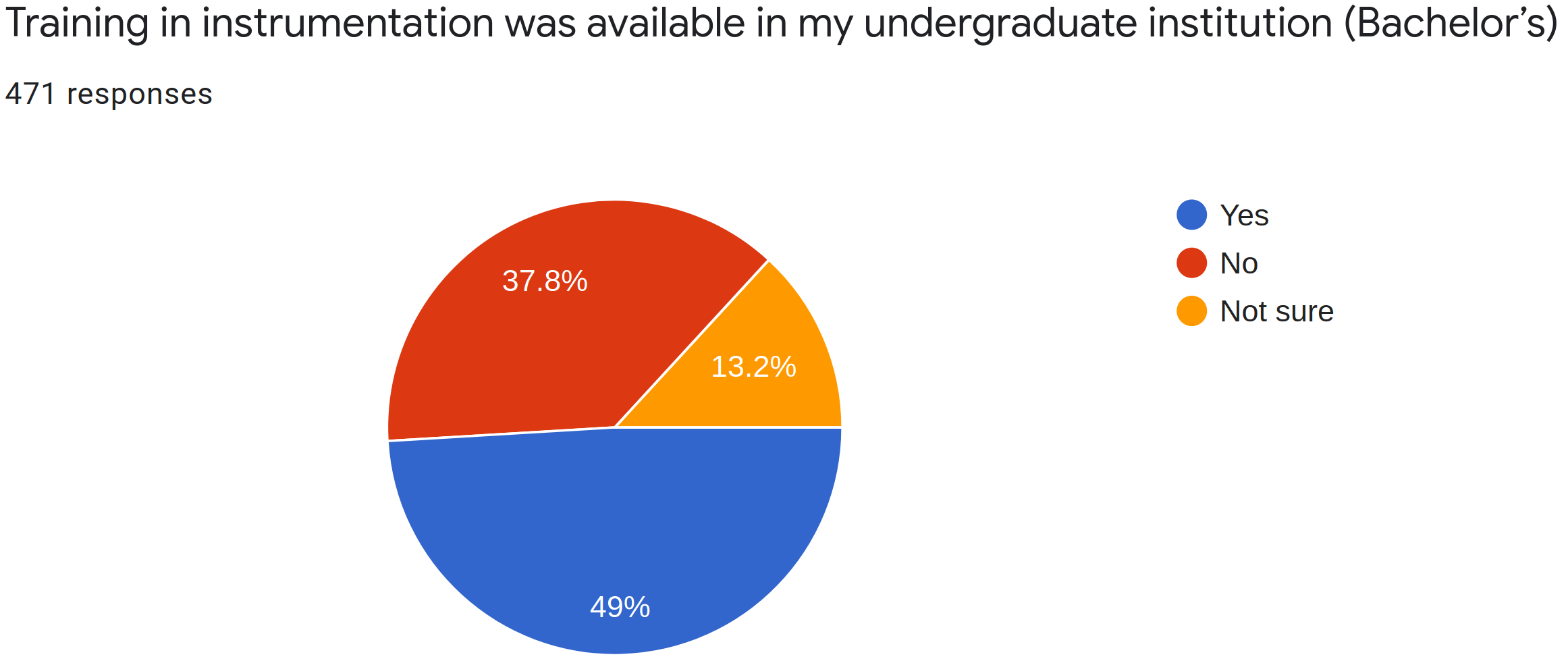}
    \caption{}
\end{figure}
\begin{figure}[h!]
    \centering
    \includegraphics[height=0.25\textheight,width=0.9\textwidth,keepaspectratio]{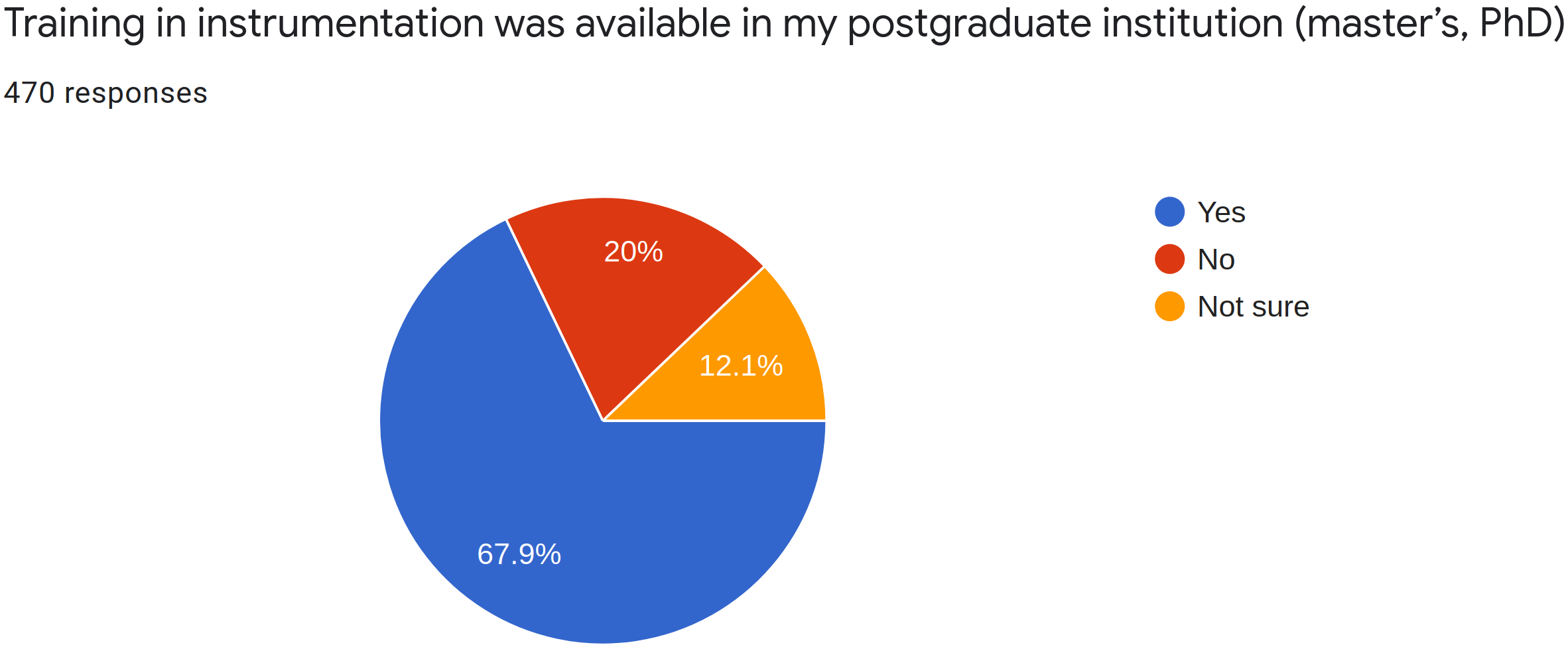}
    \caption{}
\end{figure}

\begin{figure}[h!]
    \centering
    \includegraphics[height=0.25\textheight,width=0.9\textwidth,keepaspectratio]{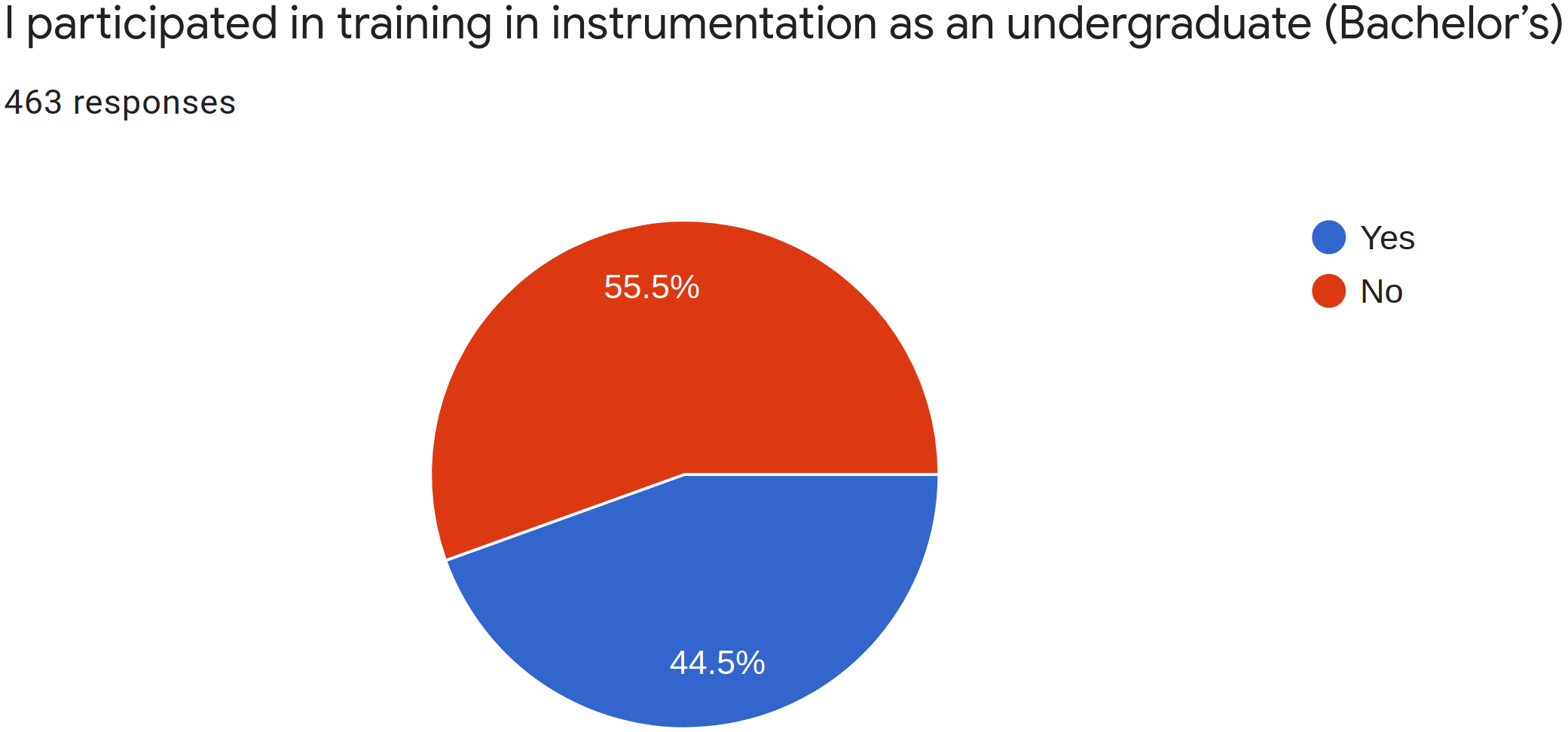}
    \caption{}
\end{figure}
\begin{figure}[h!]
    \centering
    \includegraphics[height=0.25\textheight,width=0.9\textwidth,keepaspectratio]{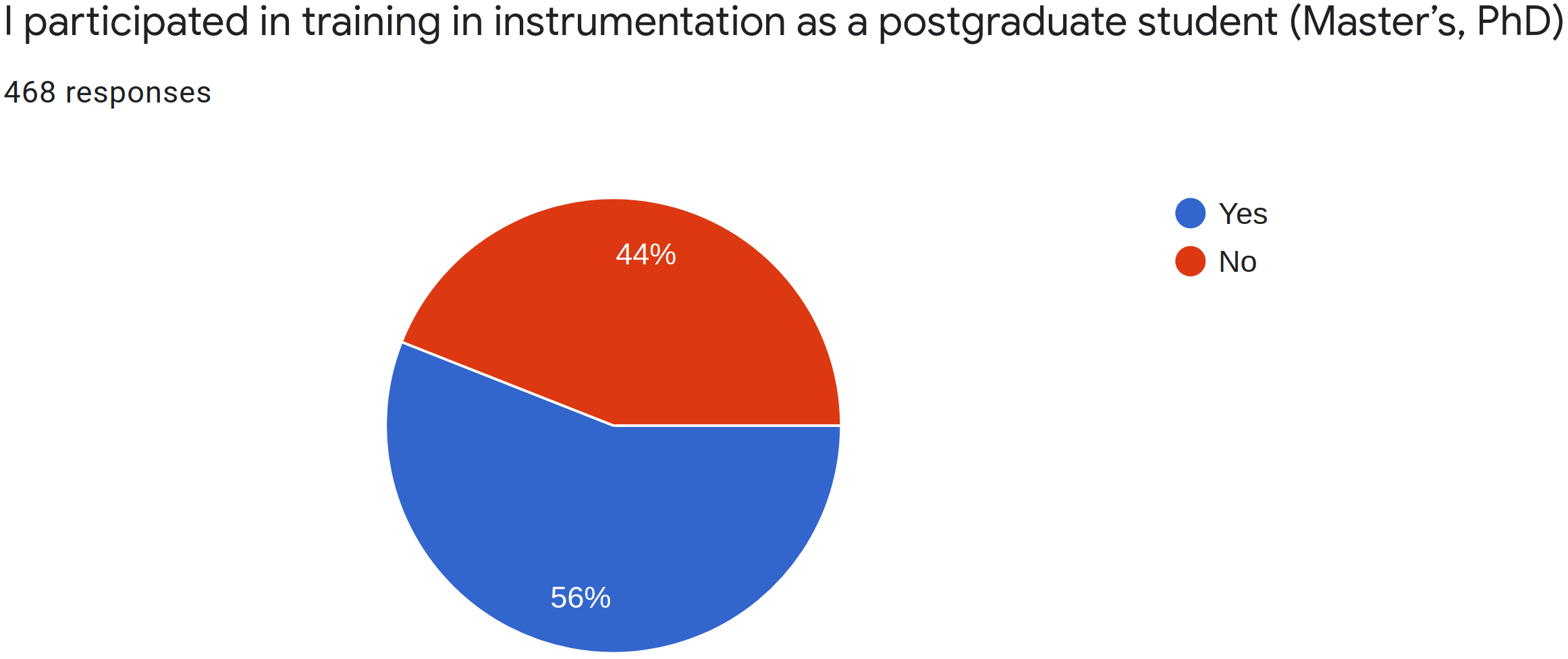}
    \caption{}
\end{figure}
\begin{figure}[h!]
    \centering
    \includegraphics[width=\textwidth]{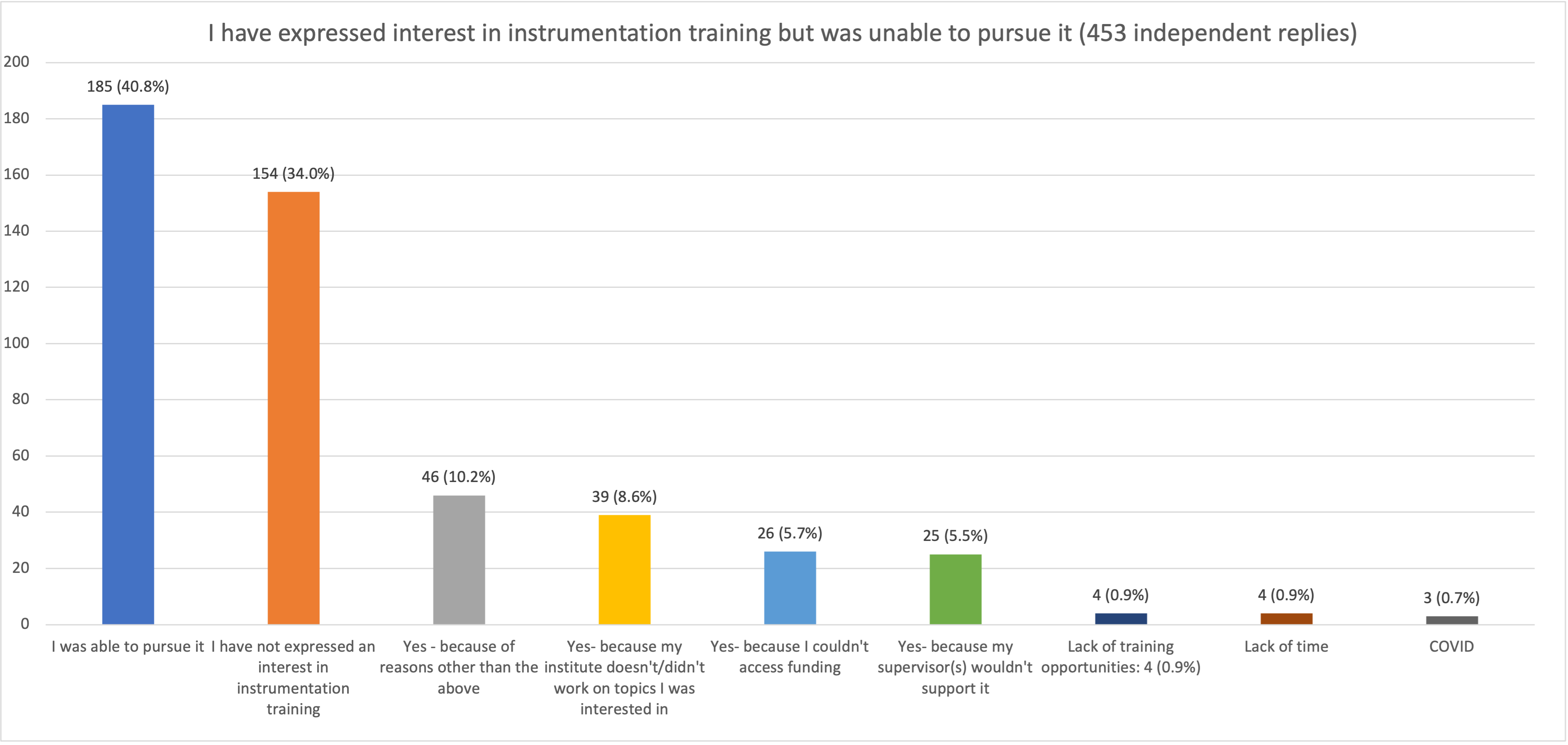}
    \caption{}
\end{figure}

\begin{figure}[h!]
    \centering
    \includegraphics[height=0.25\textheight,width=0.9\textwidth,keepaspectratio]{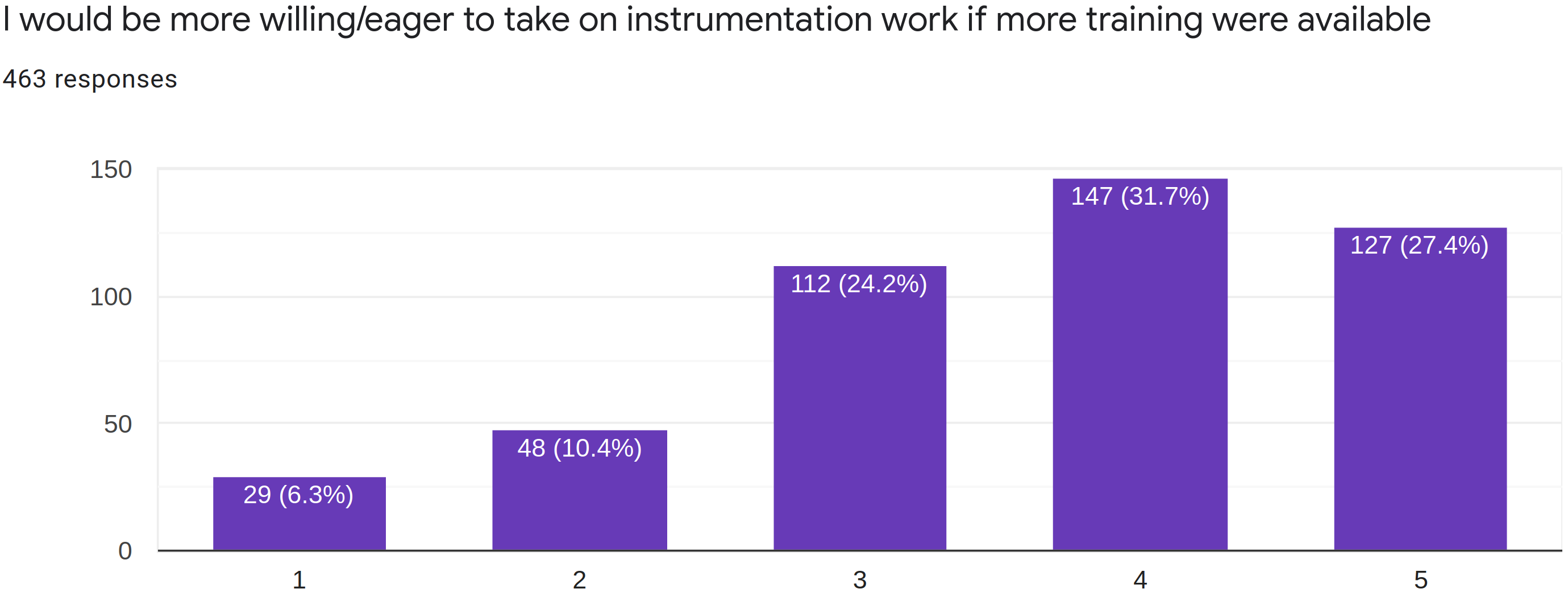}
    \caption{}
\end{figure}
\begin{figure}[h!]
    \centering
    \includegraphics[width=\textwidth]{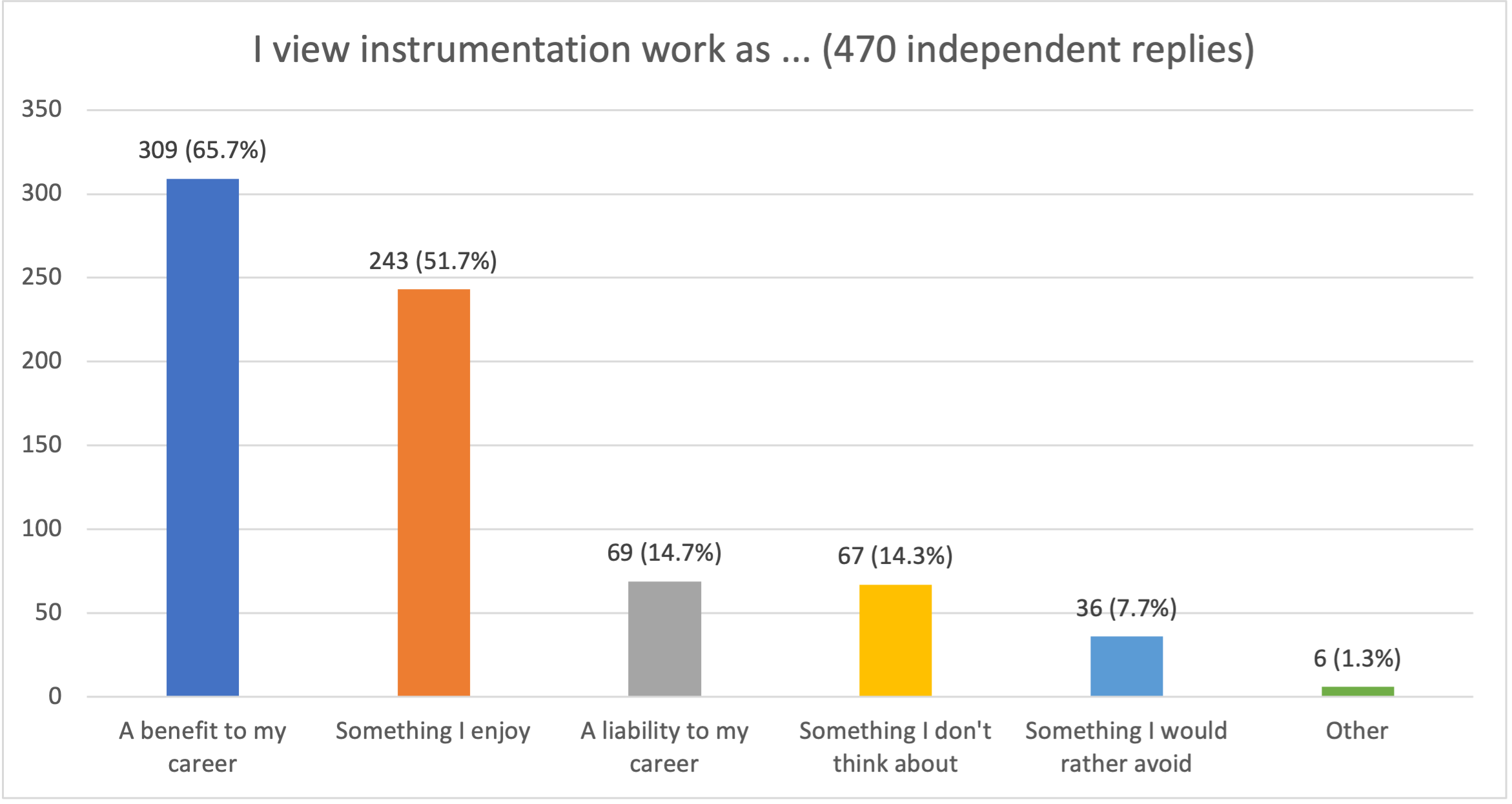}
    \caption{}
\end{figure}
\begin{figure}[h!]
    \centering
    \includegraphics[width=\textwidth]{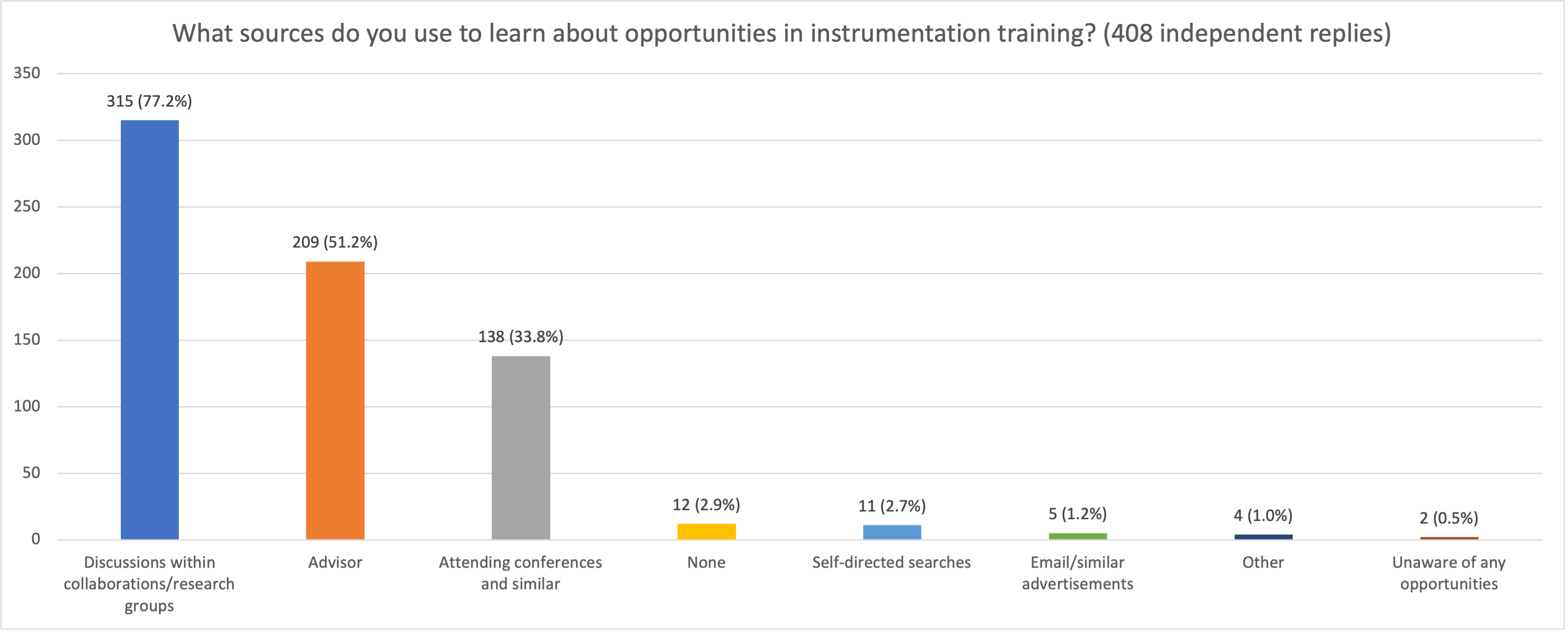}
    \caption{}
\end{figure}

\begin{figure}[h!]
    \centering
    \includegraphics[height=0.25\textheight,width=0.9\textwidth,keepaspectratio]{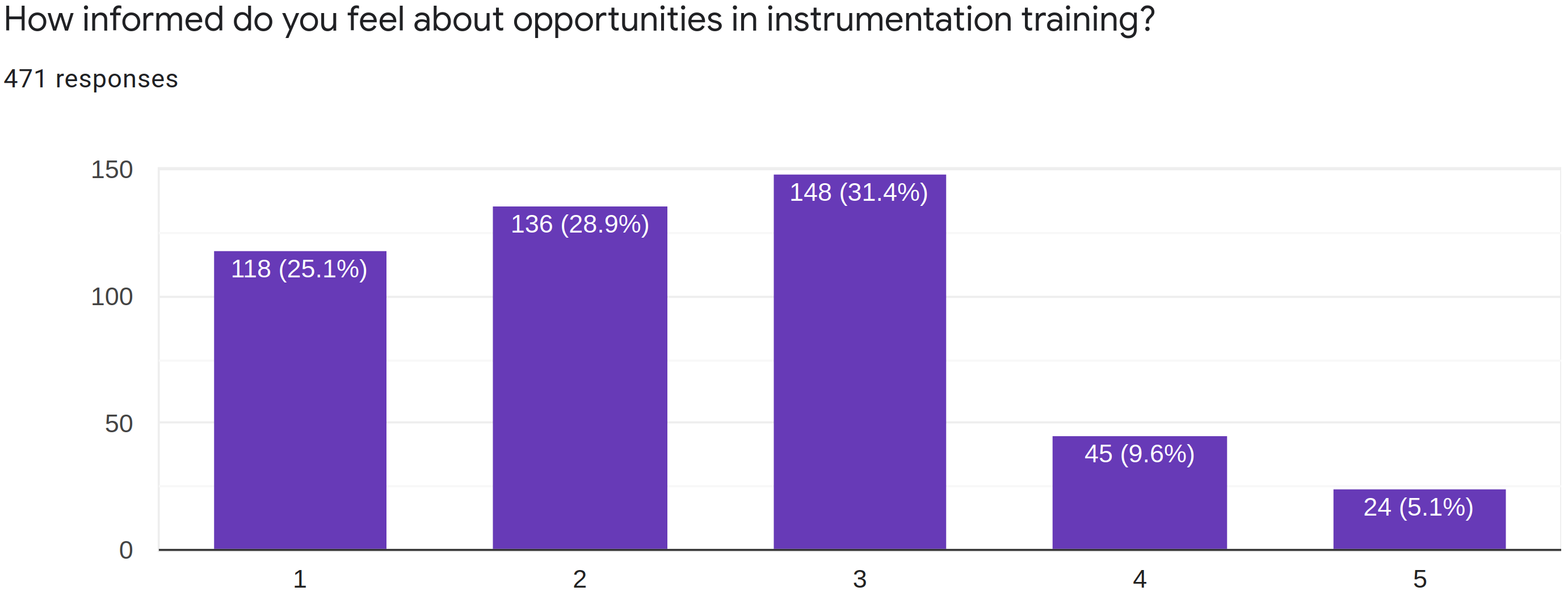}
    \caption{}
\end{figure}
\begin{figure}[h!]
    \centering
    \includegraphics[height=0.25\textheight,width=0.9\textwidth,keepaspectratio]{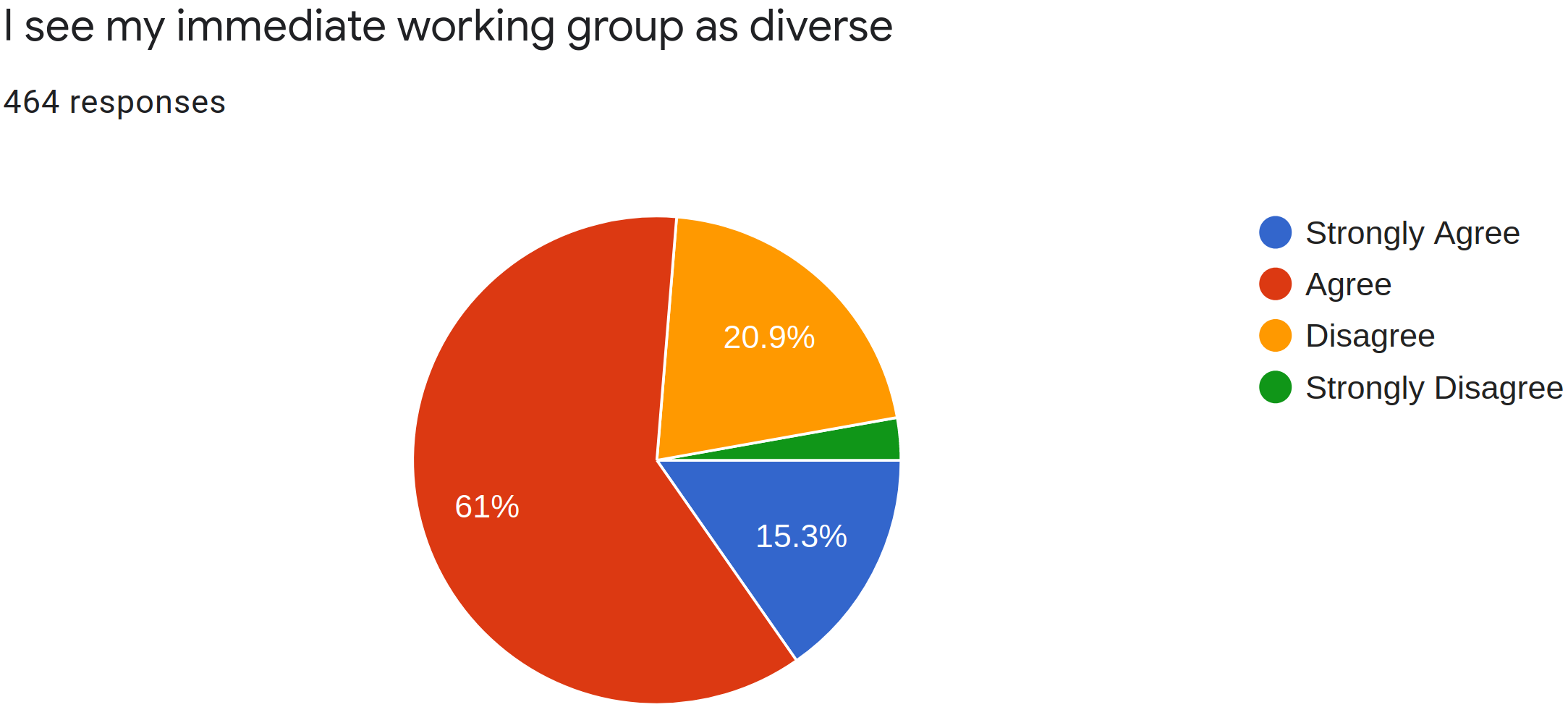}
    \caption{}
    \label{fig:informed}
\end{figure}
\begin{figure}[h!]
    \centering
    \includegraphics[height=0.25\textheight,width=0.9\textwidth,keepaspectratio]{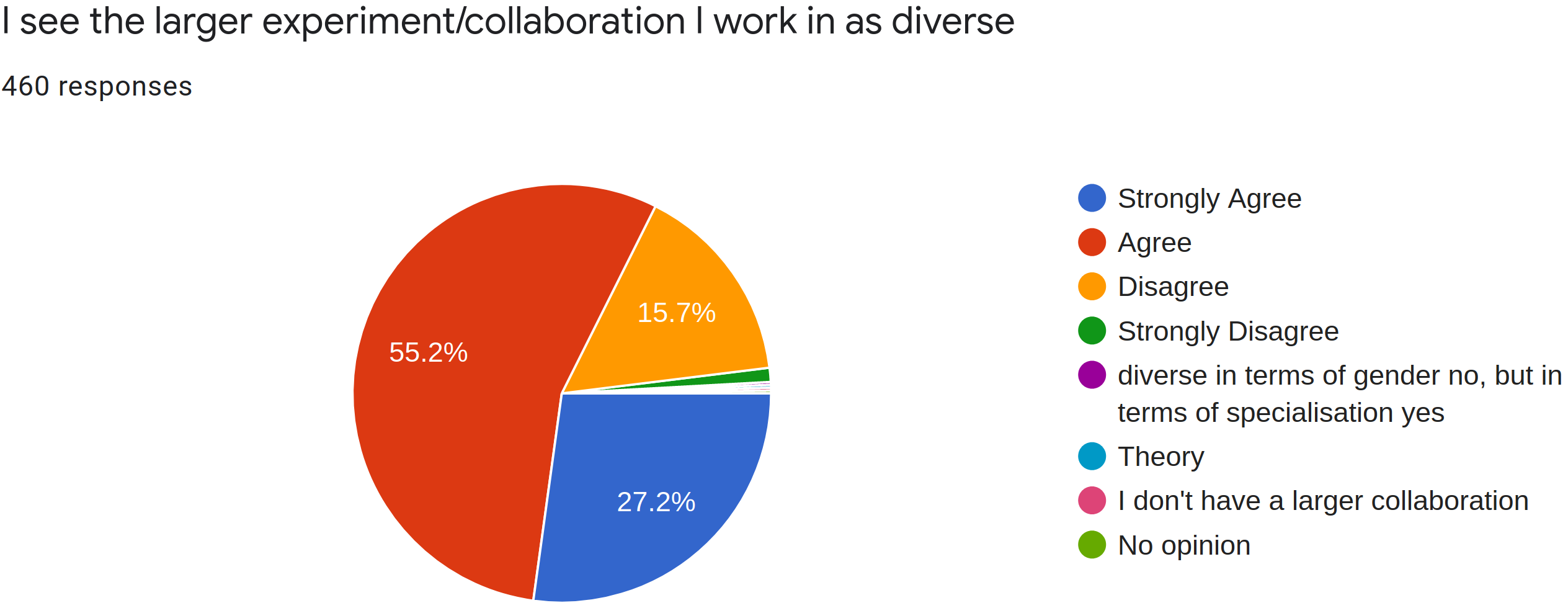}
    \caption{}
\end{figure}
\begin{figure}[h!]
    \centering
    \includegraphics[height=0.25\textheight,width=0.9\textwidth,keepaspectratio]{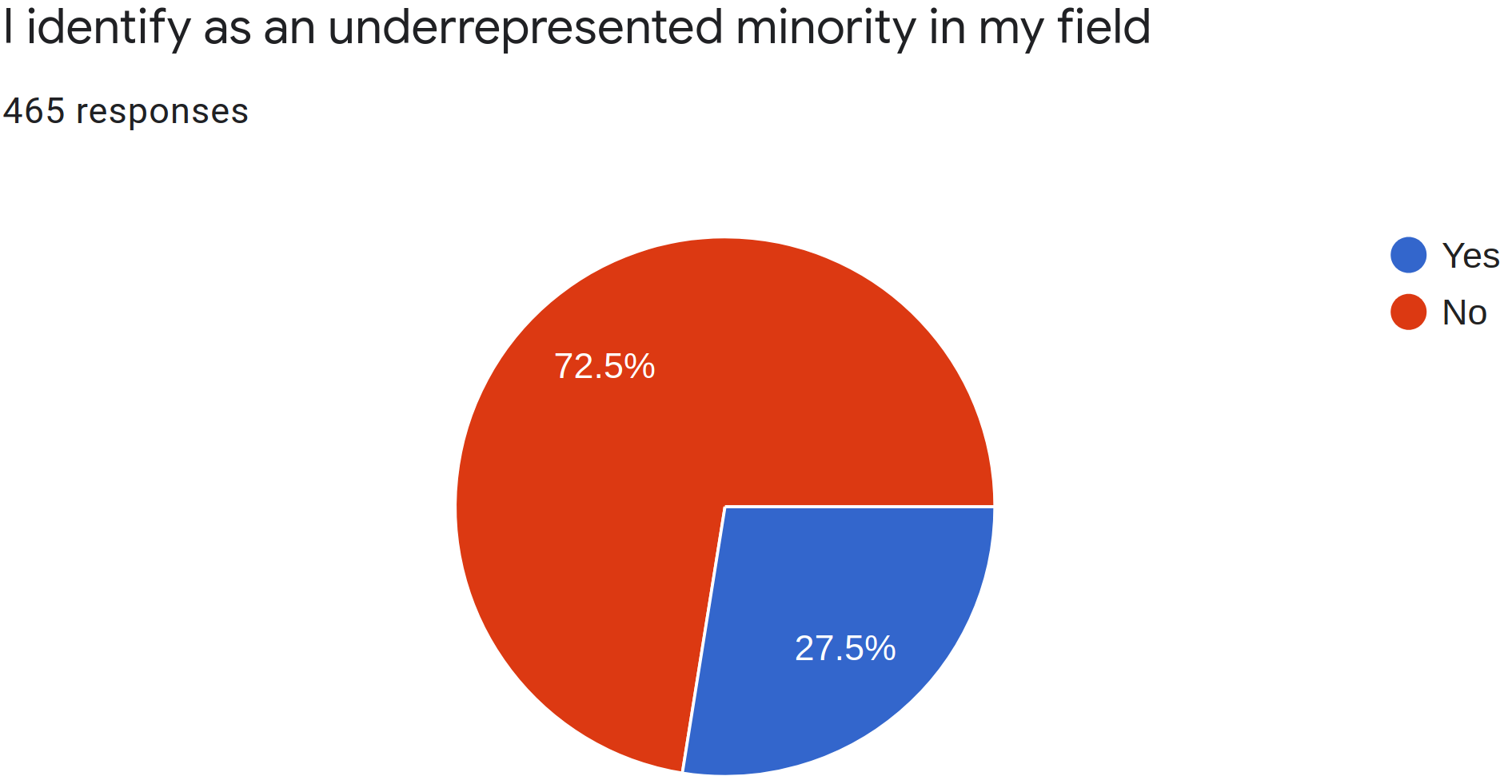}
    \caption{}
\end{figure}

\begin{figure}[h!]
    \centering
    \includegraphics[height=0.25\textheight,width=0.9\textwidth,keepaspectratio]{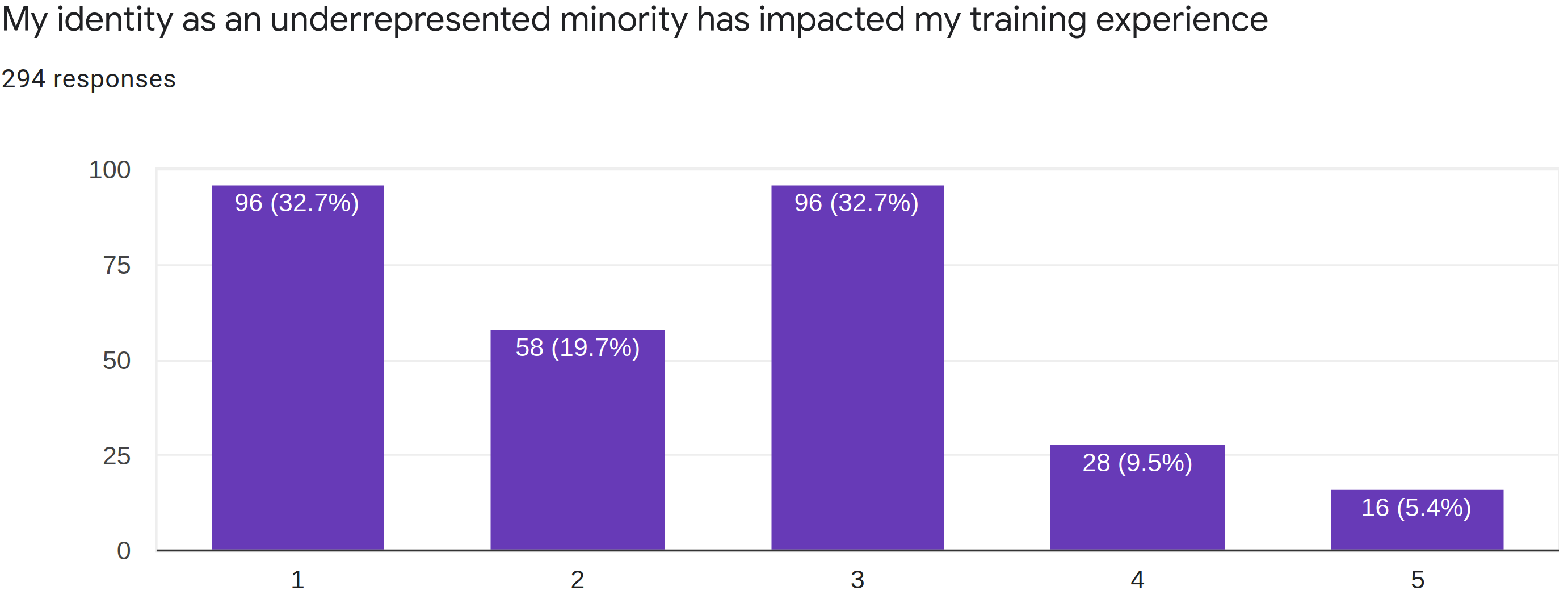}
    \caption{}
\end{figure}
\begin{figure}[h!]
    \centering
    \includegraphics[width=\textwidth]{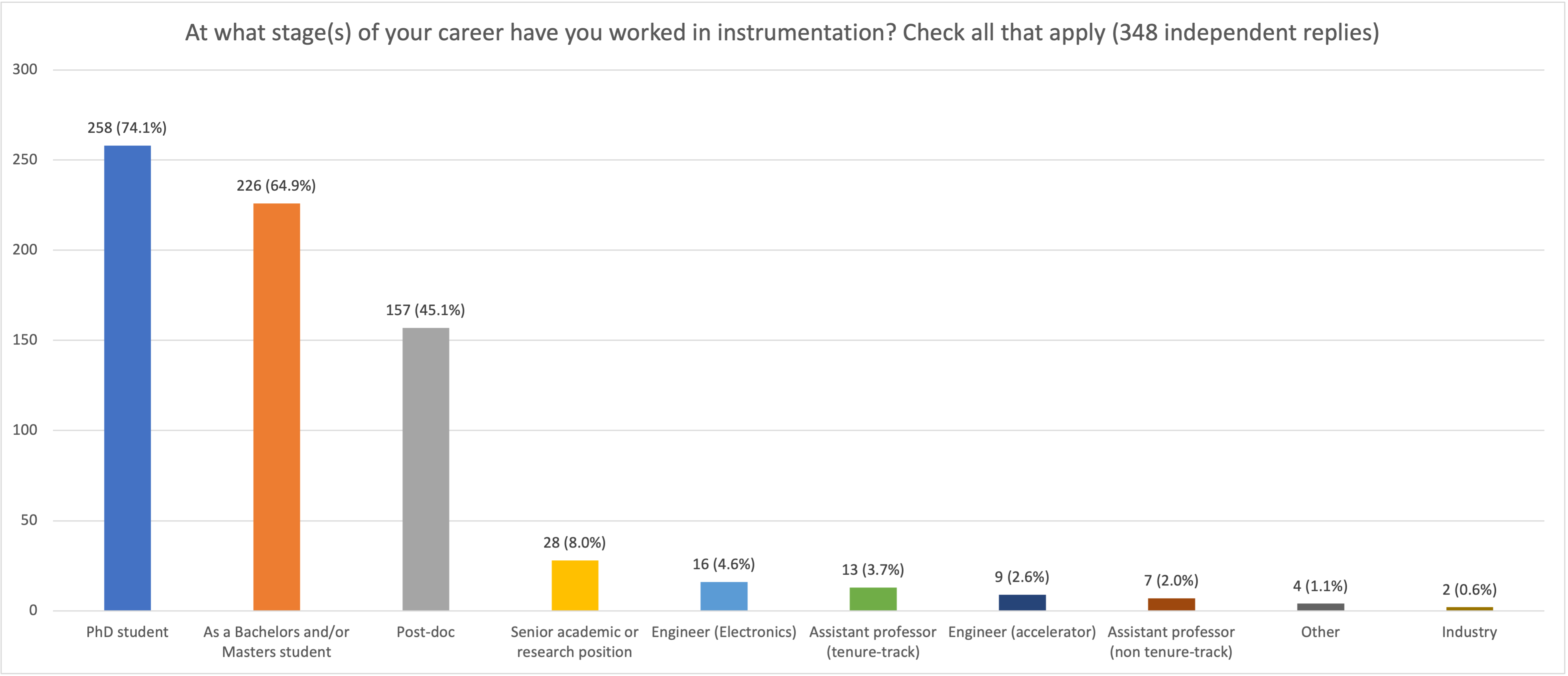}
    \caption{}
\end{figure}
\begin{figure}[h!]
    \centering
    \includegraphics[width=\textwidth]{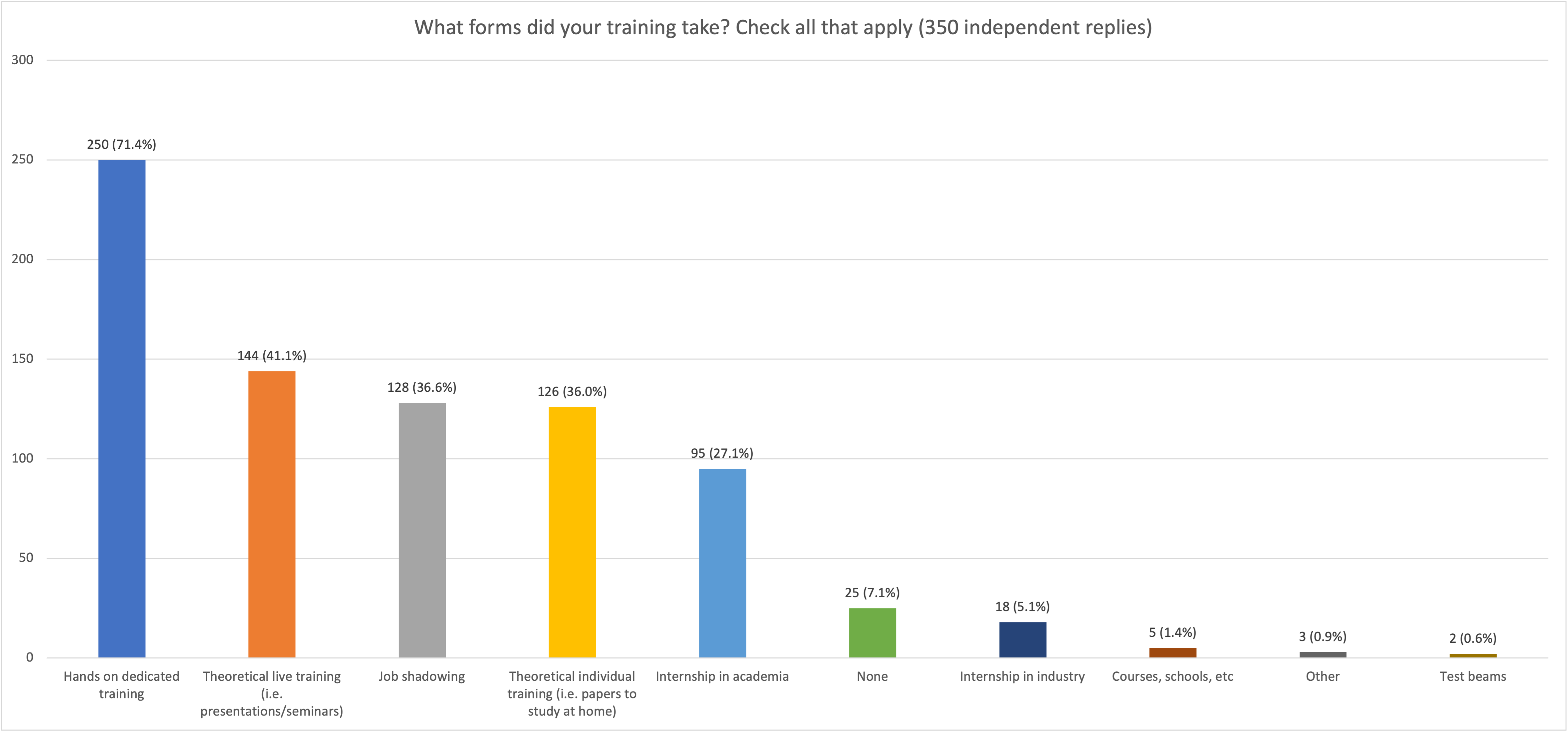}
    \caption{}
\end{figure}

\begin{figure}[h!]
    \centering
    \includegraphics[height=0.25\textheight,width=0.9\textwidth,keepaspectratio]{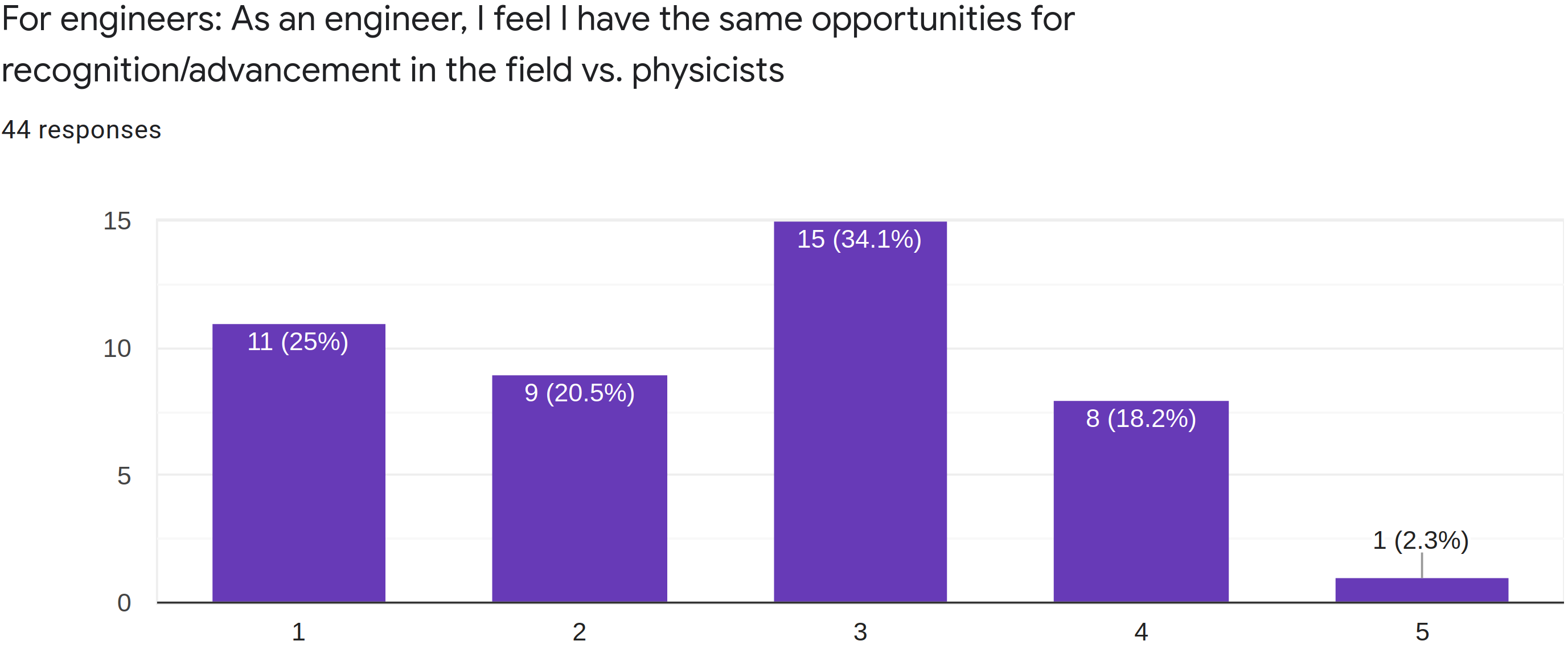}
    \caption{}
\end{figure}
\begin{figure}[h!]
    \centering
    \includegraphics[height=0.25\textheight,width=0.9\textwidth,keepaspectratio]{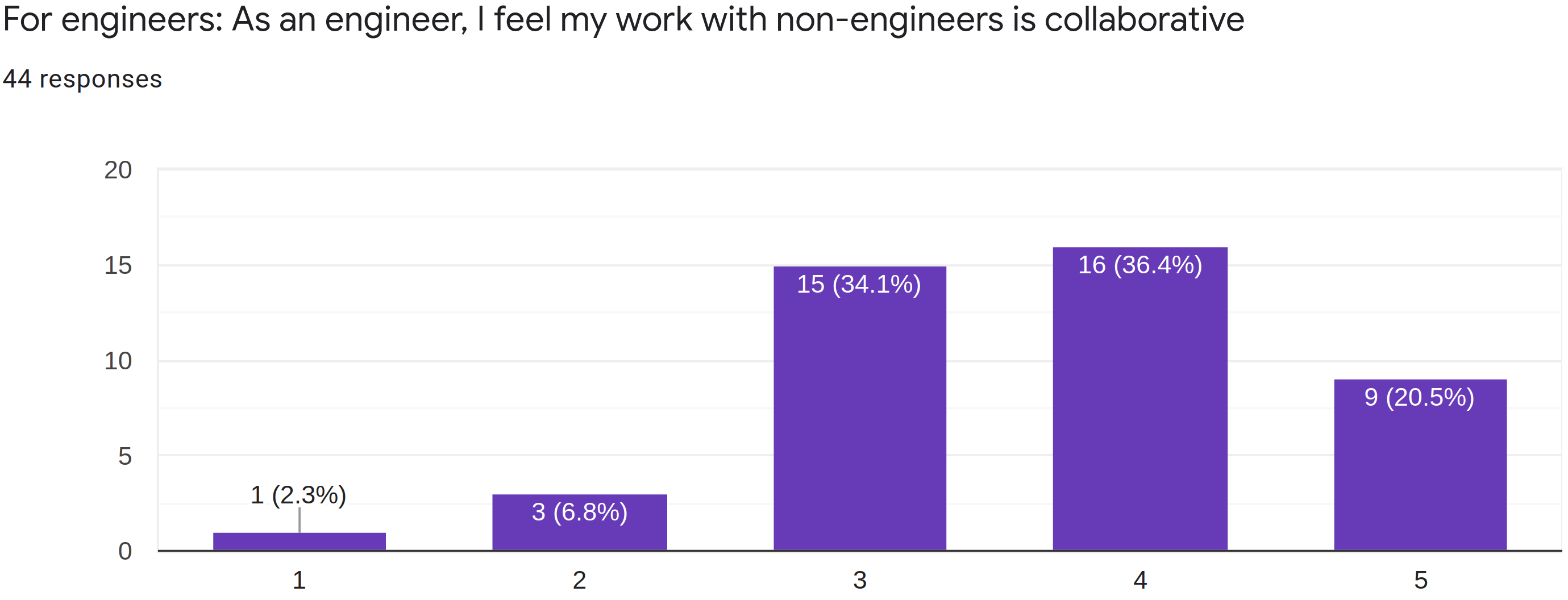}
    \caption{}
\end{figure}
\begin{figure}[h!]
    \centering
    \includegraphics[width=\textwidth]{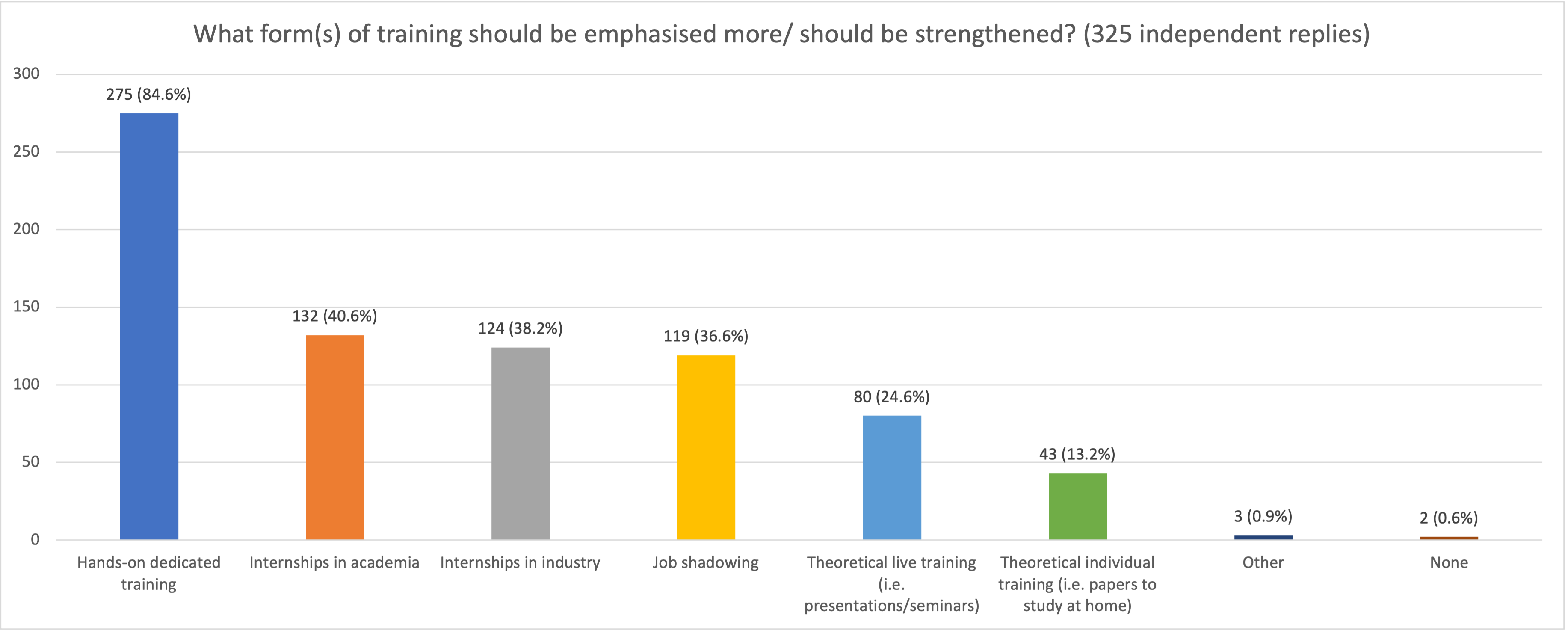}
    \caption{}
\end{figure}

\begin{figure}[h!]
    \centering
    \includegraphics[height=0.25\textheight,width=0.9\textwidth,keepaspectratio]{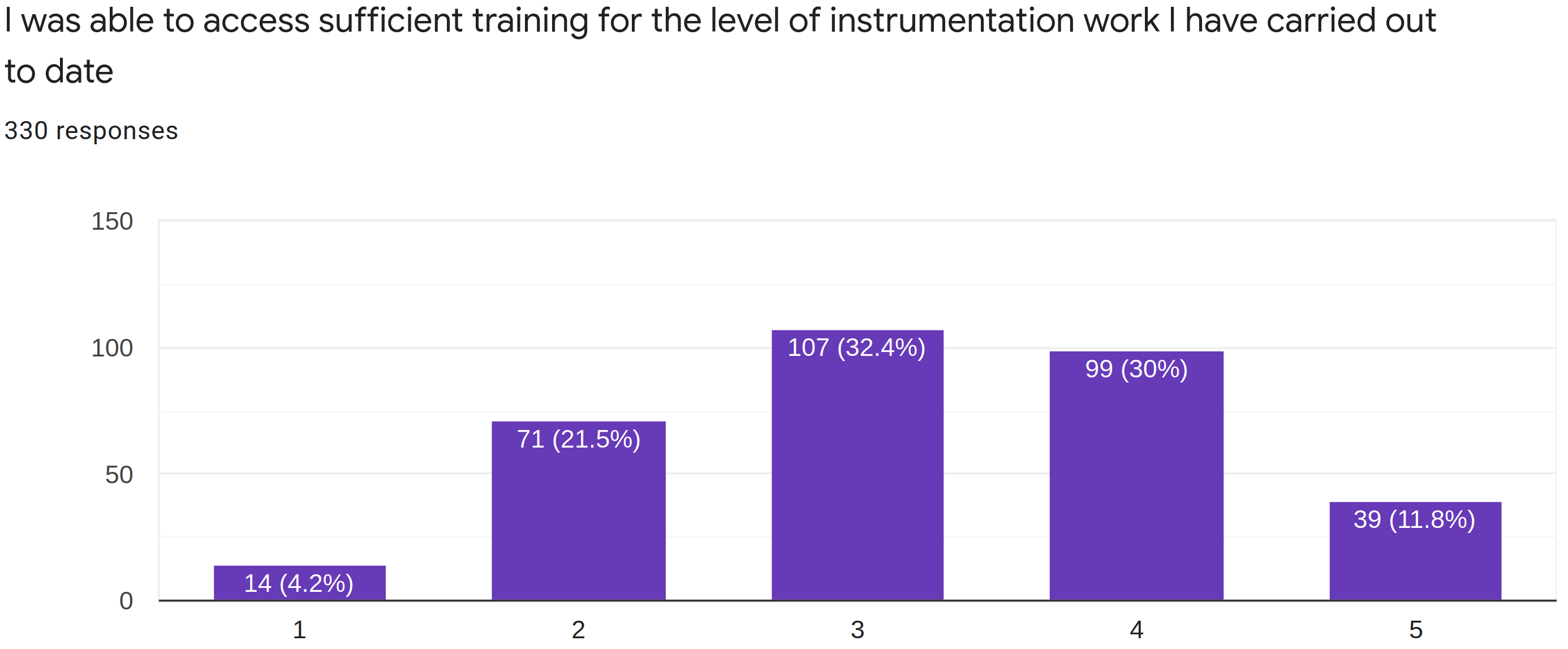}
    \caption{}
    \label{fig:access_to_training}
\end{figure}
\begin{figure}[h!]
    \centering
    \includegraphics[width=\textwidth]{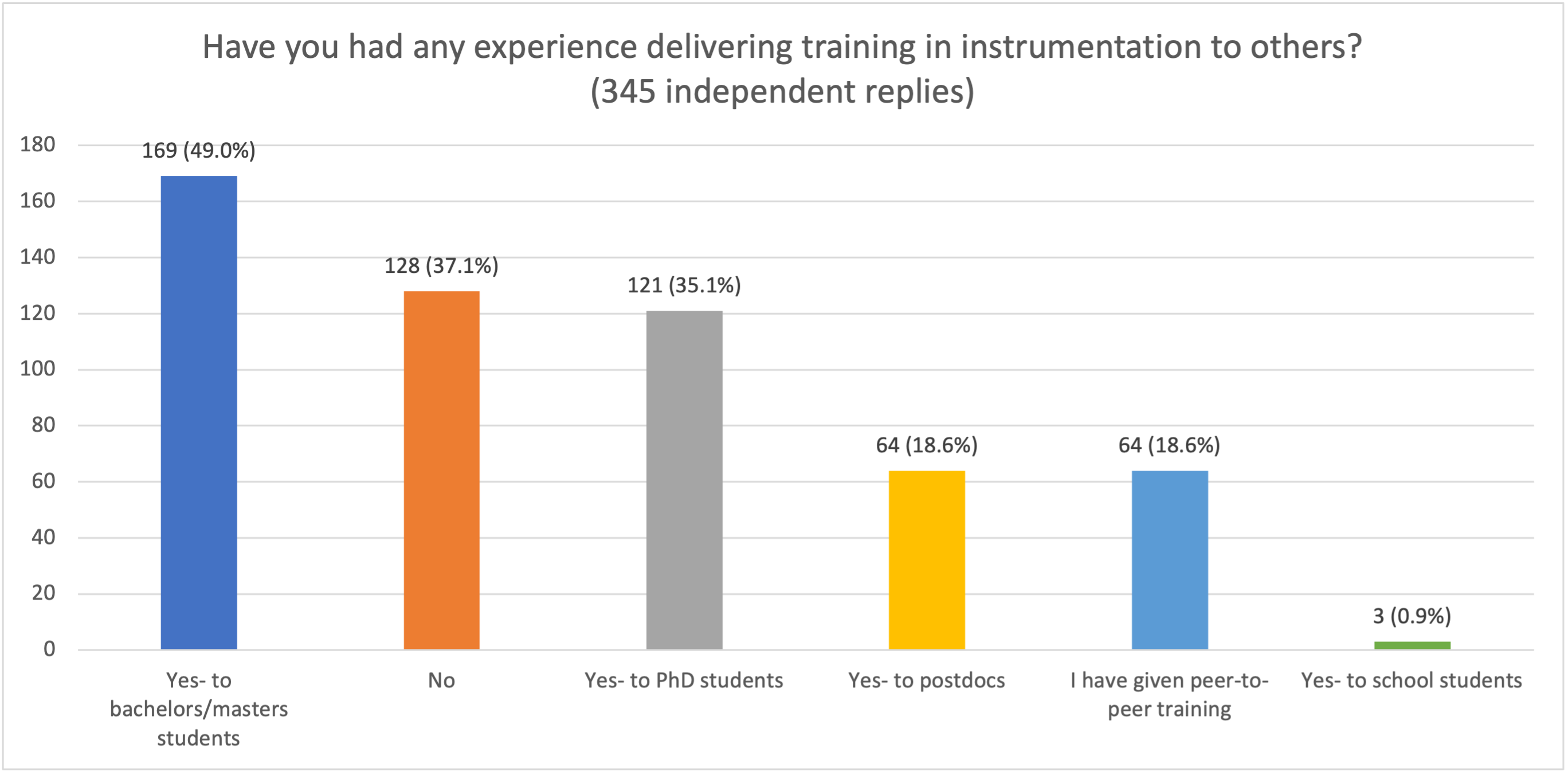}
    \caption{}
    \label{fig:sufficient}
\end{figure}
\begin{figure}[h!]
    \centering
    \includegraphics[height=0.25\textheight,width=0.9\textwidth,keepaspectratio]{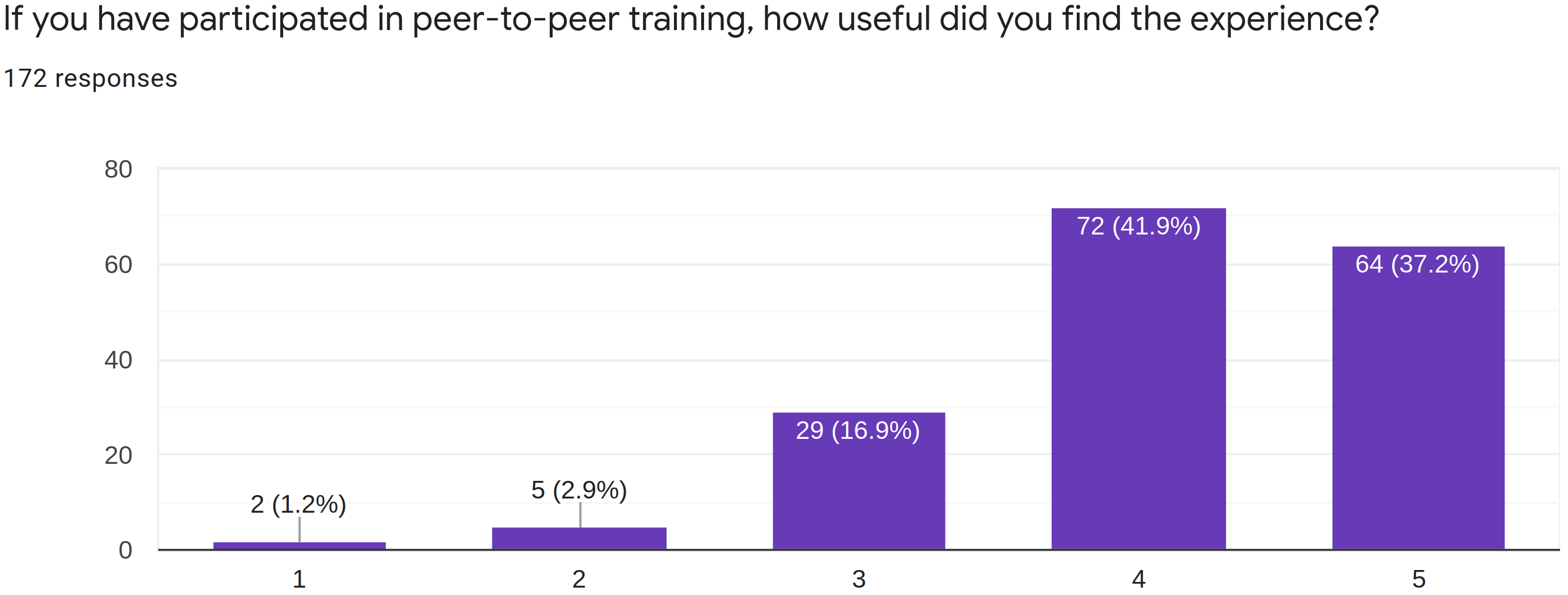}
    \caption{}
\end{figure}
\begin{figure}[h!]
    \centering
    \includegraphics[height=0.25\textheight,width=0.9\textwidth,keepaspectratio]{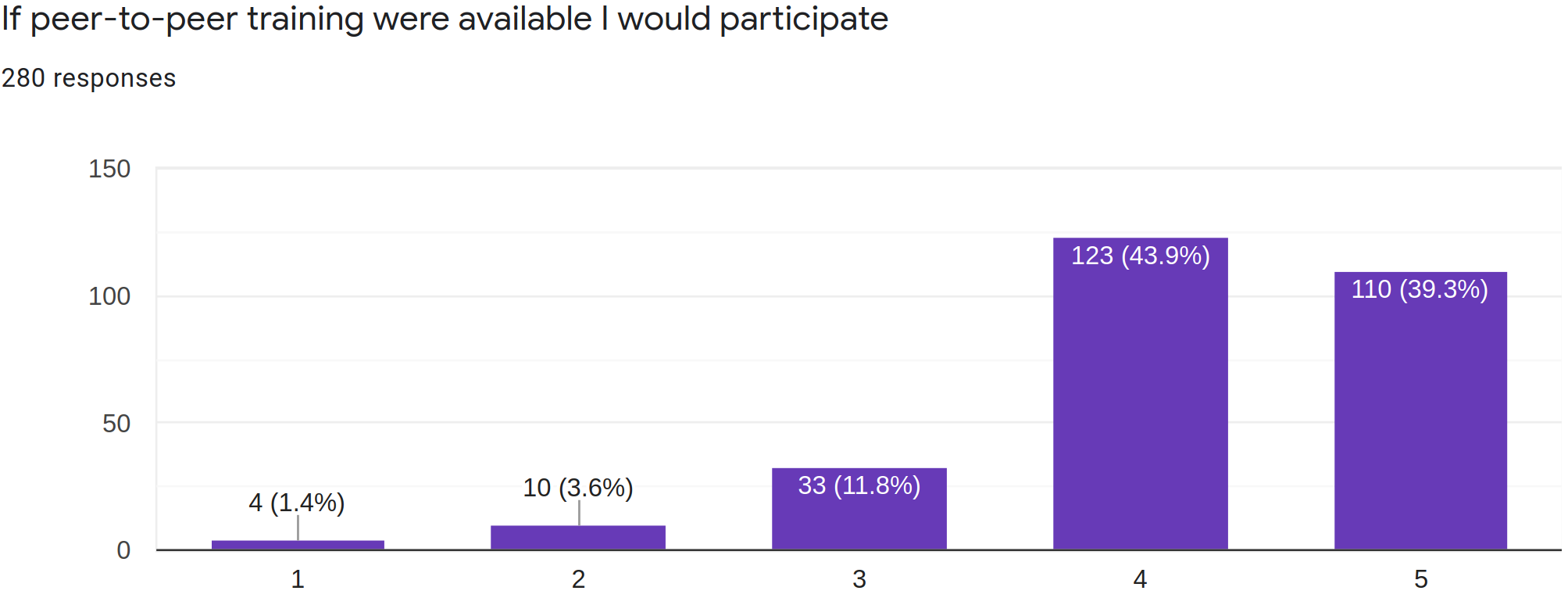}
    \caption{}
    \label{fig:peer}
\end{figure}

\begin{figure}[h!]
    \centering
    \includegraphics[height=0.25\textheight,width=0.9\textwidth,keepaspectratio]{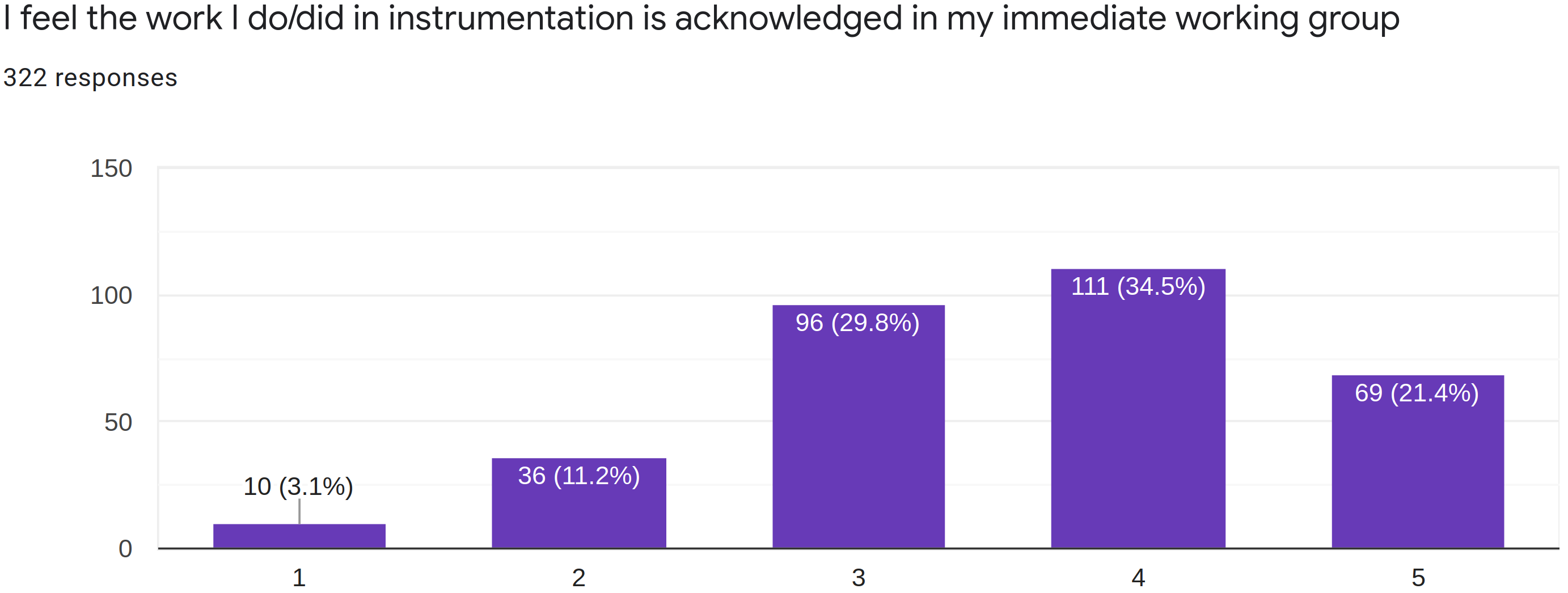}
    \caption{}
    \label{fig:app-ack-WG}
\end{figure}
\begin{figure}[h!]
    \centering
    \includegraphics[height=0.25\textheight,width=0.9\textwidth,keepaspectratio]{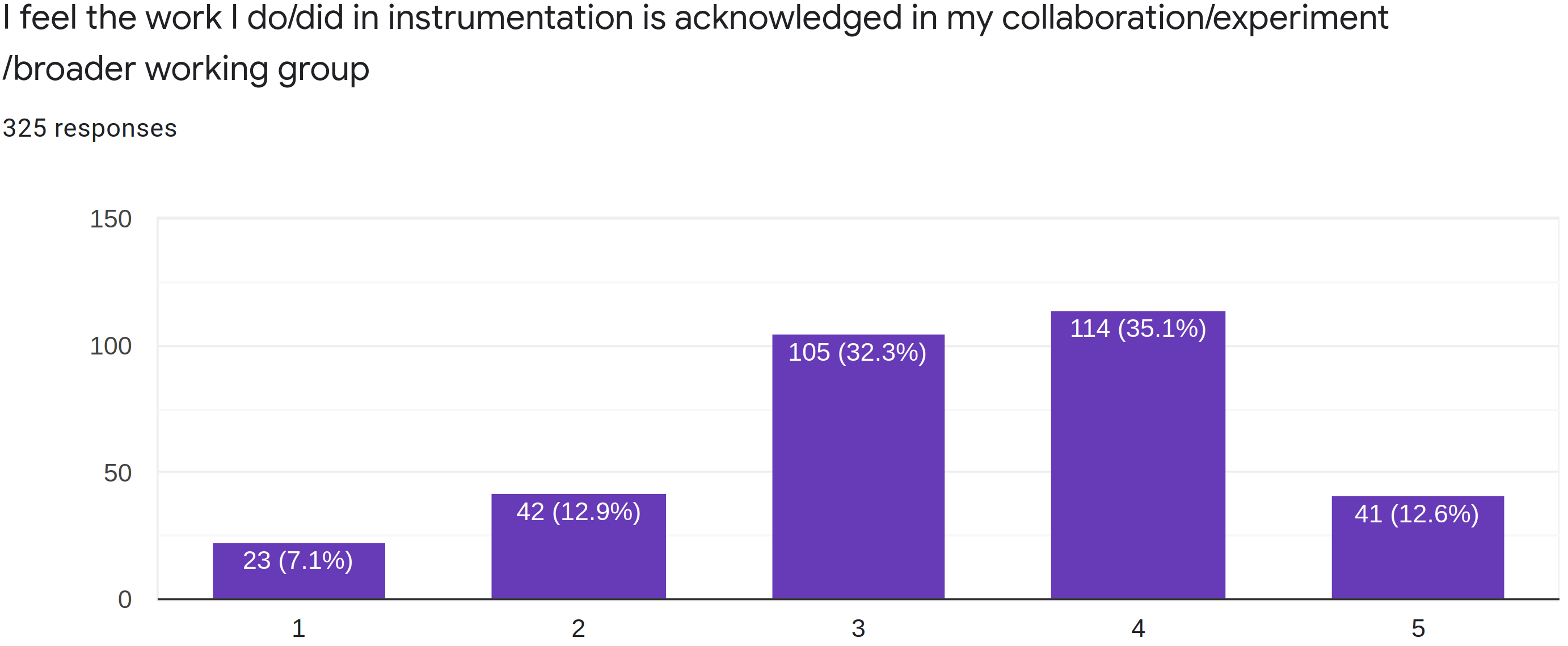}
    \caption{}
    \label{fig:app-ack-glob}
\end{figure}
\begin{figure}[h!]
    \centering
    \includegraphics[height=0.25\textheight,width=0.9\textwidth,keepaspectratio]{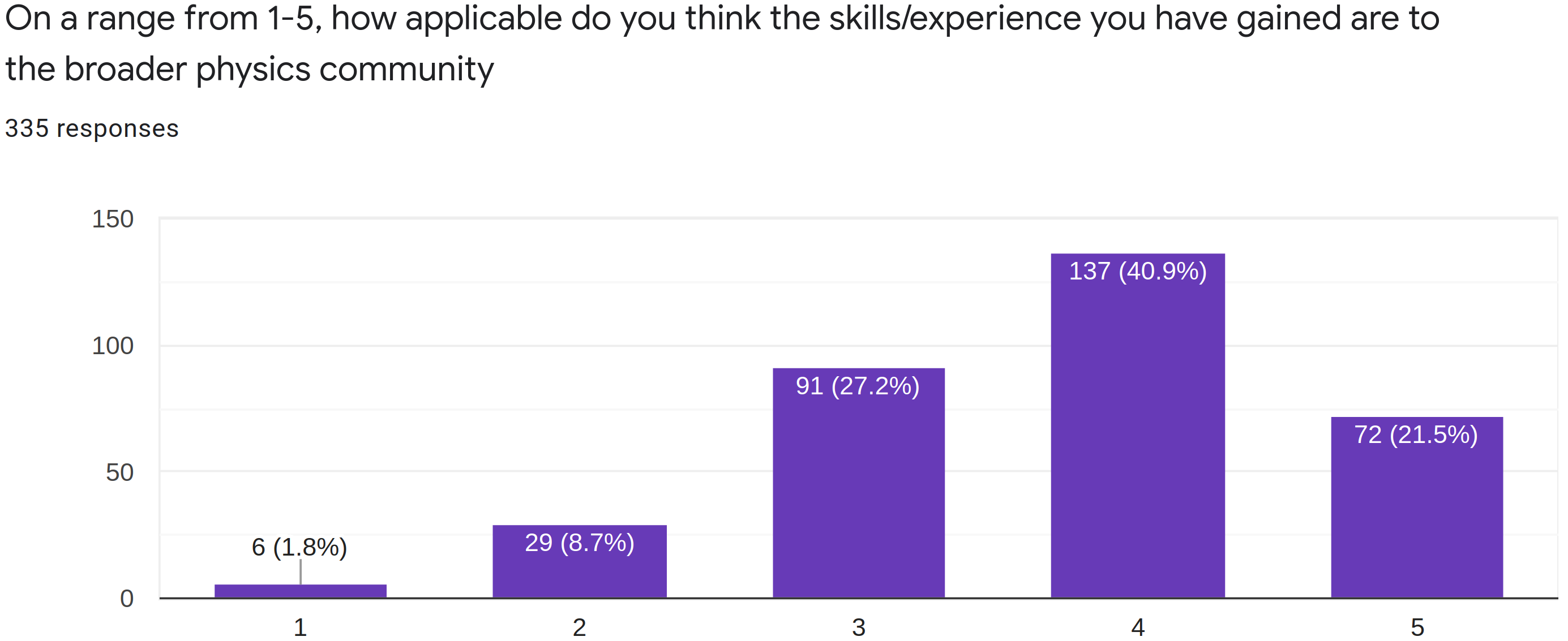}
    \caption{}
    \label{fig:application_to_broader_community}
\end{figure}

\begin{figure}[h!]
    \centering
    \includegraphics[height=0.25\textheight,width=0.9\textwidth,keepaspectratio]{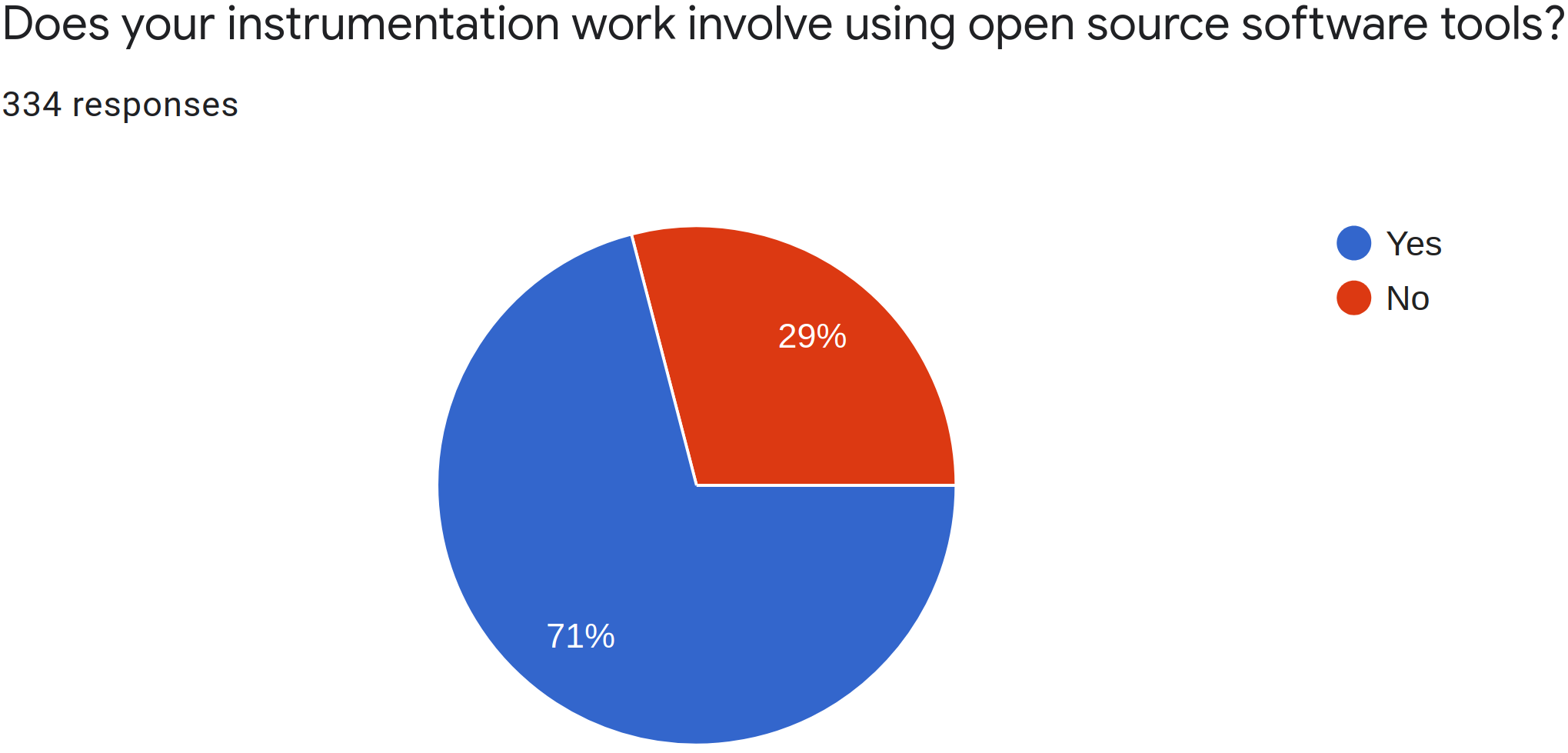}
    \caption{}
    \label{fig:open_source}
\end{figure}
\begin{figure}[h!]
    \centering
    \includegraphics[height=0.25\textheight,width=0.9\textwidth,keepaspectratio]{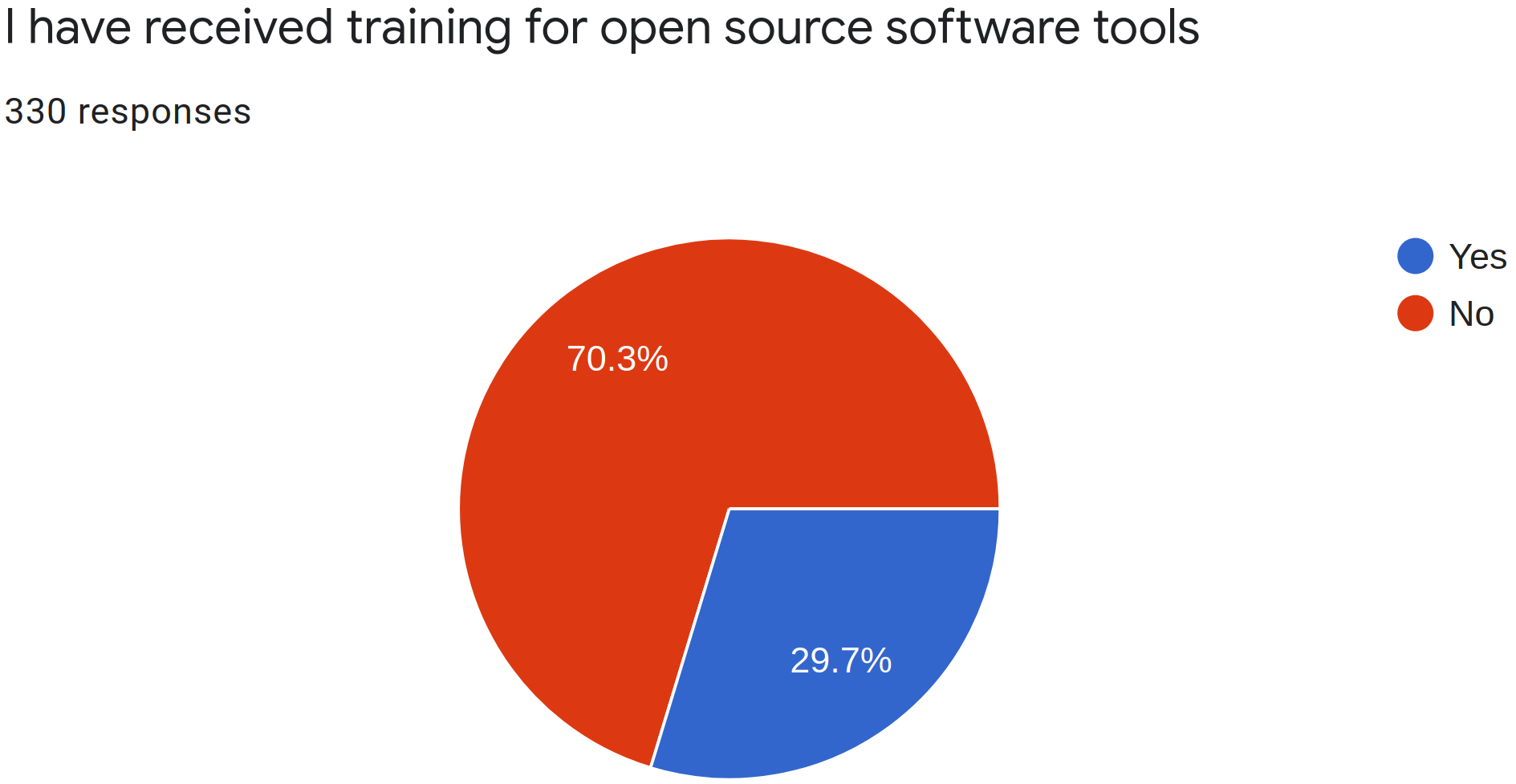}
    \caption{}
    \label{fig:open_source_training}
\end{figure}
\begin{figure}[h!]
    \centering
    \includegraphics[height=0.25\textheight,width=0.9\textwidth,keepaspectratio]{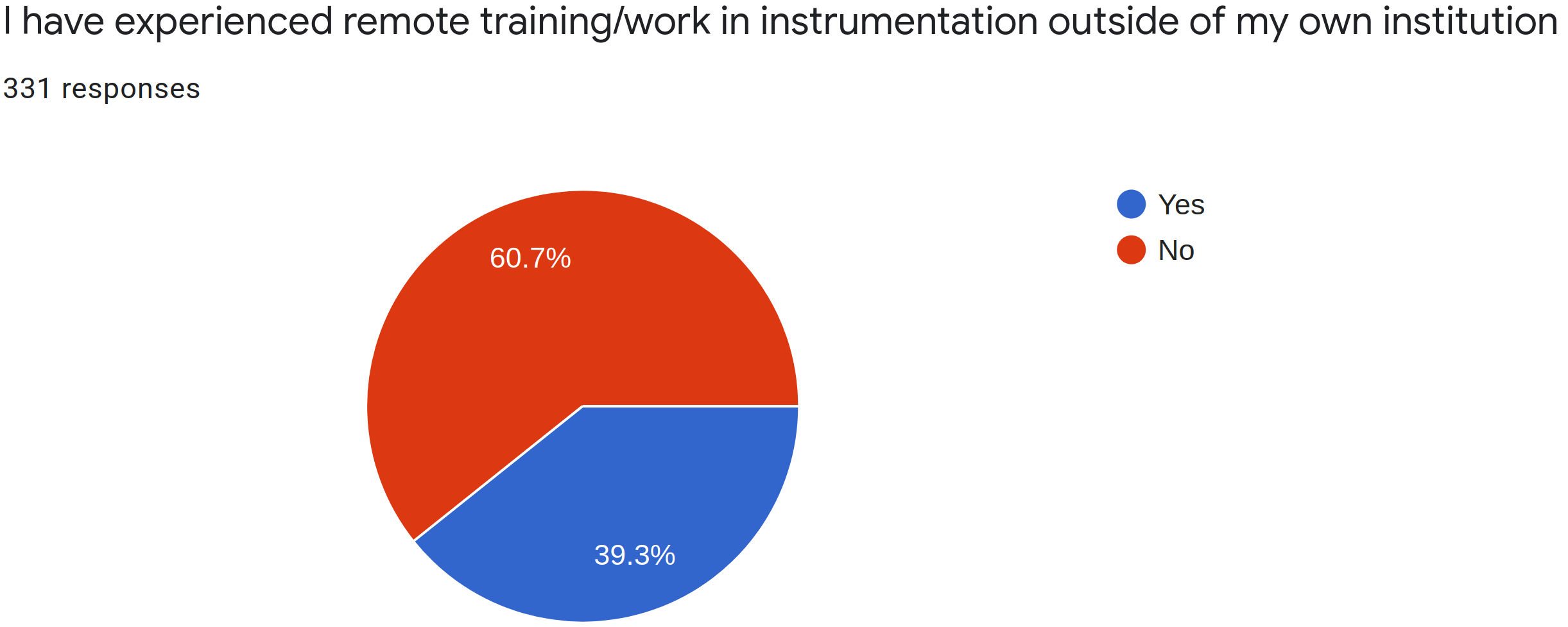}
    \caption{}
\end{figure}

\begin{figure}[h!]
    \centering
    \includegraphics[height=0.25\textheight,width=0.9\textwidth,keepaspectratio]{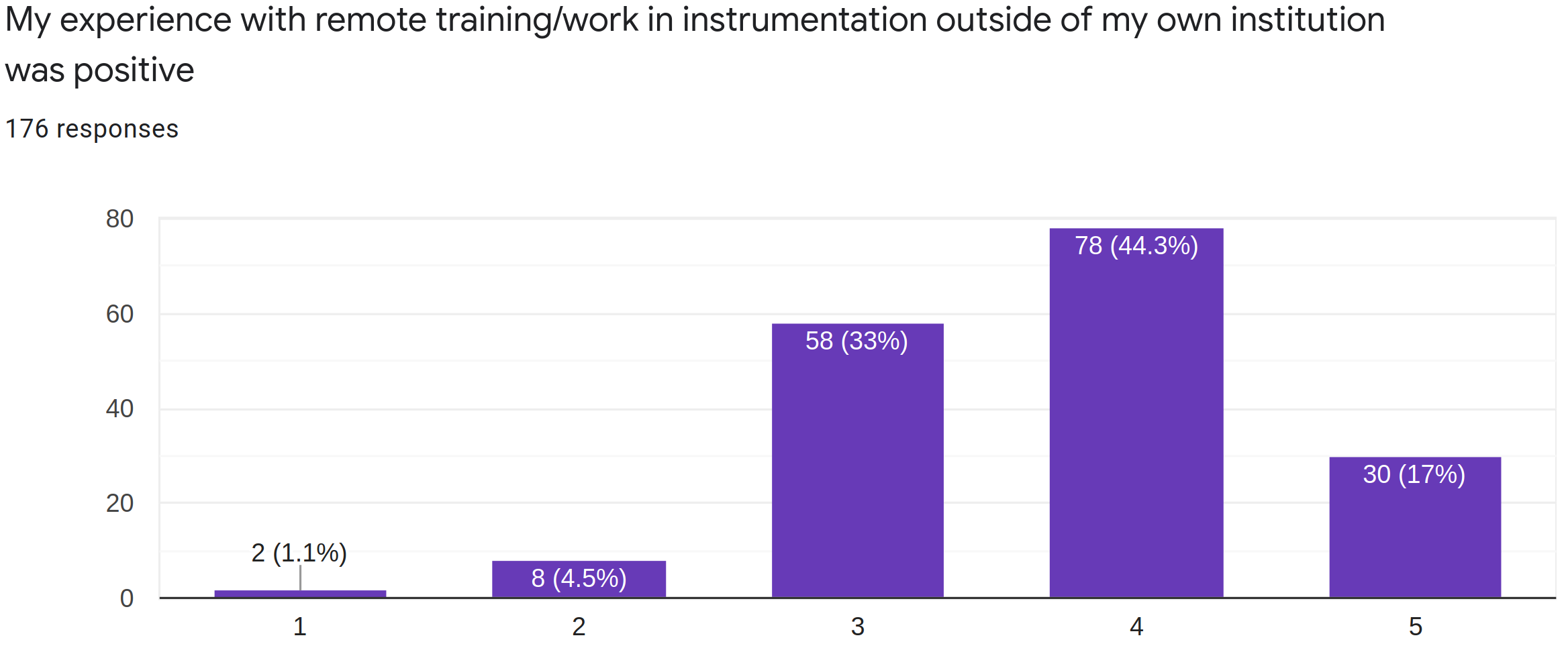}
    \caption{}
    \label{fig:remote}
\end{figure}
\begin{figure}[h!]
    \centering
    \includegraphics[height=0.25\textheight,width=0.9\textwidth,keepaspectratio]{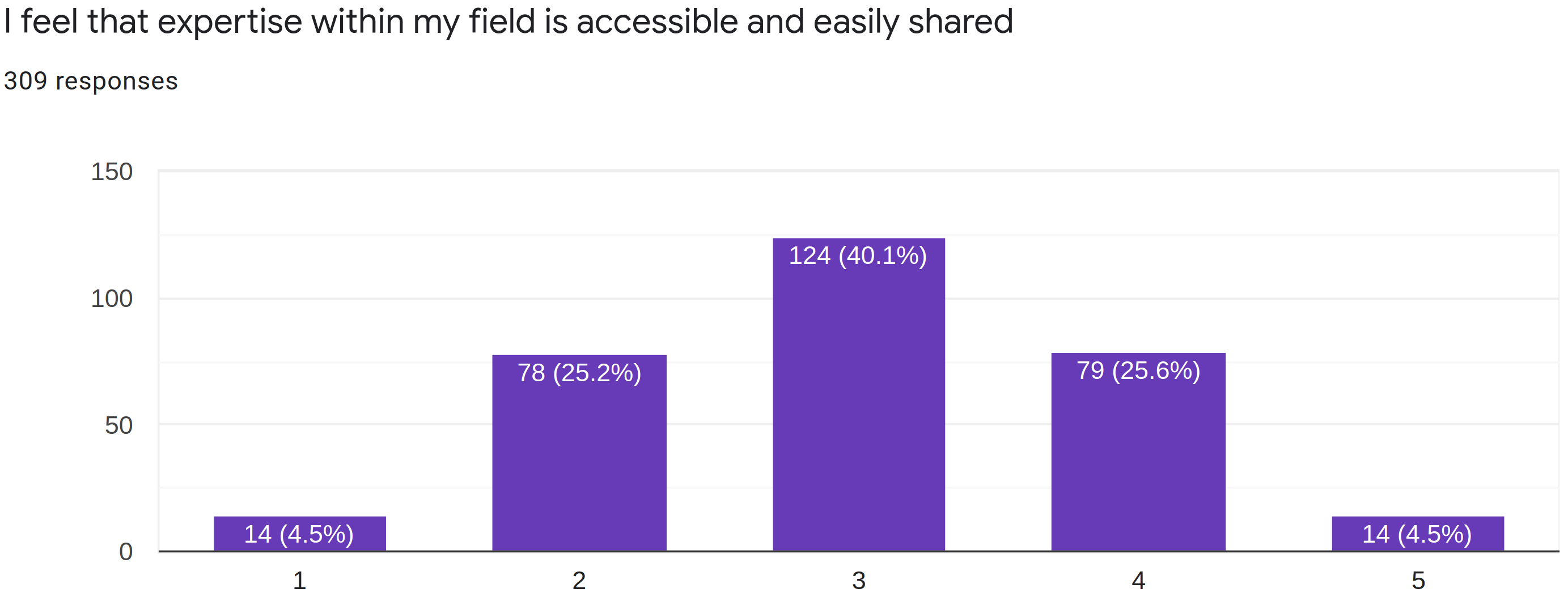}
    \caption{}
    \label{fig:expertise}
\end{figure}
\begin{figure}[h!]
    \centering
    \includegraphics[height=0.25\textheight,width=0.9\textwidth,keepaspectratio]{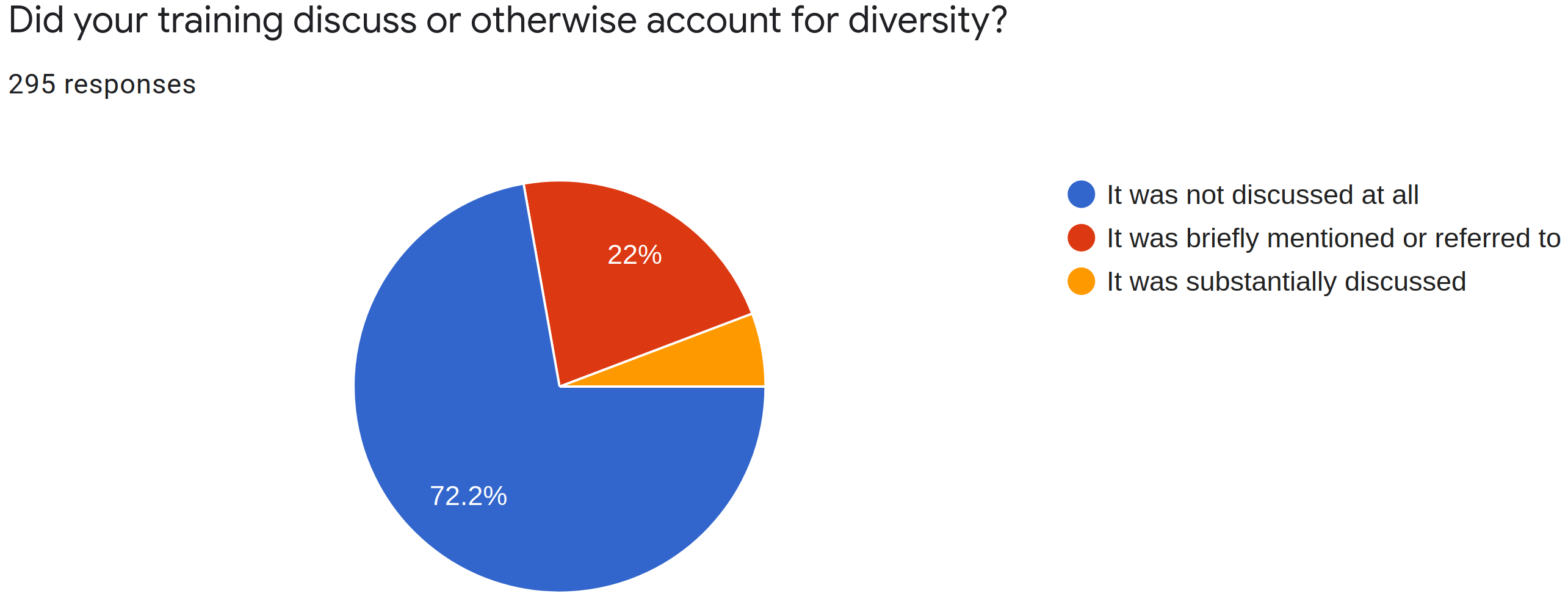}
    \caption{}
\end{figure}

\begin{figure}[h!]
    \centering
    \includegraphics[height=0.25\textheight,width=0.9\textwidth,keepaspectratio]{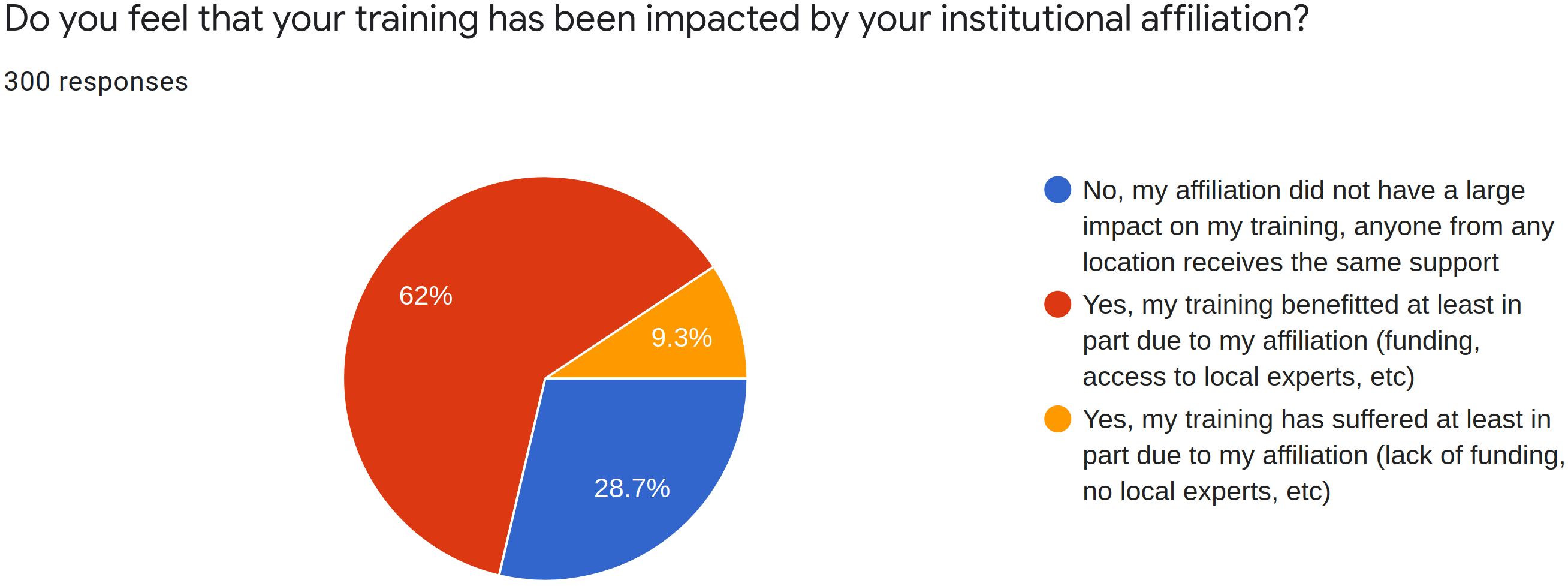}
    \caption{}
\end{figure}
\begin{figure}[h!]
    \centering
    \includegraphics[height=0.25\textheight,width=0.9\textwidth,keepaspectratio]{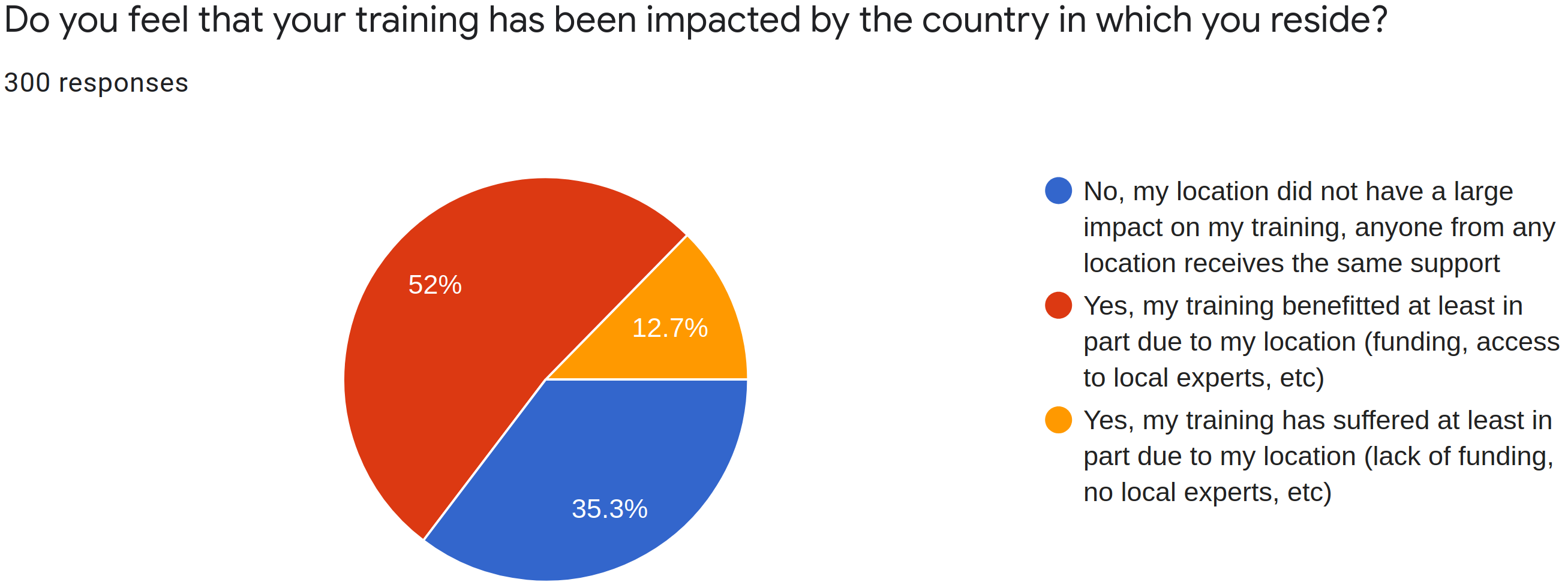}
    \caption{}
\end{figure}
\begin{figure}[h!]
    \centering
    \includegraphics[height=0.25\textheight,width=0.9\textwidth,keepaspectratio]{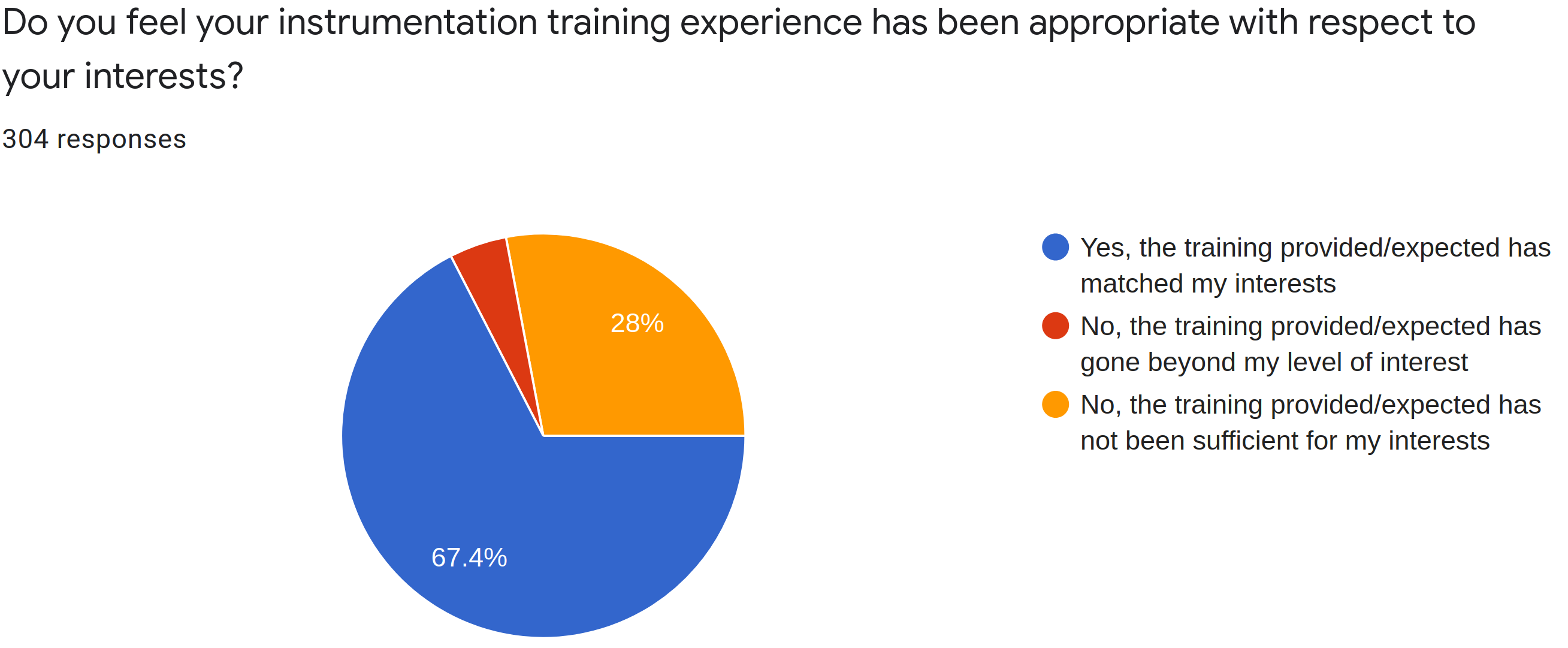}
    \caption{}
    \label{fig:training_experience_appropriate_to_interests}
\end{figure}

\begin{figure}[h!]
    \centering
    \includegraphics[height=0.25\textheight,width=0.9\textwidth,keepaspectratio]{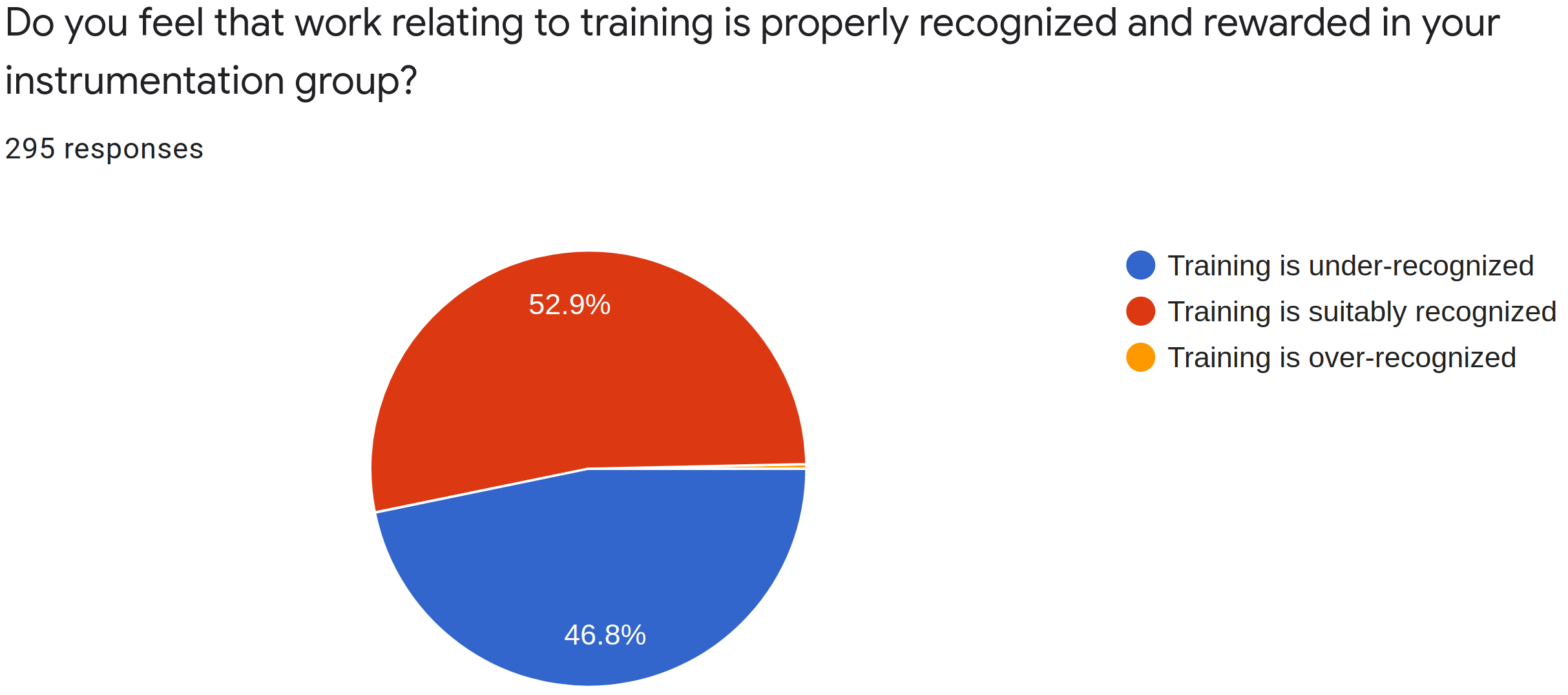}
    \caption{}
    \label{fig:recognized}
\end{figure}
\begin{figure}[h!]
    \centering
    \includegraphics[height=0.25\textheight,width=0.9\textwidth,keepaspectratio]{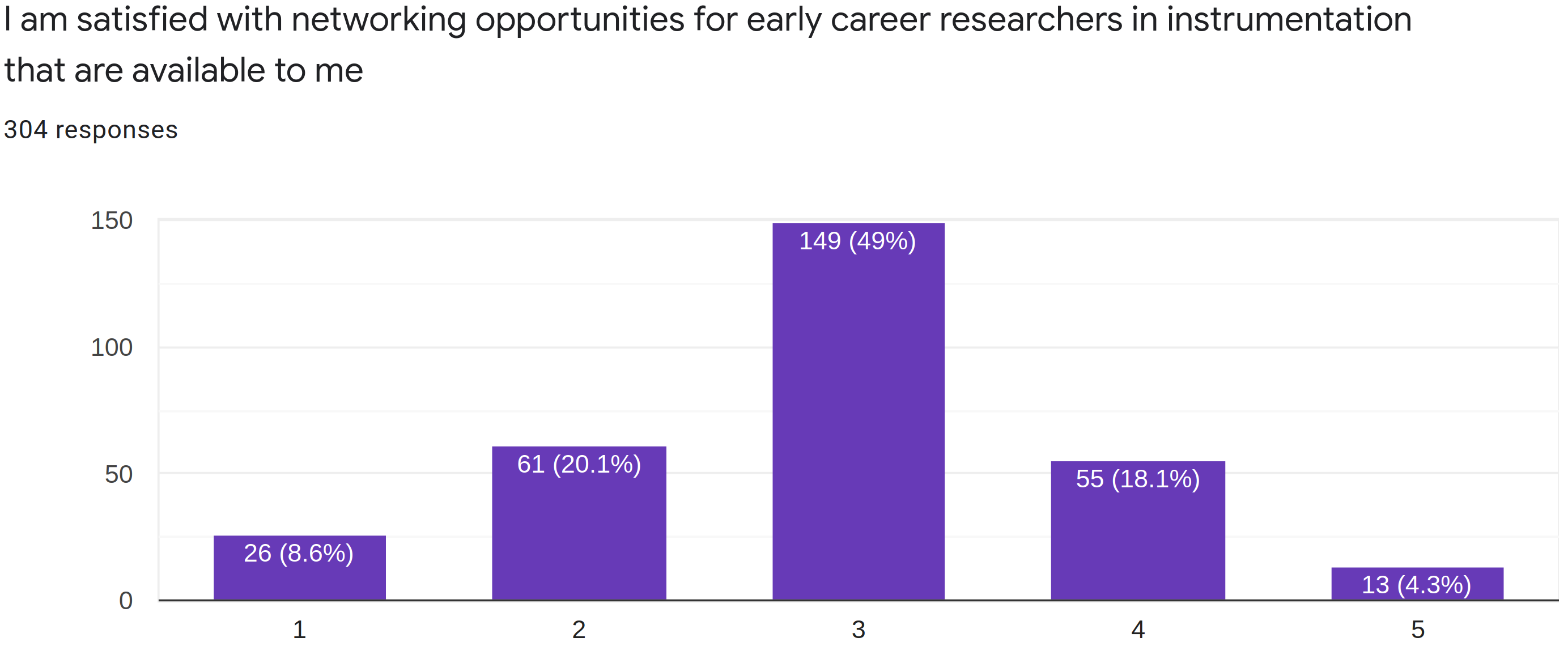}
    \caption{}
\end{figure}

\clearpage
\addcontentsline{toc}{section}{Appendix B}
\section*{Appendix B: Raw survey cross-analysis results}

The following plots are raw numbers from the cross-analysis of survey replies.  These are companion plots to those which are shown in the body of the report, showing the actual number of replies in each category instead of the fraction of replies.


\begin{figure}[h!]
\centering
\begin{subfigure}[b]{.45\linewidth}
\includegraphics[width=\linewidth,height=0.25\textheight,keepaspectratio]{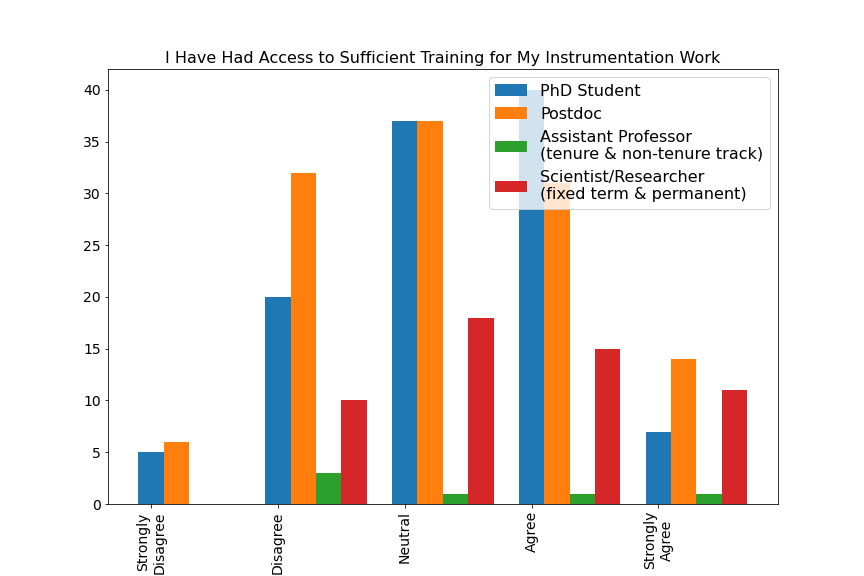}
\caption{}
\end{subfigure}
\begin{subfigure}[b]{.45\linewidth}
\includegraphics[width=\linewidth,height=0.25\textheight,keepaspectratio]{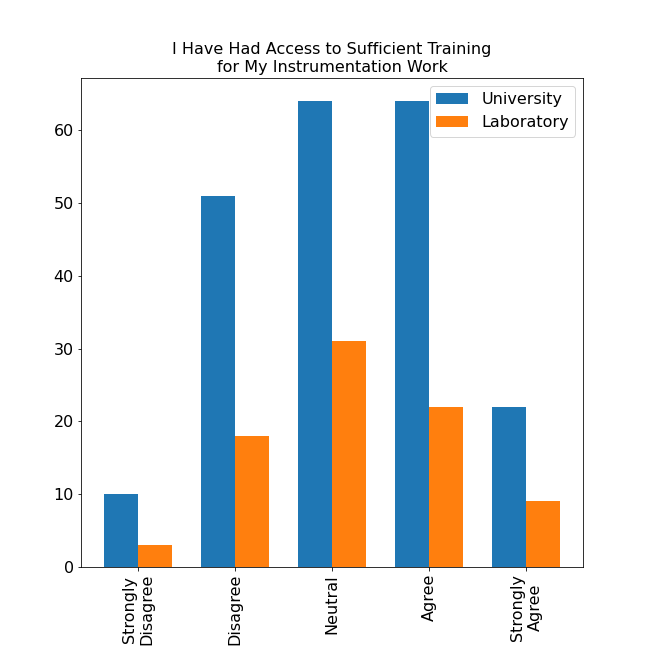}
\caption{}
\end{subfigure}
\caption{}
\label{fig:raw-begin}
\end{figure}

\begin{figure}[h!]
\centering
\begin{subfigure}[b]{.45\linewidth}
\includegraphics[width=\linewidth,height=0.25\textheight,keepaspectratio]{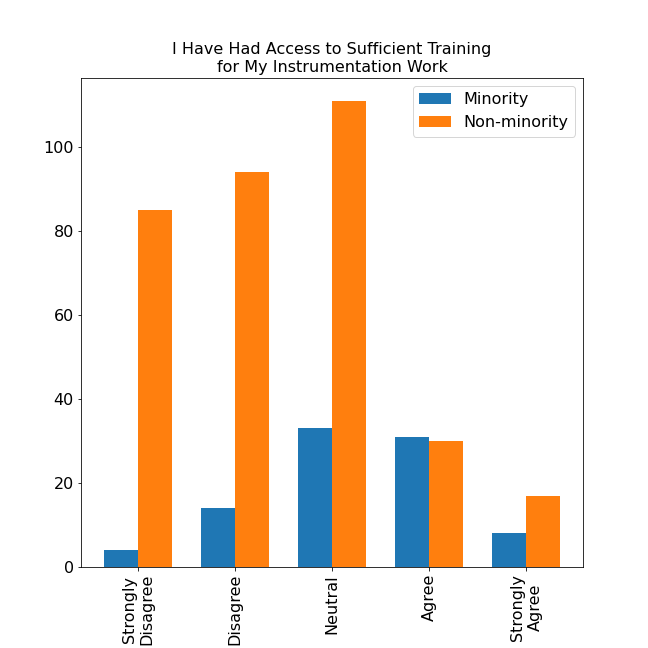}
\caption{}
\end{subfigure}
\begin{subfigure}[b]{.45\linewidth}
\includegraphics[width=\linewidth,height=0.25\textheight,keepaspectratio]{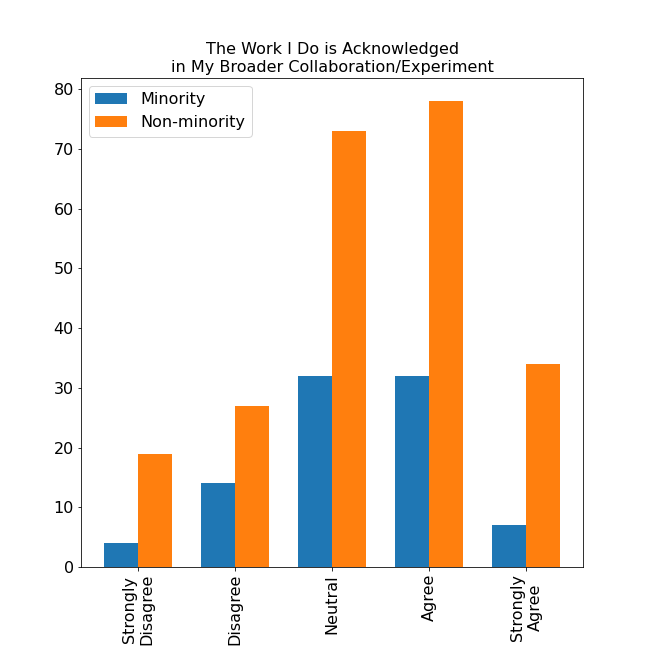}
\caption{}
\end{subfigure}
\caption{}
\end{figure}

\begin{figure}
\centering
\begin{subfigure}[b]{.45\linewidth}
\includegraphics[width=\linewidth,height=0.25\textheight,keepaspectratio]{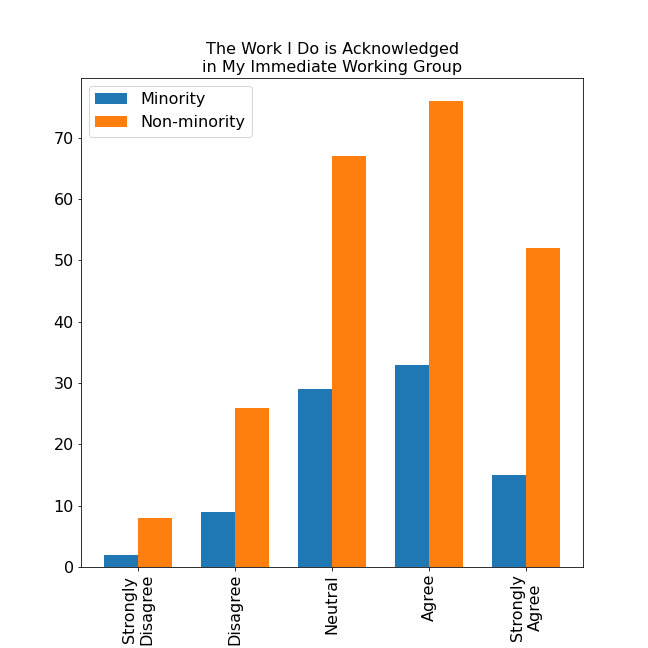}
\caption{}
\end{subfigure}
\begin{subfigure}[b]{.45\linewidth}
\includegraphics[width=\linewidth,height=0.25\textheight,keepaspectratio]{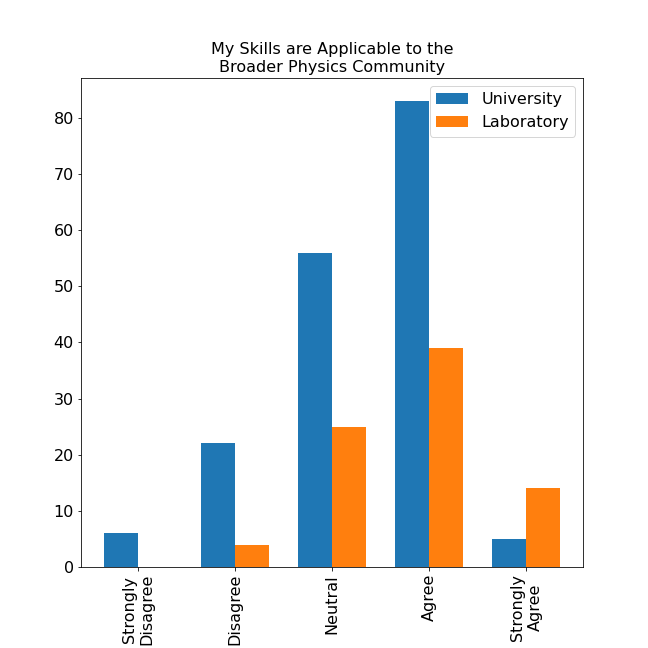}
\caption{}
\end{subfigure}
\caption{}
\end{figure}

\begin{figure}[h!]
\centering
\begin{subfigure}[b]{.45\linewidth}
\includegraphics[width=\linewidth,height=0.25\textheight,keepaspectratio]{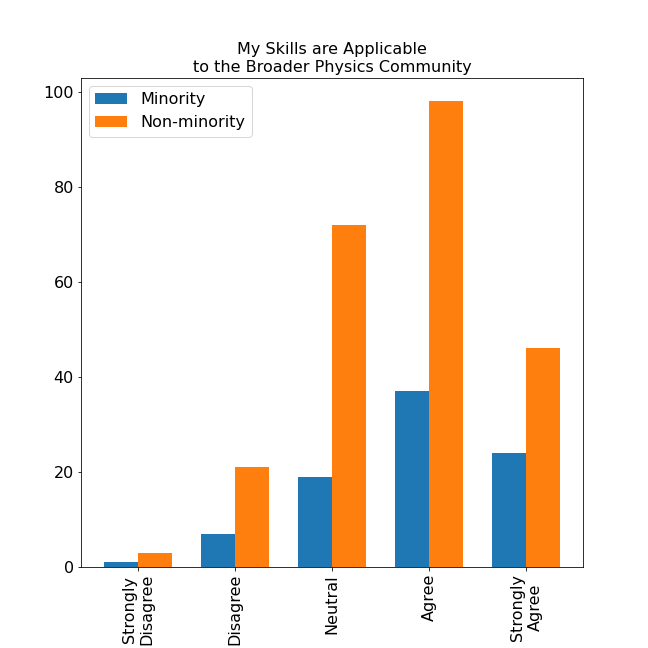}
\caption{}
\end{subfigure}
\begin{subfigure}[b]{.45\linewidth}
\includegraphics[width=\linewidth,height=0.25\textheight,keepaspectratio]{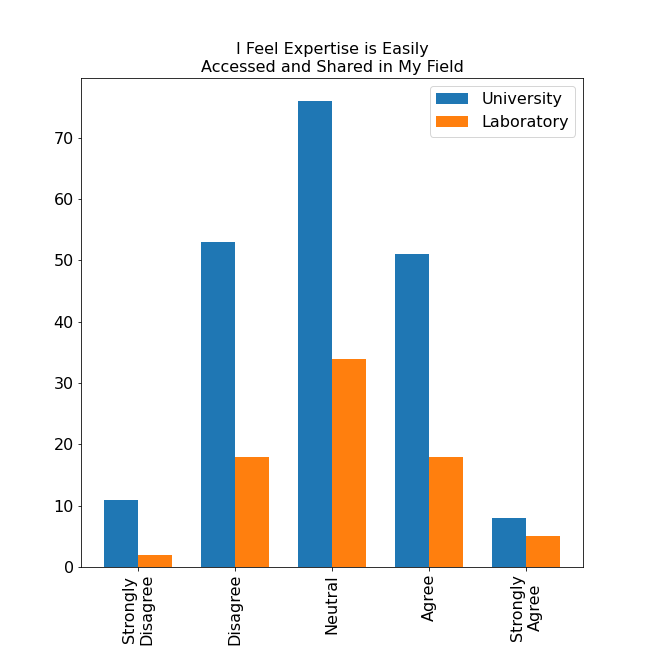}
\caption{}
\end{subfigure}
\caption{}
\end{figure}

\begin{figure}[h!]
\centering
\begin{subfigure}[b]{.45\linewidth}
\includegraphics[width=\linewidth,height=0.25\textheight,keepaspectratio]{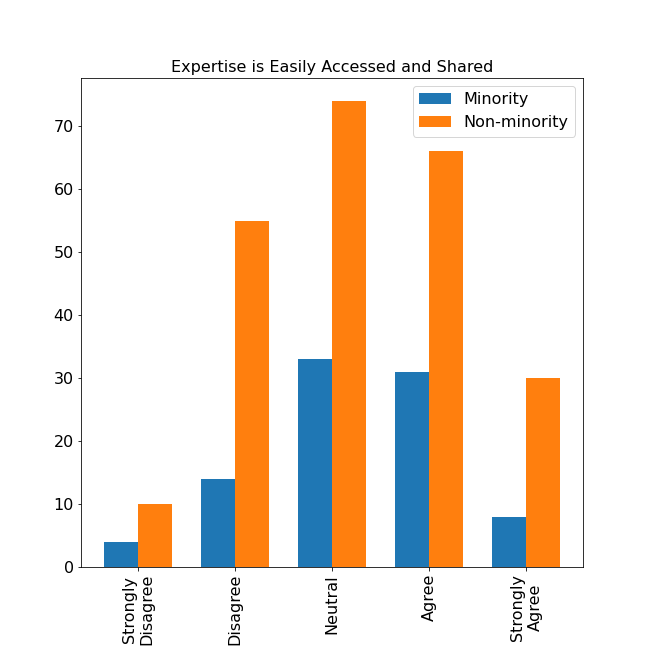}
\caption{}
\end{subfigure}
\begin{subfigure}[b]{.45\linewidth}
\includegraphics[width=\linewidth,height=0.25\textheight,keepaspectratio]{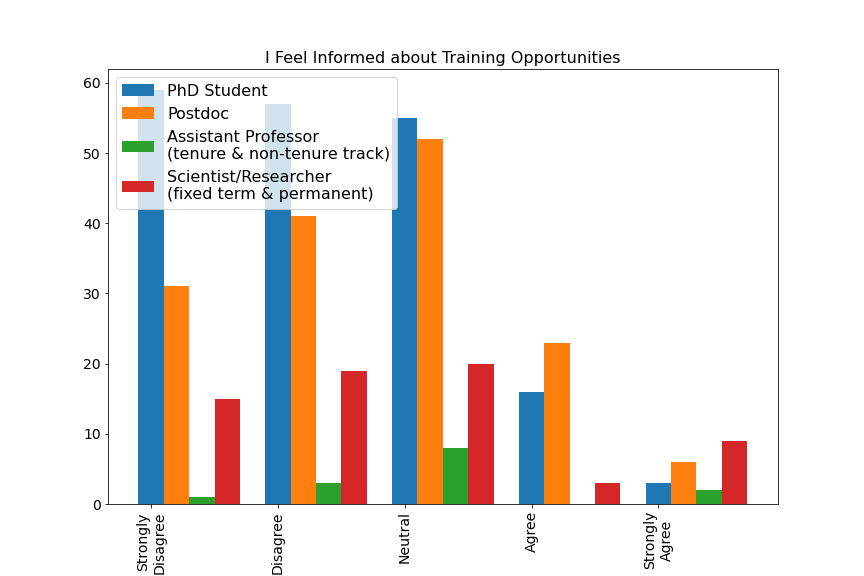}
\caption{}
\end{subfigure}
\caption{}
\end{figure}

\begin{figure}[h!]
\centering
\begin{subfigure}[b]{.45\linewidth}
\includegraphics[width=\linewidth,height=0.25\textheight,keepaspectratio]{RawCrossAna/informed-career-raw.png}
\caption{}
\end{subfigure}
\begin{subfigure}[b]{.45\linewidth}
\includegraphics[width=\linewidth,height=0.25\textheight,keepaspectratio]{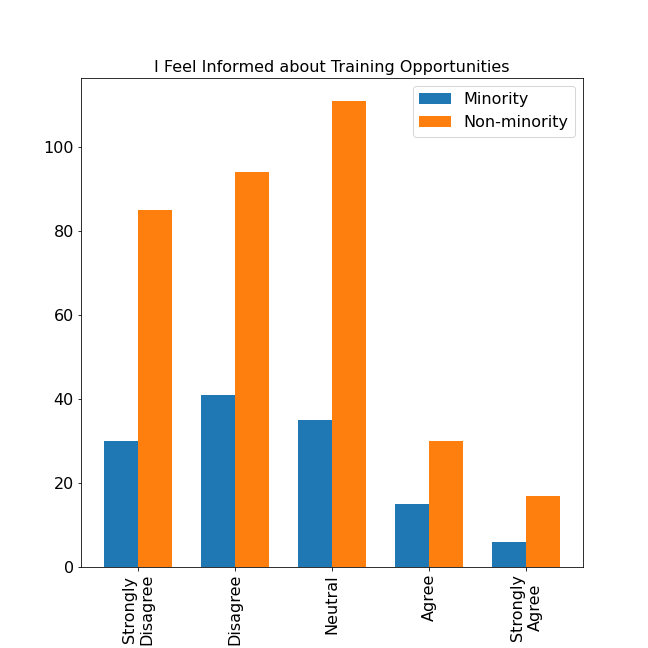}
\caption{}
\end{subfigure}
\caption{}
\end{figure}

\begin{figure}[h!]
\centering
\begin{subfigure}[b]{.45\linewidth}
\includegraphics[width=\linewidth,height=0.25\textheight,keepaspectratio]{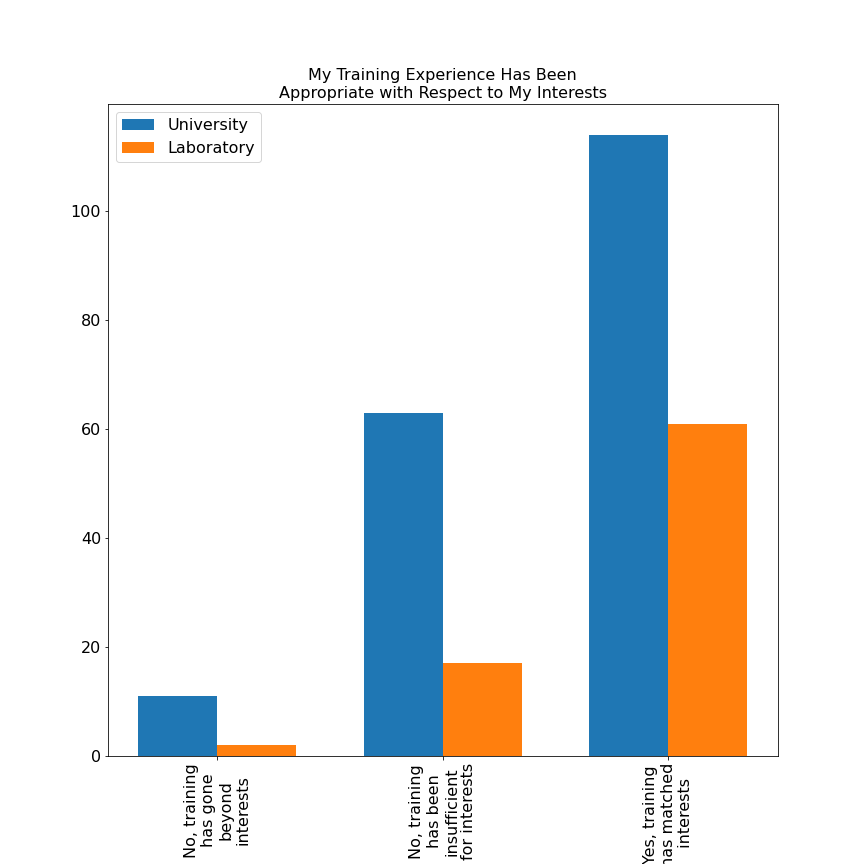}
\caption{}
\end{subfigure}
\begin{subfigure}[b]{.45\linewidth}
\includegraphics[width=\linewidth,height=0.25\textheight,keepaspectratio]{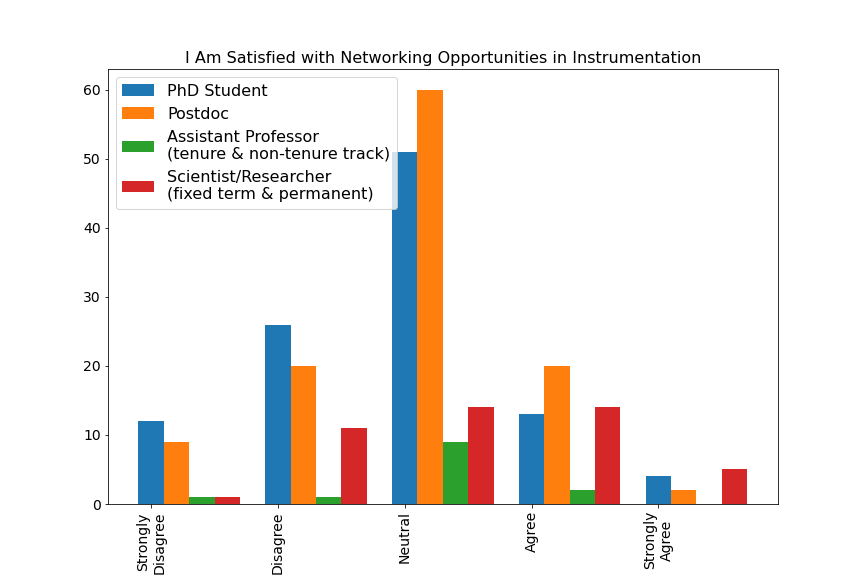}
\caption{}
\end{subfigure}
\caption{}
\end{figure}

\begin{figure}[h!]
\centering
\begin{subfigure}[b]{.45\linewidth}
\includegraphics[width=\linewidth,height=0.25\textheight,keepaspectratio]{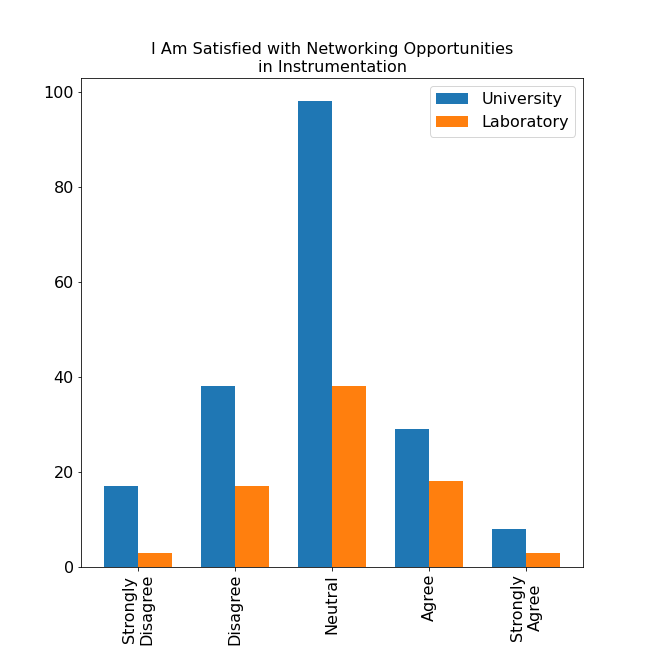}
\caption{}
\end{subfigure}
\begin{subfigure}[b]{.45\linewidth}
\includegraphics[width=\linewidth,height=0.25\textheight,keepaspectratio]{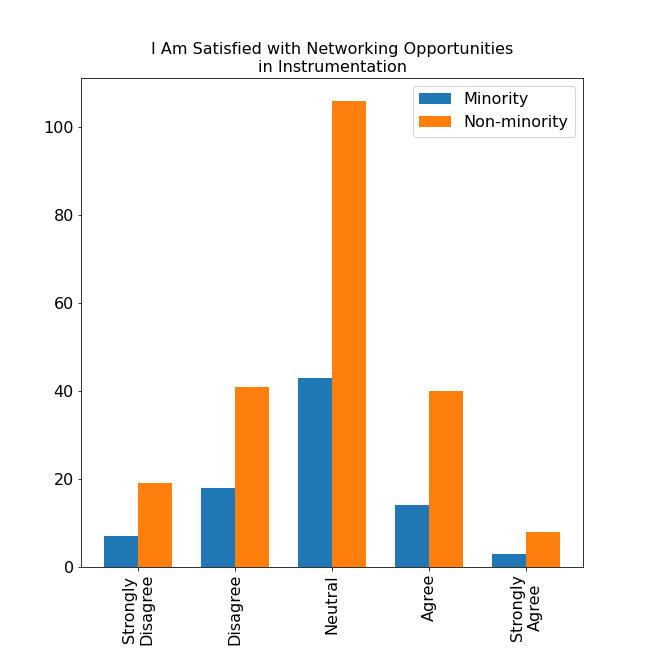}
\caption{}
\end{subfigure}
\caption{}
\end{figure}

\begin{figure}[h!]
\centering
\begin{subfigure}[b]{.45\linewidth}
\includegraphics[height=0.25\textheight,width=\linewidth,keepaspectratio]{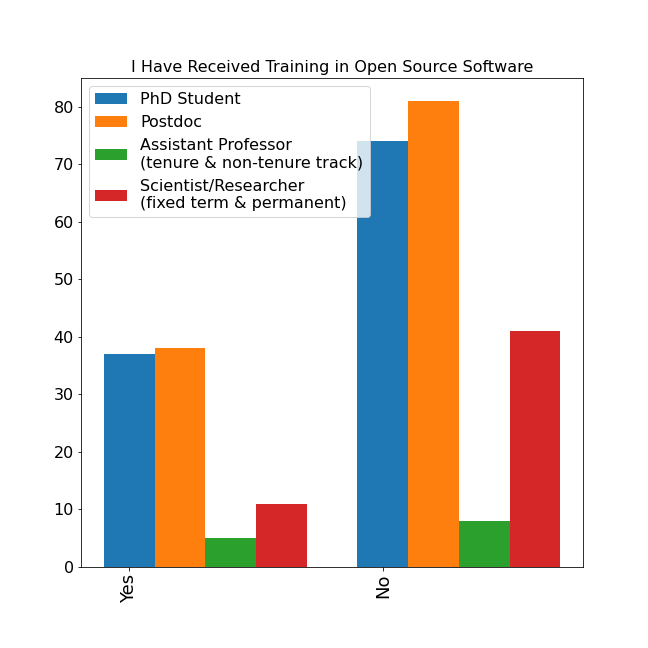}
\caption{}
\end{subfigure}
\begin{subfigure}[b]{.45\linewidth}
\includegraphics[height=0.25\textheight,width=\linewidth,keepaspectratio]{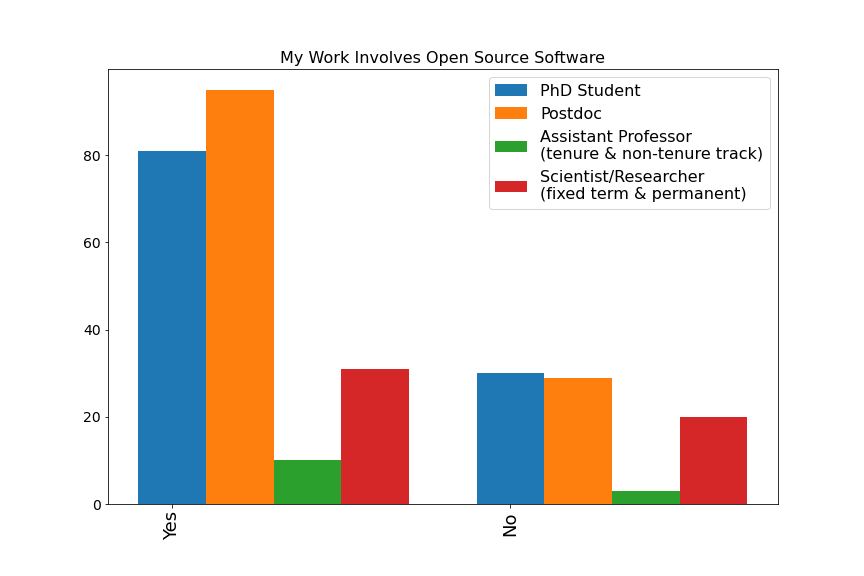}
\caption{}
\end{subfigure}
\caption{}
\end{figure}

\begin{figure}[h!]
\centering
\begin{subfigure}[b]{.45\linewidth}
\includegraphics[height=0.25\textheight,width=\linewidth,keepaspectratio]{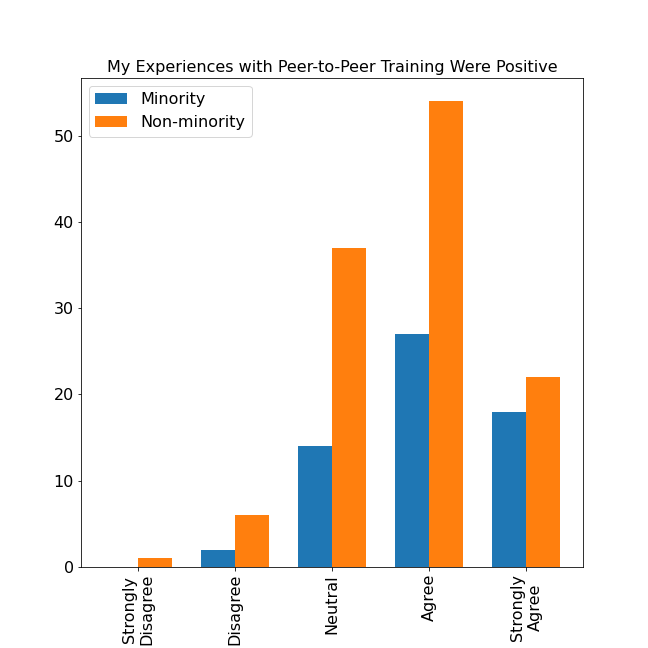}
\caption{}
\end{subfigure}
\begin{subfigure}[b]{.45\linewidth}
\includegraphics[height=0.25\textheight,width=\linewidth,keepaspectratio]{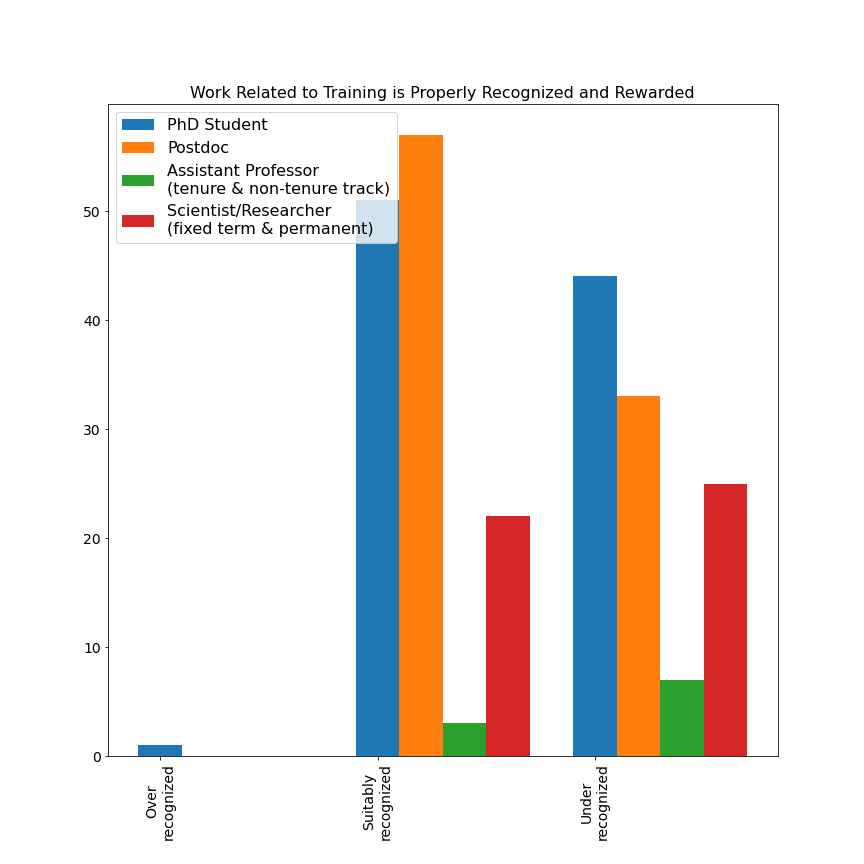}
\caption{}
\end{subfigure}
\caption{}
\end{figure}

\begin{figure}[h!]
\centering
\begin{subfigure}[b]{.45\linewidth}
\includegraphics[height=0.25\textheight,width=\linewidth,keepaspectratio]{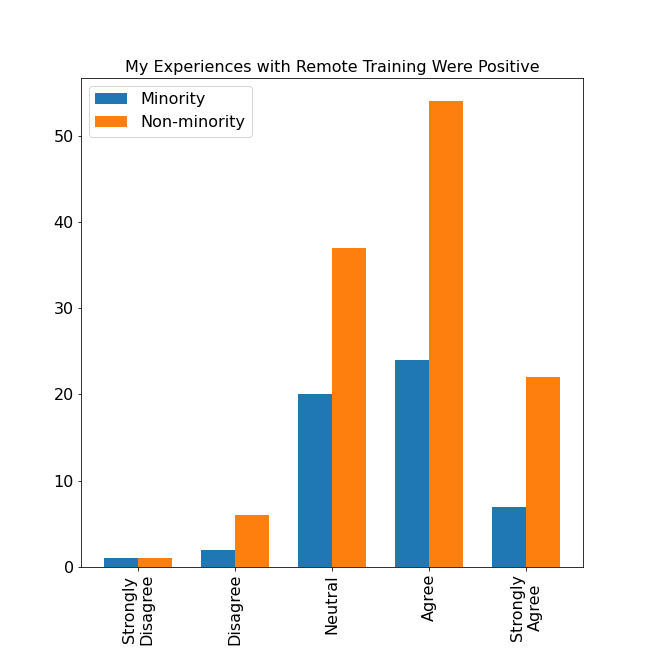}
\caption{}
\end{subfigure}
\begin{subfigure}[b]{.45\linewidth}
\includegraphics[height=0.25\textheight,width=\linewidth,keepaspectratio]{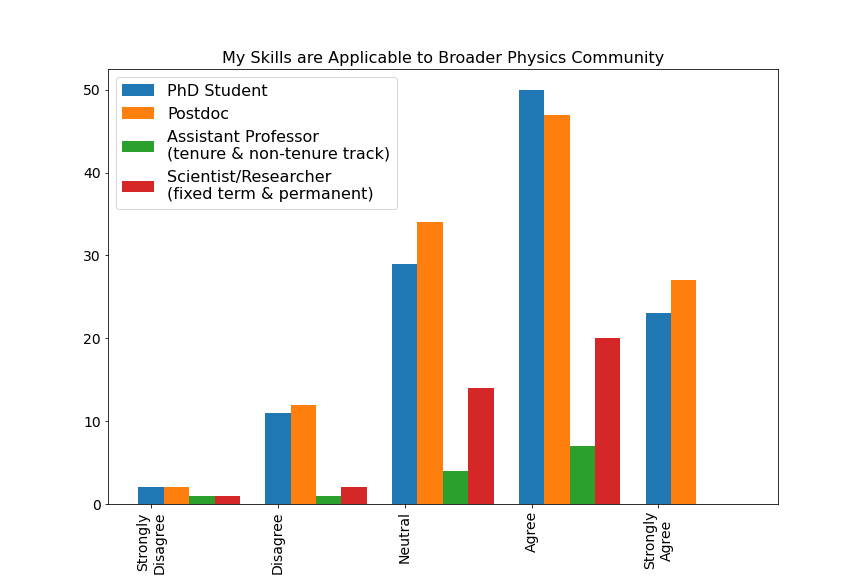}
\caption{}
\end{subfigure}
\caption{}
\end{figure}

\begin{figure}[h!]
\centering
\begin{subfigure}[b]{.45\linewidth}
\includegraphics[width=\linewidth,height=0.25\textheight,keepaspectratio]{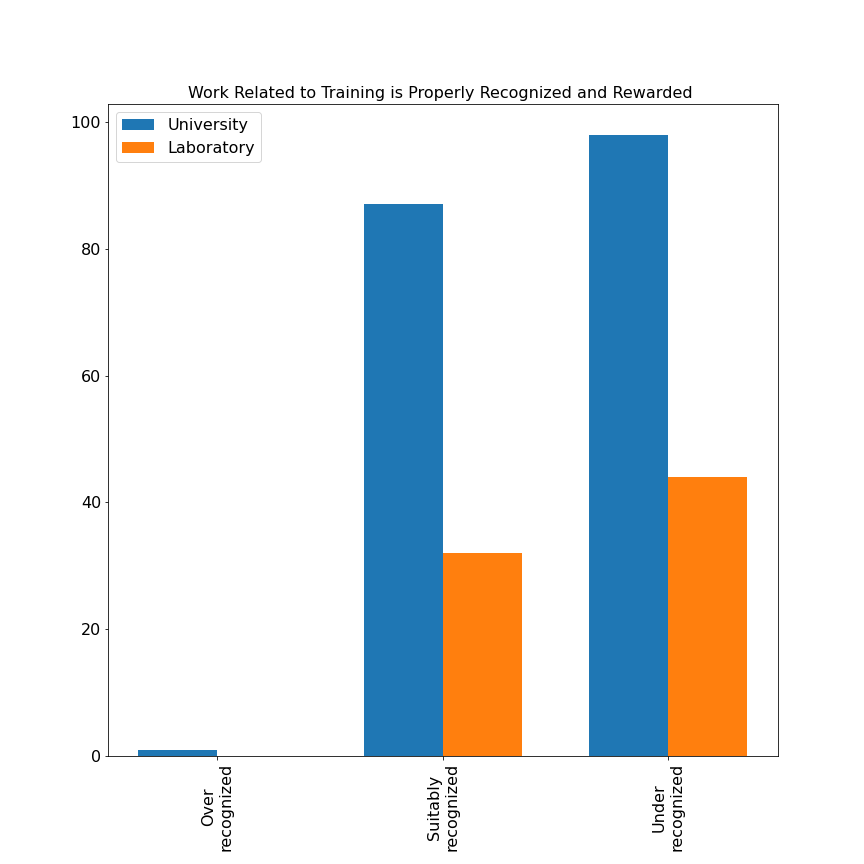}
\label{fig:raw-end}
\end{subfigure}
\caption{}
\end{figure}

\end{document}